\documentclass[11pt,a4paper]{article}
\pdfoutput=1
\usepackage{jheppub}
\usepackage{braket}
\usepackage[english]{babel}
\usepackage[utf8]{inputenc}
\usepackage{amsmath}
\usepackage{array, url}
\usepackage{fancyhdr}
\usepackage{amsfonts}
\usepackage{amsthm}
\usepackage{mathtools}
\usepackage{bm}
\usepackage{titlesec}
\usepackage{graphicx}
\usepackage{color}
\usepackage{tensor}
\usepackage{mathrsfs}
\usepackage{caption}
\usepackage{subcaption}
\usepackage{etoolbox}
\usepackage{tikz}
\usetikzlibrary{positioning}
\usepackage{comment}
\usepackage{tensor}
\usepackage[title]{appendix}
\usepackage{enumerate}

\numberwithin{equation}{section}

\usepackage{amsmath}

\DeclareMathOperator\tr{Tr}

\DeclareMathOperator\csch{csch}

\usepackage{xcolor}

\renewcommand*\thesection{\arabic{section}}
\definecolor{mathematica1}{rgb}{0.368417, 0.506779, 0.709798}
\definecolor{mathematica2}{rgb}{0.880722, 0.611041, 0.142051}
\definecolor{mathematica3}{rgb}{0.560181, 0.691569, 0.194885}
\definecolor{mathematica4}{rgb}{0.922526, 0.385626, 0.209179}
\definecolor{mathematica6}{rgb}{0.772079, 0.431554, 0.102387}
\definecolor{pink}{rgb}{1, 0.5, 0.5}

\newcommand{\IR}{\scriptscriptstyle\text{IR}}
\newcommand{\UV}{\scriptscriptstyle\text{UV}}

\newcommand{\be}{\begin{equation}}
\newcommand{\ee}{\end{equation}}

\usepackage{tabu}

\title{Holographic confining theories on space-times with constant positive curvature}

\author[\flat]{Jani~Kastikainen,}
\author[\dagger,\sharp,*]{Elias~Kiritsis,}
\author[\dagger]{Francesco~Nitti}

\affiliation[\flat]{Institute for Theoretical Physics and Astrophysics and Würzburg-Dresden Cluster of Excellence
ct.qmat, Julius-Maximilians-Universität Würzburg, Am Hubland, 97074 Würzburg, Germany}
\affiliation[\dagger]{Université de Paris Cit\'e, CNRS, Astroparticule et Cosmologie, F-75013 Paris, France}
\affiliation[\sharp]{Crete Center for Theoretical Physics, Institute for Theoretical and
Computational Physics, Department of Physics
University of Crete, Heraklion, Greece}
\affiliation[*]{Arnold Sommerfeld Center for Theoretical Physics,
Ludwig-Maximilians-Universit\"at M\"unchen, 80333 M\"unchen, Germany}

\emailAdd{jani.kastikainen@uni-wuerzburg.de}
\emailAdd{kiritsis@apc.in2p3.fr}
\emailAdd{nitti@apc.in2p3.fr}

\preprint{
\begin{flushright}
CCTP-2025-2\\
ITCP-2025/2
\end{flushright}
}

\abstract{Varying the curvature, quantum phase  transitions are investigated in holographic confining  QFTs defined on a fixed constant positive curvature background. We find a competition between two branches of solutions and a phase transition as one varies the space-time curvature. The low-curvature phase has the same kind of IR geometry as the flat-space solution, while the high-curvature phase  has a regular interior. We argue that, depending on   the leading asymptotic exponent of the scalar potential,  the transition may be  first-order or  higher-order.}

\begin{document}
\maketitle

\section{Introduction}

The AdS/CFT correspondence provides  us with a tool to describe strongly coupled (holographic), large-$N$ quantum field theories (QFTs) using higher-dimensional general relativity.  This approach gave a new perspective and new insights on various aspects of quantum field theories in flat space-time (confinement \cite{Witten:1998zw,Maldacena:1998im,Rey:1998ik}, renormalization group flows \cite{Freedman:1999gp,Bianchi:2001de}, entanglement entropy \cite{Ryu:2006bv,Ryu:2006ef,Hubeny:2007xt}, to cite only a few examples) as well as  for theories  defined on curved  space-times.

In particular, understanding QFTs in curved space-time is  interesting for both theory and phenomenology as  new infra-red effects can arise triggered by the curvature, which may be relevant (for example) in cosmology. One paradigmatic example is the fate of de Sitter space-time when coupled to a QFT with  gapless modes \cite{Tsamis:1996qm}. Another is the description of various cosmological phase transition or crossovers when, during  the universe expansion, the temperature crosses  various thresholds (like the electroweak scale or the QCD scale). Some of these questions can be  tackled within perturbation theory, like in the  electroweak case. Others, like in the case of QCD, need some non-perturbative tool to perform an analysis that goes beyond some very generic expectations.

The holographic duality provides such a tool, since on the gravity side there is  no restriction on the geometry on which the dual field field theory is defined: the boundary metric is an initial condition to the bulk equations. Although the curvature of such metric does not affect much the UV region of the solution (i.e. the region near the boundary) it is expected to have a large effects in the interior geometry, and this has been seen in several past studies.

One of they main interests of holography is that it provides a   dual gravitational description  of QCD-like theories (and supersymmetric versions thereof) which exhibit   confinement in four-dimensional flat space-time.  In this work we shall be interested in the interplay,  between confinement  and space-time curvature.

More precisely, we shall consider holographic models in which  the dual field theory  exhibits confinement in flat space-time, and   investigate how  turning on a  (constant) positive space-time curvature modifies the IR dynamics. Since curvature is a relevant deformation, one  expects it will modify, and eventually interfere with,  the geometrical features  which were  responsible for confinement of the dual QFT  in flat space-time. While for small curvature (with respect to the confinement scale) we expect similar features as in flat space-time, what happens at large curvatures is not obvious, except  for the fact that we expect the curvature to dominate over the physics responsible for confinement.

The holographic description of confining gauge theories in flat space-time is a well-explored problem in  the context of top-down string theory or supergravity models like the ten-dimensional Witten--Sakai--Sugimoto setup \cite{Witten:1998zw,Sakai:2004cn} or the effective five-dimensional models dual to deformations of N=4 SYM theory \cite{Girardello:1999bd}.

In parallel, bottom-up five-dimensional  phenomenological models have been developed which borrow important features and the basic dictionary from top-down holography and adapt them to come closer to real-world QCD. Improved Holographic Yang--Mills   \cite{Gursoy:2007er,Gursoy:2007cb} and V-QCD \cite{Jarvinen:2011qe} are the most encompassing and realistic models in this class.

In all such holographic models,  confinement can be essentially traced to the fact that the holographic dual space-time has an ``endpoint'' (which may be regular or singular) in the IR,  gaps the IR spectrum and provides an area law for the holographic Wilson loop.\footnote{In bottom-up holographic models the Wilson loop area law depends on  additional  assumptions on how the string frame is related to the Einstein frame \cite{Gursoy:2007er}.} This endpoint may be caused by an internal circle shrinking smoothly to zero size, like in higher-dimensional models;  or by the whole space-time shrinking fast enough to zero size into a singularity, as it is the case in all five-dimensional  models mentioned above.

 The presence of the singularity may cause some worry. It is, however, a broadly accepted  point of view that certain space-time singularities are acceptable in holography (see e.g. \cite{gubser_curvature_2000}). Underlying this belief is that such singularities are ``resolvable" either by Kaluza--Klein (KK) states or stringy states. Many examples exist where such singularities are resolved by adding back KK states and lifting the solutions to the appropriate higher dimensions\footnote{There is another issue that appears in such ``resolvable" singularities, namely to what extend one can reliably compute spectra and correlators, without having the precise resolution. This has been studied in \cite{Gursoy:2007cb,Charmousis:2010zz,Kiritsis:2016kog}, where the notion of a ``repulsive resolvable singularity" was defined and explored. In such repulsive singularities, correlators and spectra can be computed without the need for the actual resolution.},    \cite{Gouteraux:2011ce,Gouteraux:2011qh}.

Here, we ask the question of how this picture changes when the dual field theory is considered on a positive curvature space-time,  specifically    when  the curvature is much larger than the scale $\Lambda$ which controls the IR physics in flat space-time. We shall explore the question  whether there is a clearly identifiable phase transition between a small curvature and large curvature regime.
A  hint a phase transition may  occur is the fact that de Sitter space-time can be considered as describing a  thermal ensemble, with temperature set by the curvature, and we know that, in flat space, a thermal phase  transition occurs in confining theories between a confined and a deconfined phase.

However this analogy  may be misleading.  In the Euclidean theory,  on a sphere, there is no standard notion of Wilson test for confinement, as one cannot have arbitrarily large distances. Moreover, the massive spectra that appear in flat space because of confinement, are also masked by the spectral  gap we expect when the theory is put on a sphere.
However, one can refine these notions, by using as parameter   $\alpha \Lambda$, where $\alpha$ is the sphere radius and  $\Lambda$  is the characteristic scale of the theory in flat space. When $\alpha \Lambda \gg 1$ we expect the Wilson loop  to exhibit similar behavior as in flat space (i.e. an area law) over distances $L$ such that $\alpha \gg L \gg \Lambda^{-1}$). For  small  $\alpha \Lambda$ instead, the Wilson loop (or any other observable) cannot probe distances larger than the confinement scale, and in this case we expect the physics to be dictated by the curvature and mass gap is dominated by the curvature rather than the strong dynamics. Whether there is a phase transition with an identifiable  order parameter between the two regimes in the QFT is far from obvious.

One of the main results of this paper is that, when phrased in terms of the dual geometry,  the distinction between the two regimes is well defined and a phase transition does generically occur in the class of models we investigate.

Previous investigations have shown that there is indeed a phase transition of the kind discussed above in simple confining holographic models obtained from CFTs compactified on $S^1$ \cite{Marolf:2010tg} and on $S^2$ or $T^2$ \cite{Blackman:2011in} when the bulk theory is Einstein gravity with a negative cosmological constant. As we shall discuss,  some of these models (notably the $S^2$ case) fall into the realm of our analysis. Here, we extend those results and  find that, for theories which confine in flat space-time,  a phase transition at positive curvature is generic and  it can be first-order (like in the works mentioned above), or higher-order.

 In this paper, we  address these questions in general bottom-up holographic theories in $d+1$ dimensions, in which the bulk dynamics is governed by Einstein gravity coupled to a single  real scalar field $\varphi$:
\begin{equation}
    I_{\text{bulk}} = M_{\text{p}}^{d-1}\int_{\mathcal{M}} d^{d+1}X\sqrt{g}\,\biggl[ R-\frac{1}{2}\,(\partial \varphi)^2-V(\varphi) \biggr]\,.
    \label{int1}
\end{equation}
We  consider potentials which correspond to a confining holographic theory  {\em on flat space}. We subsequently search for semiclassical saddle points, when the same theory is considered  on a sphere (or de Sitter). When multiple saddle points are present, we analyze which one is dominant in the path integral and analyze the  corresponding phase transitions as a function of the curvature. Throughout the paper,  by the term  ``confining theory'' we always mean a theory which is confining on flat space

These models of the type \eqref{int1}  may be considered as toy-models in which all the boundary $d$-dimensional QFT dynamics is encoded in the RG flow of a single relevant operator dual to $\varphi$. Despite their simplicity, these models capture all the features which characterize confining QFTs in flat space: Wilson loop area law, gapped spectra and deconfinement phase transitions at finite temperature. They constitute the backbone for more complete (and more realistic) bottom-up holographic theories.

In the context  described by the action \eqref{int1},  there is a very simple universal criterion for whether the bulk theory admits solutions which are dual to a confining QFT in flat space: confinement occurs if (and only if)  the scalar field reaches  infinity  in the bulk (this is the singular ``end-of-space'' mentioned earlier)  and that,  in that asymptotic limit, the  potential $V(\varphi)$ diverges to $-\infty$  {\em faster} than $e^{2b_{\text{c}}\varphi}$ with $b_{\text{c}} = 1/\sqrt{2\,(d-1)}$ \cite{Gursoy:2007er,Kiritsis:2016kog}, but slower than $b_{\text{G}} \equiv \sqrt{d}\,b_{\text{c}} $.

In this work, motivated by string theory, we  assume that the potential, asymptotically, takes the form of a simple exponential:
\begin{equation} \label{int2}
V(\varphi)\sim V_\infty\, e^{2b\varphi}\rightarrow -\infty\,, \qquad \varphi \to \infty \,,
\end{equation}
with $V_\infty <0$. Then, the confining regime corresponds to:
\be \label{int3}
\textrm{confinement in flat space} \qquad \Leftrightarrow \qquad b_{\text{c}} < b < b_{\text{G}}\,,
\ee
where $b_{\text{c}} = 1/\sqrt{2\,(d-1)}$ and $b_{\text{G}} \equiv \sqrt{d}\,b_{\text{c}} $. When  $b $ saturates the lower bound \eqref{int3}, we are in an intermediate situation where the flat-space spectrum is continuous but gapped, and the thermal phase transition becomes at least second-order (infinite-order is also possible) depending on the growth of the subleading terms of the potential (see \cite{Gursoy:2010jh,Betzios:2018kwn,Betzios:2017dol} where this case is discussed in detail). The case $b=b_{\text{c}}$ demands special techniques and will be discussed elsewhere \cite{JL}.

This work furthers  the program that was initiated in \cite{Ghosh:2017big}, which started a systematic investigation of  the effects of space-time curvature on holographic QFTs:
\begin{enumerate}
\item In \cite{Ghosh:2017big,Ghodsi:2022umc}, the focus was on holographic RG flows between a UV and an IR fixed point. In  this case, adding positive curvature had the effect that the  IR fixed point is never reached, but the flow stops at a regular endpoint in the Euclidean signature, (or a regular horizon with zero area in the Lorentzian signature, see e.g. \cite{Ghosh:2021lua}).
 \item  More recently, \cite{Ghodsi:2024jxe} has dealt with theories in the confining class, but  on {\em negatively} curved space-time. In this case, there can be  no regular endpoint, but either the scalar runs to infinity as in the flat case, or the whole  space-time has a {\em bounce} and it develops a second asymptotic boundary.
\end{enumerate}
Here, we deal  with general confining theories on {\em positively}  curved space-times and as we shall observe, we find a mixture of features from both cases described above. In the next subsection  we briefly summarize the main results of our analysis.

\subsection{Summary of results}\label{subsec:sum}

It is convenient to introduce a  {\em dimensionless} curvature parameter ${\cal R}$ which measures the curvature in units of  another relevant scale in the problem,
\be
{\cal R} \equiv R_{\text{UV}}\,\Lambda^{-2}\,,
\ee
where $R_{\text{UV}}$ is the curvature seen by the boundary QFT and $\Lambda$ may be taken as the IR non-perturbative energy scale of the flat space theory or, equivalently (as we do in the rest of the paper), as the mass scale of the deformation driving the RG flow in the UV.\footnote{We assume the theory has a conformal fixed point in the UV and that the bulk field $\varphi$ is dual to a relevant deformation.}

\paragraph{IR classification of solutions.}
As we  review in Section  \ref{sec:reducedtheory}, with an exponential potential, for zero space-time curvature, the bulk geometry is always singular. Generic singular solutions (which we call type 0)  have no holographic interpretation. However,   there is a different branch of singular solutions which have  acceptable  IR singularities. These are {\em special} solutions, with one integration constant less than  the generic solutions.  A similar distinction can be made when the field theory lives on a sphere. We refer to these special, acceptable singular solutions (by an abuse of language) as ``regular'' (in quotes), since they require a relation between integration constants, in the same way as truly regular solutions.

\begin{figure}[h!]
\begin{center}
\includegraphics[width=5cm]{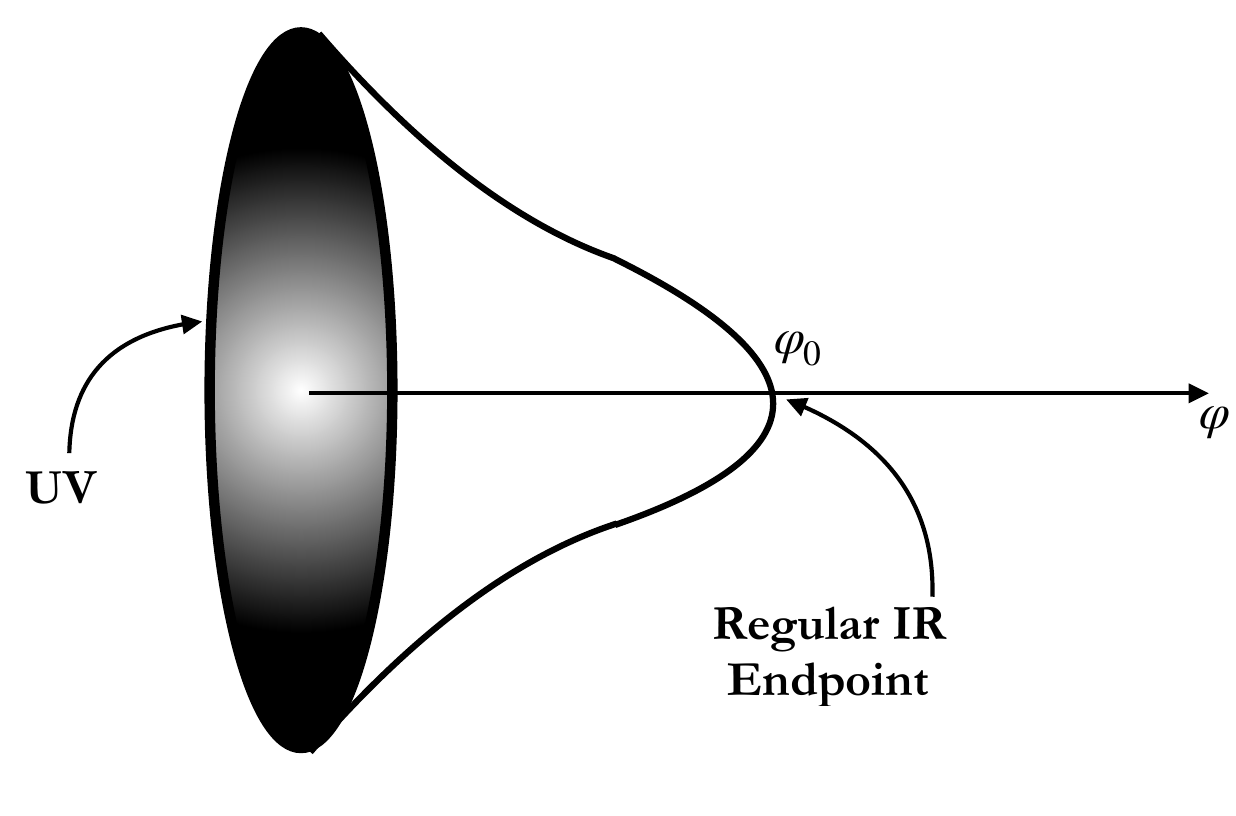} \hspace{1cm} \includegraphics[width=5cm]{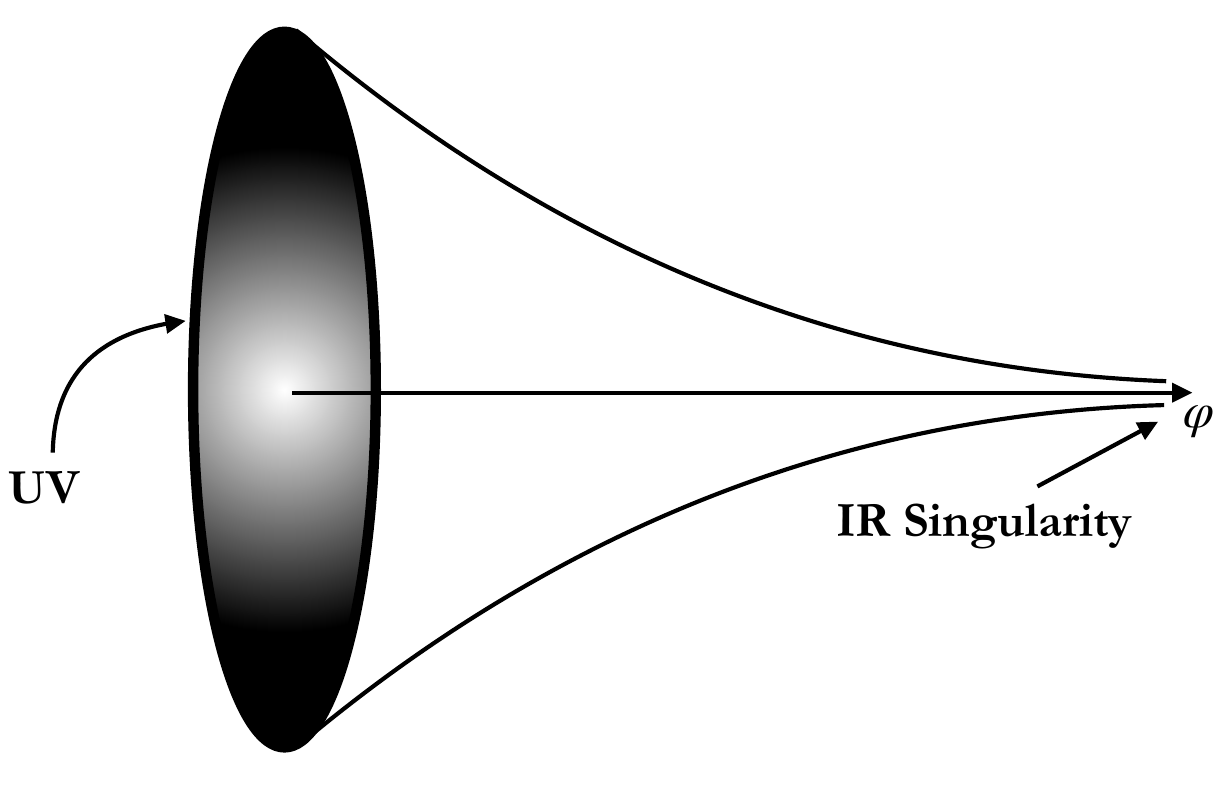}\\
Type III \hspace{6cm} Type I-II
\caption{A sketch of the solutions with regular endpoints (left) and with an acceptable (``regular'')  IR singularity (right)} \label{Geomfig}
\end{center}
\end{figure}

In the case of dual field theories on a positive curvature manifold, the space of acceptable bulk solutions on the gravity side can be divided into three different classes according to how they terminate in the IR.
\begin{enumerate}
\item There are two branches of ``regular'' solutions, which we call {\bf type I} and {\bf type II}, in which  $\varphi \to  \infty$ at the endpoint. These are sketched on  the right side of Figure \ref{Geomfig}. The two branches are distinguished by the detailed behavior at infinity. More importantly,  type I solutions exist only for  a {\em fixed value} of the curvature parameter, ${\cal R}= {\cal R}_{\text{I}}$, whereas type II solutions come in a continuous family with curvatures varying from  zero up to a finite maximum value. This  may be equal or larger than ${\cal R}_{\text{I}}$.  In the limit ${\cal R} \to 0$,  type II solutions go over smoothly to the ``regular'' solution with ${\cal R}=0$.

\item There exists a third class of solutions, called {\bf type III} and sketched in the left side of Figure \ref{Geomfig},  in which the IR endpoint is (truly) regular and it is reached as $\varphi \to \varphi_0$, where $\varphi_0$ is now a free parameter. These regular endpoints are of the same type as those described in the past for non-confining theories \cite{Ghosh:2017big,Ghosh:2021lua}.  Changing $\varphi_0$ is equivalent to changing the curvature parameter ${\cal R}$. The limit $\varphi_0\to \infty$ corresponds to  ${\cal R} \to  {\cal R}_{\text{I}}$.
\end{enumerate}

\paragraph{Curvature-driven phase transition.} We can summarize our main result, which holds for potentials with a  leading large-$\varphi$ asymptotics as in \eqref{int2} and independently of subleading terms,  as follows:
\begin{center}
\begin{minipage}[t]{0.9\textwidth}
{\em For confining  bulk potentials, there is a critical  value $\mathcal{R}_{\text{c}}$ of $\mathcal{R}$.  Above  this value,  the  Euclidean  path integral is dominated by solutions with a regular endpoint; below the critical value the preferred solutions  run to the  $\varphi =\infty$ endpoint as in flat space.}
\end{minipage}
\end{center}
This indicates a curvature-driven quantum phase transition which in the bulk appears as a transition between two distinct kinds of bulk geometries.  At low curvatures the dominant solution is of type II, and it has the same IR asymptotics as in flat space: for the Lorentzian theory, this results in a discrete and gapped spectrum of massive excitations, as in the confining flat-space theory. At large curvatures  instead, the dominant semiclassical solution is type III. In this case the Lorentzian solution has a horizon in the bulk and one can show \cite{spectra} that the spectrum of excitation is continuous (albeit still gapped,  this time by the  curvature, as one may have expected).

Phase transitions driven by curvature were already observed in Einstein-scalar theories in \cite{Ghosh:2017big,Buchel:2011cc}. In that case the transition was first-order, and it occurred  between two branches of solutions with the same qualitative IR features (a regular endpoint/horizon). Here instead, the transition occurs between geometries with very different IR properties.

\begin{figure}[h!]
\begin{center}
\includegraphics[width=12cm]{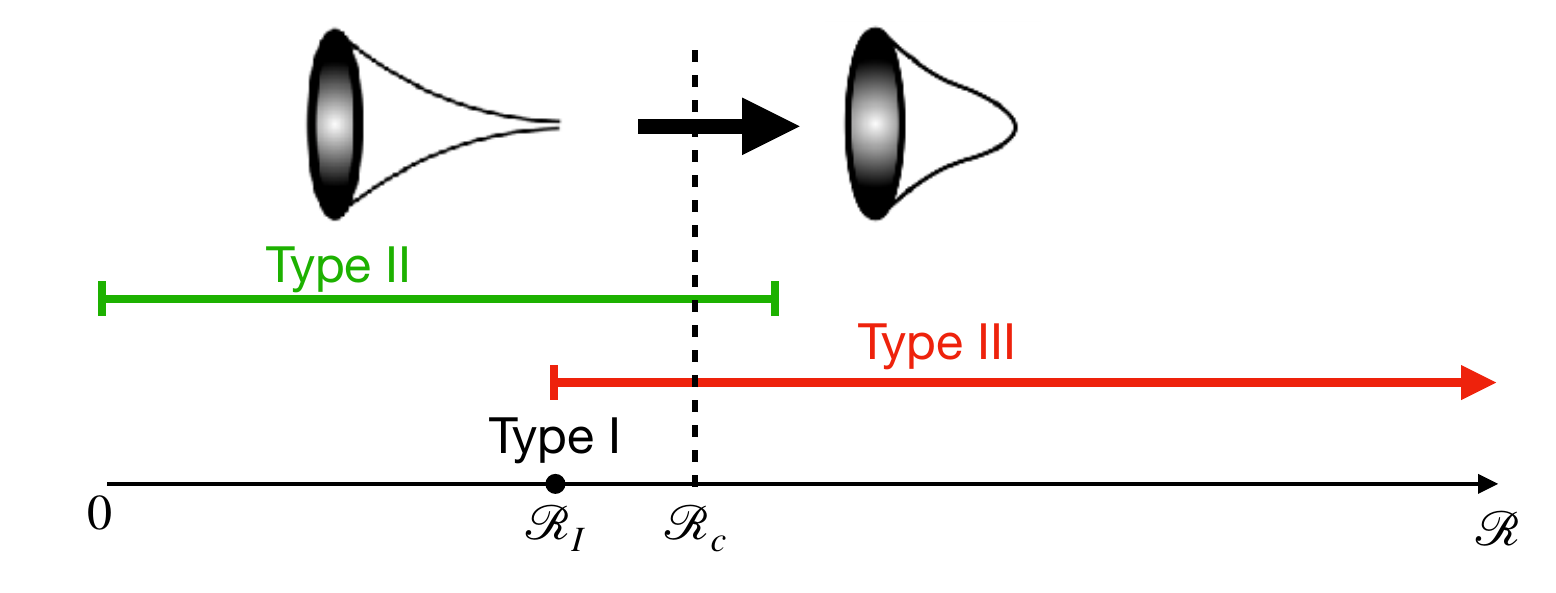} \\
1st order transition\\
\vspace{1cm}
 \includegraphics[width=12cm]{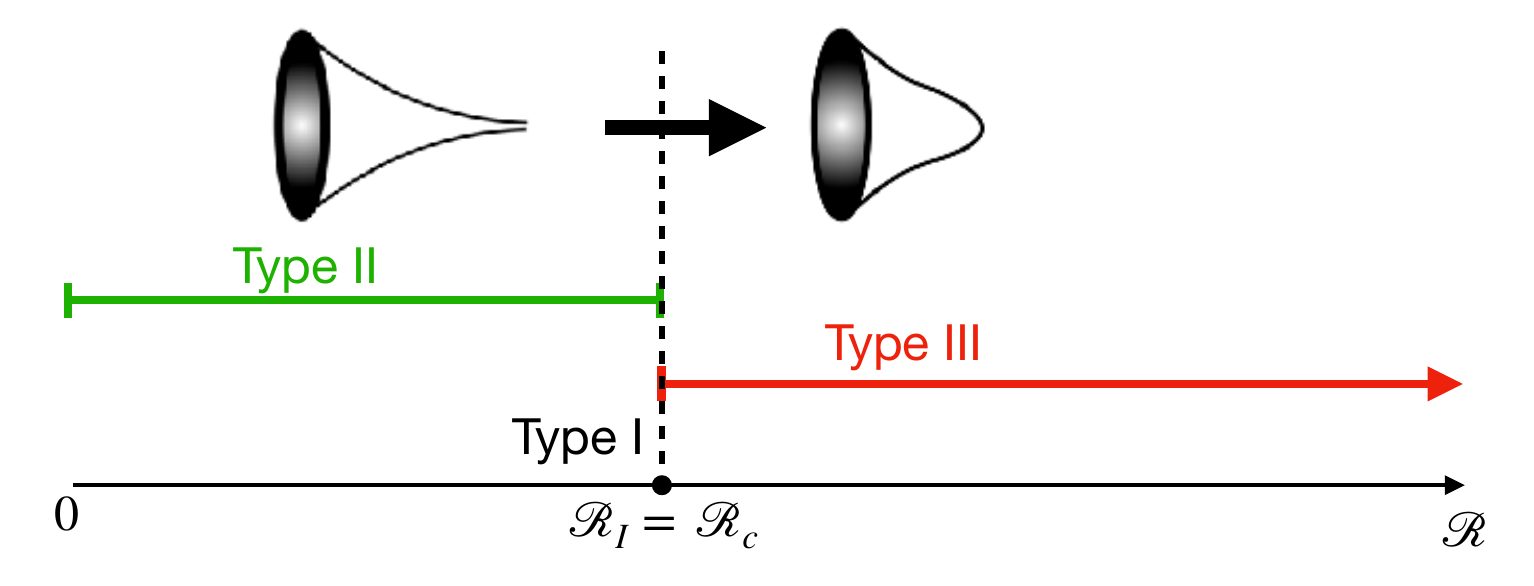}\\
2nd or higher order transition
\caption{The figures illustrate which kinds of geometries  exist as a function of  the dimensionless curvature parameter ${\mathcal R}$, and corresponding transitions, indicated by the dashed vertical line.  At large curvature, the regular solutions dominate, whereas at small curvature "regular" singular solutions dominate, limiting to  the flat-space solution at ${\mathcal R} \to 0$.} \label{PTfig}
\end{center}
\end{figure}

Depending on the value of $b$, the transition may be first-order or higher-order. In the first-order case, illustrated in the top Figure \ref{PTfig}, there is a  region (as a function of ${\mathcal R}$) where both of type II and type III coexist, and the critical curvature $\mathcal{R}_c$ is above the type I curvature $\mathcal{R}_\text{I}$.  The transition is continuous (bottom Figure \ref{PTfig}) when there is no coexistence region, and the transition occurs at $\mathcal{R}_\text{I}$.

\paragraph{Dimensional uplift.} We show that it is possible to understand the  features  of the  IR solutions and  of the phase transition in terms of (generalized) dimensional reduction from a higher-dimensional pure gravity theory: as shown in  \cite{Gouteraux:2011qh,Gouteraux:2011ce}, the IR exponential potentials \eqref{int2} in the range \eqref{int3} arise from dimensionally reducing $d+N+1$-dimensional Einstein gravity with a negative cosmological constant on an internal $N$-sphere. In the Euclidean theory, the uplift is therefore a product of $S^d \times S^N$ warped along the extra holographic direction,
\be \label{int4}
ds^2 = du^2 + e^{2A_1(u)}\, \alpha^2ds^2_{S^d} +  e^{2A_2(u)}\, \bar{\alpha}^2ds^2_{S^N} \,.
\ee
Here, $S^d$ represent Euclidean space-time and $S^N$ is the internal sphere. The scalar field in the reduced picture can be mapped to  the volume of the internal sphere, and it obtains an exponential potential of the form \eqref{int2}, with
\be \label{int5}
b = \sqrt{\frac{d+N-1}{2N\,(d-1)}}\,.
\ee
The map between the uplifted and reduced theories is limited to the  deep IR region and it breaks down as one moves towards the UV.

Gravitational dynamics of warped products of spheres \eqref{int4} have been discussed in the context of holography in \cite{Aharony:2019vgs,Kiritsis:2020bds} (see \cite{bohm1998inhomogeneous,bohm1999non,Kol:2002xz,Asnin:2006ip,Kalisch:2017bin} for earlier work). There are two distinct possibilities for the way the space-time terminates in the IR in a regular fashion:
\begin{enumerate}
\item The internal sphere $S^N$ shrinks to zero size while the space-time $S^d$ sphere stays finite. Upon dimensional reduction, this behavior maps to the type II IR solution, in which the scalar field reaches infinity. The size of $S^d$ at the endpoint is mapped to the curvature parameter in the reduced theory.
\item The space-time sphere $S^d$ shrinks to zero size and the internal sphere $S^N$ stays finite.  This maps to type III solutions in the reduced theory, where the size of the $S^N$ is mapped to $\varphi_0$, the value of the scalar field at the endpoint.
\item Both spheres shrink to zero size at the same time. This solution is singular,  has no free parameters in the IR, and it maps to the (unique) type I solution in the reduced theory. In the UV of the higher-dimensional theory, this solution corresponds to a special ratio of the radii $\alpha$, $\bar{\alpha}$ of $S^d$, $S^N$ respectively, namely \cite{Aharony:2019vgs,Kiritsis:2020bds}
\begin{equation}
\frac{\alpha}{\bar{\alpha}} = \sqrt{\frac{d-1}{N-1}}\,. \label{critratio}
\end{equation}
\end{enumerate}
It was found in \cite{Aharony:2019vgs,Kiritsis:2020bds} that for $d+N<9$ there is a discrete, infinite set of solutions with the same curvature ratio \eqref{critratio}. Only one of them is singular and reduces to the IR geometry of the  type I  solutions of Einstein-scalar theory. The others are regular, and have either $S^d$ or $S^N$ shrink to zero size, reducing to either type II or type III.

These solutions lie on a spiral in the space of integration constants and exhibit a discrete scaling of the Efimov type. The singular solution is the center of the spiral. As a result, the theory has a multiple first-order phase transitions along the spiral, as the curvature ratio crosses the critical value, in which the dominant solution changes from type II to type III or vice-versa. In practice, only one phase transition is the leading one (the one between solutions which are farthest from the center of the spiral). Dimensions $d+N> 9$ outside the Efimov regime were not considered in detail in \cite{Aharony:2019vgs,Kiritsis:2020bds} as they are above the critical dimension $d+N+1 = 10$ of superstring theory, an interesting coincidence already pointed out in \cite{Kol:2002xz}.

The dimensional reduction gives a natural interpretation of the phase transition in the reduced theory from the point of view of the  uplifted theory: it corresponds to the transition between solutions where either the space-time $S^d$ or the internal $S^N$ sphere shrinks to zero size, i.e. a conifold transition.

\paragraph{Nature of the transition.} We show that Einstein-scalar theory with an exponential potential \eqref{int2} contains a special value of the exponent $b$ called the \textit{Efimov bound}
\begin{equation}
    b = b_{\text{E}} \equiv \frac{2}{\sqrt{(d-1)(9-d)}}\,,
\end{equation}
which splits the theory into two regimes $b< b_{\text{E}}$ and $b> b_{\text{E}}$. For an exponent \eqref{int5} arising from dimensional reduction, $b > b_{\text{E}}$ coincides with the Efimov regime $d+N<9$ of the higher-dimensional theory. The behavior of the dimensionless vacuum expectation value (vev) $\mathcal{C}\propto \langle \mathcal{O} \rangle $ of the deforming operator as a function of the background curvature $\mathcal{R}$ differs in the two regimes so that the nature of the transition is also different.

\begin{enumerate}
\item  $b>b_{\text{E}}$: {\bf Efimov regime.} In the Efimov regime $b >b_{\text{E}}$, the vev $\mathcal{C} =\mathcal{C}(\mathcal{R}) $ is a multi-valued function of $\mathcal{R}$ exhibiting an Efimov spiral that circles around the critical value $(\mathcal{R}_{\text{c}},\mathcal{C}_{\text{c}})$ of the type I solution (see Figure \ref{plot:analytical_Efimov_above_spiral} below). The transition is of first-order.
\item  $b<b_{\text{E}}$: {\bf Monotonic regime.} In the monotonic regime, there is no spiral, but there are two qualitatively possible behaviors in which the order of the phase transition is different.
\begin{enumerate}[(i)]

    \item In this first case, the vev $\mathcal{C} =\mathcal{C}(\mathcal{R})$ is multi-valued and exhibits a single swing (instead of a spiral) as it passes through $(\mathcal{R}_{\text{c}},\mathcal{C}_{\text{c}})$ as in Figure \ref{plot:analytical_peaks} below. The free energy contains a swallow tail and the transition is first-order.

    \item In this second case, the vev $\mathcal{C} =\mathcal{C}(\mathcal{R})$ is single-valued and there is a single regular solution for each value of $\mathcal{R}$, and in particular, the only solution at the critical curvature $\mathcal{R}_{\text{c}}$ is the type I solution (see Figure \ref{plot:analytical_no_peaks} below). Therefore as one varies the ratio between the curvatures of the two spheres,  one continuously transitions from type II solutions to type III solution, and the transition occurs exactly at $\mathcal{R}_{\text{c}} = \mathcal{R}_{\text{I}}$ across the type I solution. In this case, we prove that the transition is at least second-order, but it may also be higher-order.
\end{enumerate}
Which case (i) or (ii) is realized is determined by the exact shape of the potential $V(\varphi)$ away from the deep IR region $\varphi\rightarrow \infty$.
\end{enumerate}
We analyze in detail the order of the transition in the monotonic regime $b_{\text{c}}<b<b_{\text{E}}$ in the second case (ii). We show that the free energy difference between type II and III solutions in the vicinity of the critical point $\mathcal{R} = \mathcal{R}_{\text{c}}$ behaves as
\begin{equation}
	\frac{F_{\text{III}}'(\mathcal{R})-F_{\text{II}}'(\mathcal{R})}{2\,\mathcal{N}} =\mathcal{F}_1\,\frac{\mathcal{R}-\mathcal{R}_{\text{c}}}{\mathcal{R}^{3}} + \mathcal{F}_\delta\,\frac{(\mathcal{R}-\mathcal{R}_{\text{c}})^{\delta}}{\mathcal{R}^{3}}+\ldots\,,\quad \mathcal{R}\rightarrow \mathcal{R}_{\text{c}}\,,
	\label{eq:Fprime_diff_expansion_summary}
\end{equation}
where the exponent of the subleading term is explicitly
\begin{equation}
	\delta = \frac{(d-1)\,b+2\sqrt{1-(b\slash b_{\text{E}})^2}}{(d-1)\,b-2\sqrt{1-(b\slash b_{\text{E}})^2}}>1
\end{equation}
and the ellipsis denotes terms that vanish faster than $(\mathcal{R}-\mathcal{R}_{\text{c}})^{\delta}$. This implies that the discontinuity in the second derivative $F_{\text{III}}''(\mathcal{R}_{\text{c}})-F_{\text{II}}''(\mathcal{R}_{\text{c}})\propto \mathcal{F}_1 $ at the critical point may be non-zero depending on the details of the potential. If the second derivative is continuous, and the transition is at least third-order, then \eqref{eq:Fprime_diff_expansion_summary} implies that
\begin{equation}
	\text{order of the transition} = 1+\lceil \delta\rceil\geq 3 \,.
\end{equation}
Therefore the transition may be higher than second-order, but cannot be infinite-order.

\paragraph{Concrete numerical example.} We explore numerically a concrete example in $d = 4$ with a potential $V(\varphi)$ that contains a single UV fixed-point (maximum) at $\varphi = 0$ and exponential asymptotics \eqref{int2} in the IR $\varphi\rightarrow \infty$. We consider two different values of $b$ that fall in the Efimov and monotonic regimes. We find numerical solutions which interpolate between a UV fixed point at a maximum of the potential, and an IR endpoint, which follows the classification above. We scan numerically the space of solutions (of type I, II and III),  find the solution(s) for every value of the parameter $\mathcal{R}$ and identify the limiting value $\mathcal{R}_{\text{c}}$. We then obtain the QFT free energy by computing the (renormalized) on-shell bulk action. We confirm the qualitative features that are discussed above:
\begin{enumerate}
\item For a value of $b$ in the Efimov regime, we find a range of curvatures with multiple solutions, and correspondingly a first-order phase transition as expected. The transition takes place between solutions at $\mathcal{R} = \mathcal{R}_{\text{c}}$ which is far away in parameter space from the type I solution with $\mathcal{R} = \mathcal{R}_{\text{I}}$ and $\varphi_0 =\infty$.
\item For a value of $b $ in the monotonic regime, we find that each curvature corresponds to a single solution: type II solutions only cover the range $0< {\cal R}< {\cal R}_{\text{I}}$, while type III  are only found for ${\cal R}> {\cal R}_{\text{I}}$. The branches meet at the type I solution with ${\cal R} = {\cal R}_{\text{I}}$, which corresponds to a critical point $\mathcal{R}_{\text{c}} =\mathcal{R}_{\text{I}} $ of (as far as the numerical error allows to discern) a second-order transition as expected. Therefore it is the case (ii) as defined above which is realized in this example. Our numerical accuracy does not allow to determine whether the transition is third- or higher-order.
\end{enumerate}
The structure of the paper is as follows: in Section \ref{sec:reducedtheory} we review the holographic description and the relevant tools for  QFTs RG-flow on curved space-times, and how they can be obtained from generalized dimensional reduction of a pure gravity theory. In Section \ref{sec:classification} we classify the solutions according to their IR behavior and we organize them in families. In Section \ref{sec:free_energy}, we calculate the renormalized free energy and also discuss the appearance of Efimov behavior in a certain range of parameters. In Section \ref{sec:numerics} we work out two numerical examples, scan the space of solutions and compute their free energy. In the first example, we find a first-order phase transition, while in the second, a second-order transition.

\section{Curved holographic RG flows in Einstein-scalar theory}\label{sec:reducedtheory}

In this section, we  introduce the holographic model we shall use to describe confining quantum field theories living on a constant positive curvature space-time. We are working in Euclidean signature, and therefore  we  take the quantum field theory to live on a $d$-dimensional sphere $S^{d}$. The bulk theory consists of a scalar field minimally coupled to Einstein gravity. The scalar field is given a potential which is assumed to have a maximum that supports and $AdS_{d+1}$ solution, and describes the UV of the dual field theory. We  show how to find backreacted solutions of the bulk theory using the first-order formulation developed in \cite{Ghosh:2017big}. We focus on scalar potentials that diverge exponentially at large field values. For appropriate range of the exponential behavior they are known to correspond to  dual field theory theories that, in flat space, are  confining  in the IR \cite{Gursoy:2007er}. We  review how such exponential scalar potentials can be obtained from dimensional reduction of pure Einstein gravity on an internal sphere. The higher-dimensional description will form the basis in classifying allowed solutions of the Einstein-scalar theory.

{Unless explicitly stated, we work in the Euclidean signature, i.e. with the dual field theory defined on a sphere. This choice does not affect the bulk field equations, and all  our solutions  carry over to Lorentzian signature (dual field theory on de Sitter space-time), although their global properties change (in particular, regular endpoints  become horizons in the Lorentzian  theory)}

\subsection{Einstein-scalar theory with an exponential potential}

We  consider Einstein-scalar theory in $d+1$ dimensions with the Euclidean action
\begin{equation}
I_{d+1} = M_{\text{p}}^{d-1}\int_{\mathcal{M}} d^{d+1}X\sqrt{g}\,\biggl[ R-\frac{1}{2}\,g^{ab}\,\partial_a\varphi\,\partial_b\varphi-V(\varphi) \biggr]\,.
\label{einsteinscalar}
\end{equation}
We write our coordinates as $ X^{a} = (u,x^{\mu}) $, where $\mu = 1,2,\ldots,d$, and look for domain wall solutions of the type
\begin{equation}
g_{ab}(X)\,dX^{a}dX^{b} = du^{2}+e^{2A(u)}\,\zeta_{\mu\nu}(x)\,dx^{\mu}dx^{\nu}\,, \quad \varphi = \varphi(u)\,,
\label{bulkansatz}
\end{equation}
described by two functions $A(u)$ and $\varphi(u)$ of the coordinate $u$. We assume that constant-$u$ slices of the metric are spheres of radius $\alpha$,
\begin{equation}
	\zeta_{\mu\nu}(x)\,dx^{\mu}dx^{\nu} \equiv \alpha^{2}ds^{2}_{S^{d}}\,,
	\label{eq:zetametric}
\end{equation}
where $ ds^{2}_{S^{d}} $ is the round metric of the unit $d$-sphere $S^{d}$. The equations of motion for $(A,\varphi)$ coming from the action \eqref{einsteinscalar} are
\begin{align}
2\,(d-1)\,\ddot{A}+\dot{\varphi}^{2}+2\kappa\, e^{-2A} &= 0\,,\label{EOM1}\\
d(d-1)\,\dot{A}^{2}-\frac{1}{2}\,\dot{\varphi}^{2}+V-d\kappa\, e^{-2A} &=0\,,\label{EOM2}\\
\ddot{\varphi}+d\,\dot{A}\,\dot{\varphi}-V' &=0\,,\label{EOM3}
\end{align}
where the dot and the prime denote derivatives with respect to $ u $ and $\varphi$ respectively. In addition, we have defined $\kappa \equiv \frac{d-1}{\alpha^{2}}$. Notice that only two of these equations are independent.\footnote{For example, the third equation can be obtained by taking a derivative of the second and substituting $\ddot{A}$ from the first}

To interpret the theory  holographically, we  assume that the potential has a maximum at $\varphi = 0$ with the expansion
\begin{equation}
	V(\varphi) = 2\Lambda-\frac{1}{2}m^{2}\varphi^{2} + \mathcal{O}(\varphi^{3})\,, \quad \varphi \rightarrow 0\,,
	\label{potentialUVfixedpoint}
\end{equation}
where $m^{2}>0$ and the cosmological constant is negative $\Lambda \equiv -\frac{d(d-1)}{2\ell^{2}}< 0$. In the vicinity of the maximum  the solution to  equations  \eqref{EOM1} - \eqref{EOM3} is approximated by:
\begin{align}
	A(u) &= -\frac{u-c}{\ell}+\bar{A}_0 - e^{2(u-c)\slash \ell} + \ldots,\quad u\rightarrow -\infty\,,\label{Aasymp}\\
	\varphi(u) &= \varphi_{-}\,\ell^{\Delta_-}\,e^{\Delta_- u\slash \ell}+\varphi_{+}\,\ell^{\Delta_+}\,e^{\Delta_+ u\slash \ell} + \ldots\,, \quad u\rightarrow -\infty\,, \label{phyasymp}
\end{align}
where $ c,\varphi_{\pm}$ are three real integration constants while the remaining constants are fixed by the equations as $\bar{A}_0 =\log{\frac{\ell}{2\alpha}} $ and
\begin{equation}
\Delta_{\pm} = \frac{1}{2}\bigl(d\pm\sqrt{d^{2}-4m^{2}\ell^{2}}\bigr)  \quad \Leftrightarrow \quad  m^{2}\ell^{2} = \Delta_{\pm}(d-\Delta_{\pm})\,. \label{deltas}
\end{equation}
The metric defined by \eqref{Aasymp} is an asymptotically locally AdS metric of curvature radius $\ell $ with the conformal boundary residing at $u\rightarrow -\infty$. The metric residing on the conformal boundary (which is the background metric of the dual QFT) is given by
\begin{equation}
    ds^2 = \lim_{u\rightarrow -\infty}e^{2u\slash \ell}\,g_{\mu\nu}(u,x)\,dx^\mu dx^\nu\,.
    \label{eq:QFTmetric}
\end{equation}
By fixing the integration constant $c$ of the scale factor in terms of $\alpha$ as $c\slash \ell = -\bar{A}_0 = -\log{\frac{\ell}{2\alpha}} $, the metric \eqref{eq:QFTmetric} coincides with the metric \eqref{eq:zetametric} of a sphere of radius $\alpha$. Note that for this choice of the boundary metric, the UV expansion of the scale factor takes the form $A(u) = -\frac{u}{\ell} - \frac{\ell^2}{4\alpha^2}\,e^{2u\slash \ell}$ where the additive $u$-independent constant vanishes. What remains are the two integration constants $\varphi_{\pm}$.

By the standard holographic dictionary, the larger exponent $\Delta_+ \equiv \Delta$ is identified with the scaling dimension $\Delta$ of the field theory scalar operator dual to $\varphi$. It follows that the coefficient $\varphi_-$ of the diverging term $\Delta_- = d-\Delta$ is proportional to the source for the scalar operator in field theory. Therefore $\varphi_-$, together with the scalar curvature $d\kappa = \frac{d(d-1)}{\alpha^{2}}$ of the boundary sphere, specify field theory data which can be combined into a single dimensionless curvature parameter
\begin{equation}
	\mathcal{R} \equiv d\kappa\,\lvert \varphi_{-}\lvert^{-2\slash \Delta_-}\,.
	\label{eq:curlyR}
\end{equation}
The  coefficient $\varphi_+$ of the subleading term in \eqref{phyasymp} is identified with the expectation value of the dual operator $ \langle \mathcal{O}\rangle = (2\Delta_+ -d)\,\varphi_+ $ which can be characterized by a single dimensionless vev parameter defined by
\begin{equation}
	\mathcal{C}\equiv \frac{\Delta_-}{d}\,\langle \mathcal{O}\rangle\,\lvert \varphi_{-}\lvert^{-\Delta_+\slash \Delta_-}\,.
	\label{eq:curlyC}
\end{equation}
From the bulk point of view, $\mathcal{R}$ and $\mathcal{C}$ appear as integration constants. All field theory dimensionless theory data are encoded in $\mathcal{R}\in (0,\infty)$ and holography provides a prediction for the vev $\mathcal{C} = \mathcal{C}(\mathcal{R})$ after imposing IR boundary conditions determined by requiring the bulk geometry to be ``regular.'' The allowed set of boundary conditions will be studied in Section \ref{sec:classification} below.

\paragraph{First-order formulation.} The set of second-order equations of motion for $(A,\varphi)$ can be written as a first-order system where $\varphi$ plays the role of the independent variable. To this end, following  \cite{Ghosh:2017big} we define the functions:
\begin{align}
W(\varphi) &= -2(d-1)\,\dot{A}\,,\label{W}\\
S(\varphi)&=\dot{\varphi}\,,\label{S}\\
T(\varphi) & = d\kappa\,e^{-2A(u)}\,,
\label{WST}
\end{align}
where the expressions on the right-hand side are evaluated at $ u = u(\varphi) $. These functions are well defined piecewise in the regions $\dot{\varphi}(u)\neq 0$ of $u$ where $ \varphi(u) $ is an invertible function.\footnote{In this paper, all relevant solutions will satisfy $S(\varphi) > 0$ over an (possibly semi-infinite) interval $\varphi \in (0,\varphi_0) $ with $S(\varphi) = 0$ at one or both of the endpoints. Therefore the first-order formulation is always well defined globally.} The equations of motion \eqref{EOM1} - \eqref{EOM3} become
\begin{align}
S^{2}-SW'+\frac{2}{d}\,T &= 0\,,\label{1st}\\
\frac{d}{2(d-1)}\,W^{2}-S^{2}-2T+2V&=0\,,\label{2nd}\\
SS'-\frac{d}{2(d-1)}\,SW-V' &=0\label{3rd}\,.
\end{align}
The first \eqref{1st} and third \eqref{3rd} equation  can be solved algebraically for $ T $ and $W$ respectively as
\begin{equation}
T = \frac{d}{2}\,(SW'-S^{2})\,,\quad W = \frac{2(d-1)}{d}\biggl(S' - \frac{V'}{S}\biggr)\,.
\label{Wtext}
\end{equation}
Substituting back to the second equation \eqref{2nd} gives
\begin{equation}
d\,S^{3}S''-\frac{d}{2}\,S^{4}-S^{2}S'^{2}-\frac{d}{d-1}\,S^{2}V+(d+2)\,SS'V'-d\,S^{2}V''-V'^{2} = 0\,.
\label{Sequation}
\end{equation}
Solving this equation for $S$ and calculating $W,T$ by using \eqref{Wtext} solves the full first-order system \eqref{1st} - \eqref{3rd}.

As a second-order equation, \eqref{Sequation} determines $S(\varphi)$, and also $W(\varphi)$, $T(\varphi)$ through \eqref{Wtext}, up to two integration constants. Integrating equation \eqref{S} yields $ \varphi(u) $ up to a third integration constant which is a constant shift in $u$. Integration of equation \eqref{W} yields naively a fourth integration constant, but it is fixed by equation \eqref{WST} in terms of the first two. These three integration of the system \eqref{W} - \eqref{WST} and \eqref{Sequation} are therefore related to the three constants $c,\varphi_{\pm}$ above. Since $c$ is determined by fixing the metric on the conformal boundary to have radius $\alpha$, only the two free constants remain, which are the dimensionless parameters $ \mathcal{R}$ \eqref{eq:curlyR} and $ \mathcal{C} $  \eqref{eq:curlyC} defined above. As shown in Appendix \ref{app:UVasymptotics}, they appear explicitly in the UV expansion of $S(\varphi)$ as (see also \cite{Ghosh:2017big})
\begin{align}
	S(\varphi) &= \frac{\Delta_-}{\ell}\,\varphi\,[1 + \mathcal{O}(\varphi)]+\frac{d}{\Delta_-}\frac{\mathcal{C}}{\ell}\,\varphi^{-1+d\slash \Delta_-}\,[1 + \mathcal{O}(\varphi)]\label{eq:S_UV_expansion_text}\\
	&\;\;\;\; +\frac{\Delta_-}{2(d-1)(2\Delta_--d+2)}\frac{\mathcal{R}}{\ell}\,\varphi^{1+2\slash \Delta_-}[1+\mathcal{O}(\varphi)]\nonumber\\
 &\;\;\;\;+\mathcal{O}(\mathcal{C}^2)+\mathcal{O}(\mathcal{R}^2)+\mathcal{O}(\mathcal{C}\,\mathcal{R})\,,\quad \varphi\rightarrow 0\,,\nonumber
\end{align}
where the higher-order terms in $\mathcal{R}$ and $\mathcal{C}$ are accompanied by powers of $\varphi$ that are subleading with respect to the terms shown. Similar UV expansions may also be obtained for $W(\varphi)$ and $T(\varphi)$, see Appendix \ref{app:UVasymptotics}.

\paragraph{IR exponential potentials.} In addition to assuming that the potential has an extremum \eqref{potentialUVfixedpoint} at $\varphi = 0$ (corresponding to the UV of the theory), we shall be focusing on potentials with exponential large-$\varphi$ asymptotics, which we parametrize as:
\begin{equation}
	V(\varphi) = V_{\infty}\,e^{2b\varphi} + V_{\infty,1}\,e^{2\gamma \varphi} + \ldots\,, \quad \varphi \rightarrow \infty\,,
	\label{exponentialpotential}
\end{equation}
where $ b> \gamma > 0 $, $ V_{\infty}< 0$ and $V_{\infty,1}\leq 0 $ are constants.\footnote{One could consider more general forms such as e.g.
\begin{equation} \label{genasympt}
V(\varphi) = V_{\infty}\,e^{2b\varphi}\left(1 + \frac{b_1}{\varphi}+ \frac{b_2}{\varphi^2} +\ldots\right) + V_{\infty,1}\,e^{2\gamma \varphi} \left(1 + \frac{c_1}{\varphi} + \frac{c_2}{\varphi^2}+\ldots\right) + \ldots\,, \quad \varphi \rightarrow \infty\,,
\end{equation}
This does not affect qualitatively our  results as long as the leading exponential  is within the confining regime \eqref{int3}, as it will be apparent from the analysis in the subsequent sections.}

\begin{figure}[t]
	\centering
	\begin{tikzpicture}
		\node (img1)  {\includegraphics{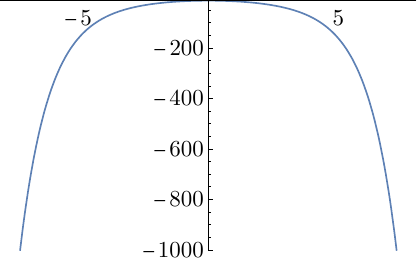}};
		\node[right=of img1, node distance=0cm, yshift=-2.0cm, xshift=-4.5cm] {$ V $};
		\node[above=of img1, node distance=0cm, anchor=center,yshift=-0.9cm,xshift=-3.8cm] {$ \varphi $};
	\end{tikzpicture}
	\caption{Plot of the potential \eqref{eq:numerics_potential}. It has a single extremum (a maximum) at $\varphi = 0$ and it is monotonically decreasing towards $\varphi\rightarrow \infty$ where it has exponential asymptotics.}
	\label{potentialplot}
\end{figure}

Further, we assume that in the intermediate region $ 0 < \varphi < \infty $ the potential is a monotonically decreasing function $ V'(\varphi) < 0 $ so that $\varphi = 0$ is the only extremum of the potential.  For such a potential,  the large-$\varphi$ limit is reached in the far IR of the theory. An explicit example of such a potential considered in this work is given by (see also \cite{Gubser:2008ny,Gubser:2008yx,Gursoy:2016ggq,Kiritsis:2016kog,Ghodsi:2024jxe})
\begin{equation}
	V(\varphi) = -\frac{d(d-1)}{\ell^{2}}+\biggl(\frac{\Delta_-(\Delta_--d)}{2\ell^{2}}-4V_{\infty}\,b^{2}\biggr)\,\varphi^{2}+4V_{\infty}\sinh^{2}{(b\varphi)}\,,
	\label{eq:numerics_potential}
\end{equation}
which has no subleading exponential $V_{\infty,1} = 0$ as $\varphi\rightarrow \infty$ (see Figure \ref{potentialplot}). This potential will be used for numerics, while all analytical calculations will be done including the possibility that $V_{\infty,1}\neq 0$.

The interest in parametrizing the leading IR asymptotics of the potential in terms of exponentials as in \eqref{exponentialpotential} is in the fact that  this parametrization gives simple classification of the IR singularity one generically finds at finite $\varphi$, whether this is acceptable in holographic terms, and what are the properties of the corresponding solution, as we review below.

First of all, one must distinguish whether the exponent lies above or below the special value
\begin{equation} \label{gubsB}
b_{\text{G}} = \sqrt{\frac{d}{2\,(d-1)}}\,,
\end{equation}
which we refer to as the {\em Gubser bound}.  For $b > b_{\text{G}}$, one can show that {\em no} solution exists (for a flat boundary $\kappa = 0$) which satisfies Gubser's criterion \cite{gubser_curvature_2000} for an acceptable singularity, that it can be made regular by a small deformation (e.g. turning on a small black hole mass).  In fact, above this bound there is no solution with  $\kappa = 0$ which has well defined large-$\varphi$ asymptotics: for a symmetric potential like the one in \eqref{eq:numerics_potential},  $\varphi$ oscillates  back and forth between $-\infty$ and $+\infty$ exploring larger and larger field values with each oscillation.  In addition, as will be reviewed in Section \ref{subsec:reduction}, only exponents below the Gubser bound have an uplift to higher-dimensional Einstein gravity (with an internal sphere $S^N$ or a torus $T^N$) \cite{Gouteraux:2011qh,Gouteraux:2011ce}. We therefore consider the range $b > b_{\text{G}}$ of exponents unacceptable for holographic theories.

For $0< b < b_{\text{G}}$ on the other hand, one can show \cite{Gursoy:2008za,Kiritsis:2016kog} that for a flat boundary geometry $\kappa = 0$, a unique special solution exists which satisfies Gubser's criterion, such that the superpotential behaves at large-$\varphi$ roughly as the square root of the potential:

\begin{equation} \label{specialasymp}
W \sim W_{\infty}\, e^{b\varphi}\,, \qquad W_{\infty} \equiv 2\, \sqrt{\frac{-V_{\infty}}{1 - 2\,(b^2- b_{\text{c}}^2)}} \,,\qquad b_{\text{c}} \equiv \frac{1}{\sqrt{2\,(d-1)}}\,.
\end{equation}
The assumption is that this solution, although singular,  is the only physically relevant one with a consistent holographic interpretation.  Requiring the IR asymptotics \eqref{specialasymp} on the solution of the bulk theory plays the same role as requiring regularity (it fixes one of the integration constants of the system). The  solution with generic integration constants (called type 0) instead has $W\propto e^{b_{\text{G}}\, \varphi}$ as $\varphi \to \infty$, regardless of the value of $b$.

Restricting the discussion to the special ``regular'' solution \eqref{specialasymp}, the parameter $b$ also determines  whether the  dual field theory, when defined on flat space (i.e. $ T(\varphi) = 0 $),  is confining \cite{Gursoy:2007er}. If we think of the dual field as a large-$N$ gauge theory,  the holographic  calculation of the Wilson loop \cite{Maldacena:1998im,Rey:1998ik}  results in an area law when \cite{Gursoy:2007er}
\begin{equation}
	b> b_{\text{c}} \equiv \frac{1}{\sqrt{2\,(d-1)}}\,.
	\label{confinementbound}
\end{equation}
We refer to the bound \eqref{confinementbound} as the {\em confinement bound} and it is always smaller than the Gubser bound $b_{\text{c}} < b_{\text{G}}$.

Above the confinement bound, linear fluctuations of the metric and scalar field  around solutions have a gapped and discrete  spectrum, which corresponds in the dual field theory to a discrete tower of ``glueballs'' with masses  $ m_n^2 \sim n^{2} $. Vice-versa, below the confinement bound, the spectrum is ungapped and continuous \cite{Gursoy:2007er}. In addition for $b> b_{\text{c}}$, the theory displays a thermal first-order phase transition towards a high-temperature black hole phase which corresponds to a deconfined state \cite{Gursoy:2008za}. In the high-temperature phase the IR singularity is cloaked by a horizon, at which the scalar field reaches a finite value $\varphi_0$ which sets the temperature of the solution. For $b<b_{\text{c}}$, the theory is in the black hole phase at all non-zero temperatures.

There are other special values of the exponent $b$ which discriminate between different qualitative types of behavior. The most important one for this paper is the {\em Efimov bound},
\begin{equation}
b_{\text{E}} = \frac{2}{\sqrt{(d-1)(9-d)}}\,, \label{efbound}
\end{equation}
which we shall derive and discuss in Sections \ref{subsec:neartypeC_solutions} and \ref{subsec:Efimov_oscillations}. For dimensions $1 < d < 8$, we have the hierarchy
\begin{equation}
    0 <b_{\text{c}}<b_{\text{E}}< b_{\text{G}}\,,
\label{ine}\end{equation}
while for $d> 8$, the Efimov bound lies above the Gubser bound $b_{\text{E}}>  b_{\text{G}}$.

Another special value that arises for $\kappa = 0$ is the {\em spectrum computability bound},
\begin{equation}
b_{\text{s}} = \sqrt{\frac{d+2}{6\,(d-1)}}\,,
\label{bsandbE}
\end{equation}
above which extra IR boundary conditions for the fluctuations must be imposed \cite{Kiritsis:2016kog}. The spectrum computability bound satisfies $b_{\text{c}} < b_{\text{s}}<b_{\text{G}}$ for all $1<d<\infty$, $b_{\text{s}} \geq b_{\text{E}}$ for $ d \leq 6$ and $b_{\text{s}} < b_{\text{E}}$ for $d> 6$. For the current work, $b_{\text{s}}$ will not play any role.

\subsection{Exponential potentials from dimensional reduction}\label{subsec:reduction}

The Einstein--scalar theory \eqref{einsteinscalar} with an exponential potential \eqref{exponentialpotential} can be obtained from pure Einstein gravity in $ d+N+1 $ dimensions by dimensional reduction on an internal $ S^{N} $ \cite{Gouteraux:2011ce,Gouteraux:2011qh}. The action is given by
\begin{equation}
	I_{d+N+1} = M_{\text{p}}^{d+N-1}\int_{\widetilde{\mathcal{M}}} d^{d+N+1}\tilde{X} \sqrt{\tilde{g}}\,(\widetilde{R} -2\widetilde{\Lambda})\,,
	\label{eq:higher_dim_action}
\end{equation}
where the higher-dimensional cosmological constant $ \widetilde{\Lambda} = -(d+N)(d+N-1)\slash(2\tilde{\ell}^2) < 0 $ defined by the curvature radius $\tilde{\ell}$ is independent of the lower-dimensional one $ \Lambda $ appearing in \eqref{potentialUVfixedpoint}. We  parametrize our coordinates as $ \tilde{X} = (\tilde{u},x^{\mu},y^{n}) $ and consider $d+N+1$-dimensional Euclidean metrics of the form
\begin{equation}
	ds^{2} = d\tilde{u}^{2} + e^{2A_1(\tilde{u})}\,\zeta_{\mu\nu}\,dx^{\mu}dx^{\nu} + e^{2A_2(\tilde{u})}\,\bar{\zeta}_{nm}\,dy^{n}dy^{m}\,,
	\label{highermetric}
\end{equation}
where $A_{1,2}(\tilde{u})$ are two independent scale factors and the two metrics $ \zeta$, $ \bar{\zeta} $ are given by
\begin{equation}
\zeta_{\mu\nu}\,dx^{\mu}dx^{\nu} = \alpha^{2}ds^{2}_{S^{d}}, \quad \bar{\zeta}_{nm}\,dy^{n}dy^{m} = \bar{\alpha}^{2}ds^{2}_{S^N}\,.
\end{equation}
Here $ \mu,\nu =1,\ldots, d $ and $ n,m = 1,\ldots, N $ run over the spatial and internal directions respectively. The metric \eqref{highermetric} is a foliation of $ \widetilde{\mathcal{M}} $ by $ S^{d}\times S^{N} $-slices.

For metrics of the form \eqref{highermetric} the equations of motion reduce to a coupled system of ordinary differential equations for $ (A_1(\tilde{u}),A_2(\tilde{u})) $.
The $ (\tilde{u},\tilde{u}) $ component of Einstein's equations is explicitly (see Appendix \ref{app:reductiondetails} for details)
\begin{equation}
	d(d-1)\left(A_1'+\frac{N}{d-1}\,A_2'\right)^{2}-\frac{N(d+N-1)}{d-1}\,(A_2')^{2} + \Bigl(2\widetilde{\Lambda} - d\kappa e^{-2A_1}-N\bar{\kappa}e^{-2A_2}\Bigr) = 0\,,
	\label{einstein1_text}
\end{equation}
while the $ (\mu,\nu) $ components are equivalent to
\begin{align}
	&d(d-1)\left(A_1'+\frac{N}{d-1}\,A_2'\right)^{2}+2(d-1)\biggl(A_1''+\frac{N}{d-1}\,A_2''\biggr)+\frac{N(d+N-1)}{d-1}\,(A_2')^{2}\nonumber\\
	&-2NA_2'\left(A_1'+\frac{N}{d-1}\,A_2'\right)+ \Bigl(2\widetilde{\Lambda} - (d-2)\kappa\, e^{-2A_1}-N\bar{\kappa}\,e^{-2A_2}\Bigr) = 0\,,
	\label{einstein2_text}
\end{align}
Here we have defined $\bar{\kappa} \equiv \frac{N-1}{\bar{\alpha}^{2}} $ so that $N\bar{\kappa}$ is the scalar curvature of the metric $\bar{\zeta}$. The $ (n,m) $ components give the same equation as the $ (\mu,\nu) $ components, but with the replacements $ A_1\leftrightarrow A_2 $, $ d \leftrightarrow N $ and $ \kappa \leftrightarrow \bar{\kappa} $.

In Appendix \ref{app:reductiondetails} we show these equations for $ (A_1(\tilde{u}),A_2(\tilde{u})) $ are equivalent to the equations of motion \eqref{EOM1} - \eqref{EOM3} for $ (A(u),\varphi(u)) $ with an exponential potential
\begin{equation}
V(\varphi) = -N\bar{\kappa}\,e^{\sqrt{\frac{2\,(d+N-1)}{N(d-1)}}\,\varphi}+2\widetilde{\Lambda}\,e^{\sqrt{\frac{2N}{(d-1)(d+N-1)}} \varphi}\,,
\label{potentialreductiontext}
\end{equation}
when the lower-dimensional fields are given in terms of the higher-dimensional ones as
\begin{equation}
A(u) = A_1(\tilde{u}) + \frac{N}{d-1}\,A_2(\tilde{u})\,, \quad \varphi(u) = -\sqrt{\frac{2N(d+N-1)}{d-1}}\,A_2(\tilde{u})
\label{lowerdimensionalfieldstext}
\end{equation}
and when the radial coordinate $\tilde{u}$ is related to the lower-dimensional counterpart via the equation
\begin{equation}
\frac{d\tilde{u}}{du} = e^{-\frac{N}{d-1}A_{2}(\tilde{u})}\,.
\label{ututext}
\end{equation}
Notice that the potential takes the exponential form \eqref{potentialreductiontext} for all values of the scalar field $\varphi $ and not only asymptotically. Therefore the potential does not have a maximum dual to a UV fixed-point.

A priori, the dimension of the internal sphere is an integer that satisfies $1 \leq N  < \infty$, however, it can be analytically continued to real values. For these values, one can observe that the first exponent in \eqref{potentialreductiontext} is always the leading one when $\varphi\rightarrow \infty$. Therefore comparing with \eqref{exponentialpotential}, we can identify the coefficients as
\begin{equation}
	V_{\infty} = -N\bar{\kappa}\,,\quad V_{\infty,1} = 2\widetilde{\Lambda}
 \label{eq:potential_coefficients}
\end{equation}
and the exponents as
\begin{equation}
	b = \sqrt{\frac{d+N-1}{2N(d-1)}}\,,\quad \gamma = \gamma_{\text{c}}\equiv \sqrt{\frac{N}{2\,(d-1)(d+N-1)}} = \frac{1}{2\,(d-1)\,b}\,.
 \label{eq:reduction_exponents}
\end{equation}
For internal dimensions $1 < N  < \infty$, the leading exponent satisfies:
\begin{equation}
	b_{\text{c}}< b< b_{\text{G}}\,,
\end{equation}
where the lower bound, corresponding to $N \rightarrow \infty$, coincides exactly with the confinement bound $b_{\text{c}}$ \eqref{confinementbound}, while the upper bound, corresponding to $ N \to 1$, coincides with the Gubser bound $b_{\text{G}}$ \eqref{gubsB}. Notice that the confinement bound can never be reached using finite-dimensional spheres and potentials with $ b > b_{\text{G}} $ do not have a higher-dimensional geometric interpretation. In addition, the subleading exponential $ \gamma_{\text{c}} = \frac{1}{2\,(d-1)\,b} = \frac{b_{\text{c}}^{2}}{b}$ coming from dimensional reduction is fixed in terms of $b$ and is not an independent parameter. The subleading exponential satisfies $\gamma_{\text{c}}< b_{\text{c}}$ when $b>b_{\text{c}}$.

The case $N=1$ needs to be treated separately, because in this case the curvature of the internal space vanishes: while one naively obtains $b=b_{\text{G}}$ by inserting $N=1$ in equation \eqref{eq:reduction_exponents}, the coefficient of the  (would-be leading) first  term in the potential \eqref{potentialreductiontext} vanishes, as $\bar{\kappa}=0$. Therefore the dimensional reduction on  $S^1$ gives a potential in the non-confining regime since the (now-leading) surviving term in the potential \eqref{exponentialpotential} has:
\begin{equation}
\gamma = \sqrt{\frac{1}{2 d\, (d-1)}} < b_{\text{c}}\,. \label{N=1}
\end{equation}
Thus, this case falls in the range of toroidal  (generalized) dimensional reductions on $T^N$ with $0<N<\infty$, where the upper limit corresponds to approaching $b_{\text{c}}$ from below (in the deconfining regime).

The case $N=1$ is special from another perspective. For $N\neq 1$,  the only way in which an internal $T^N$ can shrink regularly to zero  size (with the $d$-dimensional factor remaining finite) is a  (locally-)$AdS_{N+1}$ space-time interior,
\begin{equation} \label{Nshrink}
ds^2 = du^2 + e^{-2u}\, ds_{T^N}^2 + \alpha_0^2\, ds_{S^d}^2 +\ldots\,, \quad u \to \infty\,,
\end{equation}
for some radius $\alpha_0$. Instead, for $N=1$ the torus $T^1 = S^1$ can shrink regularly in two different ways:
\begin{equation} \label{1shrink}
ds^2 = \begin{dcases}
    du^2 + e^{-2u}\, ds_{S^1}^2 + \alpha^2_0\, ds_{S^d}^2 + \ldots\,,\quad  &u\to \infty\\
    du^2 + (u-u_0)^2\, ds_{S^1}^2 + \alpha_0^2\, ds_{S^d}^2 + \ldots\,,\quad &u\to u_0
\end{dcases}\,.
\end{equation}
Therefore, in the case of an $S^1$ compactification there are possibly two  regular IR geometries  for which the $S^d$ remains finite (the ones in \eqref{1shrink}),  whereas for toroidal $N\neq 1$ compactifications there is at most one,  \eqref{Nshrink}.

In the reduced theory, admitting multiple branches of solution is a generic feature of theories in the confining range, $b_{\text{c}} < b < b_{\text{G}}$, but not for $b<b_{\text{c}}$. So it seems that  the $N=1$ reduction is a special case, and it does not  capture the IR physics correctly. Another possibility is that the special value  in \eqref{N=1} for the leading exponent of the potential leads to multiple branches of solutions in the reduced theory. We leave the resolution of this clash to future investigation, and from now on we consider $N>1$ strictly.

\section{Classification of solutions}\label{sec:classification}

In the context of holography, lower-dimensional theories often arise from dimensional reduction of higher-dimensional ones (which may happen to be  UV complete, if the reduction comes from  a well-defined string theory background). The notion of regularity in the lower-dimensional description may therefore be more relaxed because sometimes even singular solutions uplift to regular geometries.  In what follows, we use this fact as a guide to decide whether to accept an IR singularity.

As we have seen in Section \ref{subsec:reduction},  $d+1$-dimensional Einstein-scalar theories with an asymptotically exponential potential,  $ V \sim e^{2b\varphi} $ as $\varphi \to \infty$ have an uplift to $ d+N+1 $-dimensional pure Einstein gravity with a negative cosmological constant,  with $N$ determined by $b$.  More precisely,  solutions of the lower-dimensional theory uplift to the class of solutions  \eqref{highermetric} of the higher-dimensional theory. In this situation, regularity of the uplifted solution may be used as a guideline to accept an IR singularity in the lower-dimensional Einstein-scalar theory  and  to consider the corresponding solution   ``regular'' (in the sense explained  in the introduction).

In  this section, we relate the lower-dimensional solutions with their higher-dimensional counterparts, and relate the corresponding IR parameters. As we shall see,  the IR endpoint always corresponds  to at least one of the spheres  in the uplift \eqref{highermetric}  shrinking  to zero size.  We find three classes of solutions:
\begin{itemize}
\item Type III: these are regular both in the lower- and in the higher-dimensional theory. The IR endpoint corresponds to the $d$-dimensional (space-time) sphere shrinking to zero;
\item Type II: these are singular solutions which uplift to regular geometries.  The IR endpoint corresponds to the $N$-dimensional (internal) sphere shrinking to zero;
\item Type I: these are singular solutions,  whose uplift is also singular (both spheres shrink to zero at the same point).
\end{itemize}
Based on these considerations, we should accept as ``regular'' the type II solutions. As for the type I solution, it is clear from the higher-dimensional picture  that these  are a degenerate limit of the type II and type III solutions. Since in the reduced theory the kind  of IR singularity of type I  is exactly the same as the one found in type II, there is no compelling reason to discard type I solutions, which we also consider ``regular''.
Also, the fluctuation problem around such a singularity is holographically well-defined if $b$ is below the computability bound.\footnote{See discussion  about the computability bound, equation \eqref{bsandbE}. The computability bound was derived for $\kappa = 0$, but it also applies  to type I and type II solutions for $\kappa \neq 0$,  since the spectral problem is essentially determined by the exponential behavior in the superpotential \cite{Kiritsis:2016kog}, and  for these solutions this  is the same as for  $\kappa = 0$.}

An important remark is  that the map between higher- and lower-dimensional theories only holds  in the deep IR,  where the potential is well approximated by the leading exponential $ V \sim e^{2b\varphi} $, and does not extend to the UV where the two theories are very different.  However, to discuss regularity it is sufficient to analyze the IR.

\subsection{Solutions of the higher-dimensional Einstein theory}\label{subsec:classification_higher}

We classify solutions of the higher-dimensional Einstein's equations of the form \eqref{highermetric} parametrized by two functions $ (A_1(\tilde{u}),A_2(\tilde{u})) $. We assume that the solution exists in the region $ -\infty < \tilde{u} < \tilde{u}_0 $ for some finite $ \tilde{u}_0< \infty $. There are three ways for the bulk solution $ (A_1,A_2) $ to end at $ \tilde{u} = \tilde{u}_0 $: the bulk geometry caps off (a) by the space-time sphere $ S^{d} $ shrinking to zero size, (b) by the internal sphere $ S^{N} $ shrinking to zero size or (c) by both $ S^{d} $ and $S^{N} $ shrinking at the same point. We call these geometries type I, II and III respectively.

In a type III geometry, the metric near the shrinking point takes the form
\begin{equation}
	ds^{2} = d\tilde{u}^{2}  + (\tilde{u}-\tilde{u}_0)^{2}\,ds^{2}_{S^{d}}+ \bar{\alpha}_{\scriptscriptstyle\text{IR}}^{2}\,ds^{2}_{S^{N}} + \ldots, \quad \tilde{u}\rightarrow \tilde{u}_0^{-}\,,
	\label{eq:typeA_higher_metric}
\end{equation}
which is locally $\mathbb{R}^{d+1}\times S^{N}$ and does not contain any conical singularities. There is a one-parameter family of type III geometries parametrized by the non-zero radius $ \bar{\alpha}_{\scriptscriptstyle\text{IR}}>0 $ of the internal sphere $S^{N}$ at the IR endpoint. For the scale factors, the metric \eqref{eq:typeA_higher_metric} amounts to
\begin{equation}
	\text{type III:}\quad\begin{dcases}
		A_1(\tilde{u}) = \log{\biggl(\frac{\tilde{u}_0-\tilde{u}}{\alpha}\biggr)} + \ldots\\
		A_2(\tilde{u}) =  \log{\frac{\bar{\alpha}_{\IR}}{\bar{\alpha}}} + \ldots
	\end{dcases}, \quad
	\tilde{u}\rightarrow \tilde{u}_0^{-}\,,
	\label{typeA}
\end{equation}
where ellipsis denote terms that go to zero when $ \tilde{u}\rightarrow \tilde{u}_0^{-} $.

The metric of a type II geometry, on the other hand, behaves as
\begin{equation}
	ds^{2} = d\tilde{u}^{2}  + \alpha_{\scriptscriptstyle\text{IR}}^{2}\,ds^{2}_{S^{d}}+ (\tilde{u}-\tilde{u}_0)^{2}\,ds^{2}_{S^{N}} + \ldots, \quad \tilde{u}\rightarrow \tilde{u}_0^{-}\,,
\end{equation}
which is locally $\mathbb{R}^{N+1}\times S^{d}$ without singularities. The free parameter of the type II geometry is $\alpha_{\scriptscriptstyle\text{IR}}$ which is the radius of the space-time sphere $ S^{d} $ at the IR endpoint. The corresponding scale factors are given by
\begin{equation}
	\text{type II:}\quad\begin{dcases}
		A_1(\tilde{u}) =  \log{\frac{\alpha_{\IR}}{\alpha}} + \ldots\\
		A_2(\tilde{u}) =  \log{\biggl(\frac{\tilde{u}_0-\tilde{u}}{\bar{\alpha}}\biggr)}+ \ldots
	\end{dcases}, \quad
	\tilde{u}\rightarrow \tilde{u}_0^{-}\,.
	\label{typeB}
\end{equation}
In the type I geometry, both of the spheres $S^{N}$ and $S^{d}$ shrink to zero size at the IR endpoint corresponding to $\alpha_{\IR} = \bar{\alpha}_{\IR} = 0$. This defines the type I geometry which is, hence, the $\bar{\alpha}_{\IR}\rightarrow 0$ limit of a type III or the $ \alpha_{\IR}\rightarrow 0$ limit of a type II geometry. It is characterized by scale factors
\begin{equation}
	\text{type I}:\quad\begin{dcases}
		A_1(\tilde{u}) =  \log{\biggl(\frac{\tilde{u}_0-\tilde{u}}{\alpha}\biggr)}+A_{10}^{\text{I}} + \ldots\\
		A_2(\tilde{u}) =  \log{\biggl(\frac{\tilde{u}_0-\tilde{u}}{\bar{\alpha}}\biggr)}+A_{20}^{\text{I}} + \ldots
	\end{dcases}, \quad
	\tilde{u}\rightarrow \tilde{u}_0^{-},
	\label{typeC}
\end{equation}
which correspond to a metric of the form
\begin{equation}
	ds^{2} = d\tilde{u}^{2}  + (\tilde{u}-\tilde{u}_0)^{2}\,e^{2A_{10}^{\text{I}}}\,ds^{2}_{S^{d}}+ (\tilde{u}-\tilde{u}_0)^{2}\,e^{2A_{20}^{\text{I}}}\,ds^{2}_{S^{N}} + \ldots, \quad \tilde{u}\rightarrow \tilde{u}_0^{-}.
\end{equation}
This metric has a conical singularity at $ \tilde{u} = \tilde{u}_0 $ and divergent curvature invariants for all values of $ A_{10}^{\text{I}}, A_{20}^{\text{I}} $.

\begin{figure}
	\centering
	\begin{center}
		\textbf{Classification of solutions of Einstein gravity on} $\text{AdS}_{d+1}\times S^{N}$
	\end{center}
	{\tabulinesep=1.6mm
		\begin{tabu} {|c|c|c|c|c|}
			\hline
			Geometry& Topology & Regularity & $ A_1(\tilde{u}) $ & $ A_2(\tilde{u}) $ \\
			\hline
			type III & $S^{d}$ shrinks & regular &  $\displaystyle{\log{\biggl(\frac{\tilde{u}_0-\tilde{u}}{\alpha}\biggr)}} $& $ \displaystyle{ \log{\frac{\bar{\alpha}_{\IR}}{\bar{\alpha}}} } $ \\
			\hline
			type II & $S^{N}$ shrinks& regular & $\displaystyle{ \log{\frac{\alpha_{\IR}}{\alpha}}}$&$ \displaystyle{ \log{\biggl(\frac{\tilde{u}_0-\tilde{u}}{\bar{\alpha}}\biggr)}} $ \\
			\hline
			type I & $S^{d}$ and $S^{N}$ shrink & singular & $\displaystyle{\log{\biggl(\frac{\tilde{u}_0-\tilde{u}}{\alpha}\biggr)}}$&$ \displaystyle{\log{\biggl(\frac{\tilde{u}_0-\tilde{u}}{\bar{\alpha}}\biggr)} } $\\
			\hline
	\end{tabu}}
	\caption{Types of IR boundary conditions near the point $ \tilde{u}_0 $ in the $ d+N-1 $-dimensional theory when the internal space is a sphere $ S^{N} $.}
	\label{higherdimensionaltypes}
\end{figure}

Type II and III geometries are regular (no conical singularities and finite curvature invariants) at the point $\tilde{u} = \tilde{u}_0$ where one of the spheres smoothly caps off while the type I geometry is singular. All three kinds of geometries, including the singular type I geometry, appear as solutions of Einstein's equations \eqref{eq:appEinstein1} - \eqref{eq:appEinstein3} computed in Appendix \ref{app:reductiondetails}. The type I solution  may be written explicitly (see also \cite{Aharony:2019vgs})
\begin{align}
	A_1^{\text{I}}(\tilde{u}) &= \log{\biggl[\frac{\tilde{\ell}}{\alpha}\sinh{\biggl(\frac{\tilde{u}_0-\tilde{u}}{\tilde{\ell}}\biggr)}\biggr]}+\frac{1}{2}\log{\biggl(\frac{d-1}{d+N-1}\biggr)}\,,\nonumber\\
	A_2^{\text{I}}(\tilde{u}) &= \log{\biggl[\frac{\tilde{\ell}}{\bar{\alpha}}\sinh{\biggl(\frac{\tilde{u}_0-\tilde{u}}{\tilde{\ell}}\biggr)}\biggr]}+\frac{1}{2}\log{\biggl(\frac{N-1}{d+N-1}\biggr)}\,.\label{eq:typeChigherdimensional_text}
\end{align}
Type II and III solutions cannot be found analytically for all values of $\tilde{u}$, but the first few terms in the expansion of $A_{1,2}(\tilde{u})$ around $\tilde{u} = \tilde{u}_0$ are derived in Appendix \ref{app:highersolutions} and given in equations \eqref{eq:typeBexplicitexp} and \eqref{eq:typeAexplicitexp} respectively. We also point out that the type II and III solutions are related by a $\mathbb{Z}_2$-transformation of the Einstein's equations \eqref{eq:appEinstein1} - \eqref{eq:appEinstein3} that simultaneously exchanges $ A_1\leftrightarrow A_2 $, $ d\leftrightarrow N $, $\alpha\leftrightarrow \bar{\alpha}$ and $\alpha_{\IR}\leftrightarrow \bar{\alpha}_{\IR}$.

To summarize, there are two types of regular $ d+N+1 $-dimensional solutions where either the $ S^{d} $ or the $ S^{N} $ shrink to zero size in a smooth manner: the former is called a type III solution while the latter a type II solution.\footnote{This distinction makes sense  if we have the mindset of dimensional reduction, in which we think of $S^d$ as space-time and $S^N$ as the internal sphere} At the interface of the two regular families of solutions lies a unique type I solution for which both $ S^{d} $ and $ S^{N} $ shrink to zero size at the same point. This geometry is singular, with diverging curvature invariants, but still a solution of the equations of motion. The IR behavior of these solutions are listed in Table \ref{higherdimensionaltypes}.

\paragraph{Dimensional reduction.} We now reduce the above type I, II and III geometries over the internal $S^{N}$ to $d+1$ dimensions. First consider type III solutions and let $ u_0 \equiv u(\tilde{u}_0) $ denote the endpoint of the geometry in the radial coordinate $ u $ of the Einstein-scalar theory. Using \eqref{lowerdimensionalfieldstext}, the higher-dimensional type III geometry \eqref{typeA} translates to (see Appendix \ref{app:highersolutionsandreduction})
\begin{equation}
	\text{type III:}\quad\begin{dcases}
		A(u) = \log{\biggl(\frac{u_0-u}{\alpha}\biggr)}  + \mathcal{O}(u_0-u)^{2}\\
		\varphi(u) =  \varphi_0  + \mathcal{O}(u_0-u)^{2}
	\end{dcases}, \quad
	u\rightarrow u_0^{-}\,,
	\label{typeAreduced}
\end{equation}
where $ \varphi_0 \equiv \varphi(u_0) $ is explicitly\footnote{Here we have written $ d $ and $ N $ in terms of $ b $ and $ b_{\text{c}} $ via equations \eqref{confinementbound} and \eqref{eq:reduction_exponents}.}
\begin{equation}
	\varphi_0 =\frac{b}{b^{2}-b_{\text{c}}^{2}}\,\log{\frac{\bar{\alpha}}{\bar{\alpha}_{\scriptscriptstyle \text{IR}}}}\,.
	\label{varphi0kappaIR}
\end{equation}
Equation \eqref{varphi0kappaIR} implies that the IR radius $ \bar{\alpha}_{\IR} $ of the internal sphere $ S^{N} $ determines the endpoint value of the scalar field in the lower-dimensional description. Notice that this identification between $ \varphi_0 $ and $ \bar{\alpha}_{\scriptscriptstyle \text{IR}} $ is only valid when the endpoint $ \varphi_0 $ lies in the regime where the lower-dimensional potential is well approximated by an exponential $ V \sim e^{2b\varphi} $.

The scale factor $A(u)$ in \eqref{typeAreduced} corresponds to the $ d+1 $-dimensional metric
\begin{equation}
	ds^{2} = du^{2} + (u-u_0)^{2}\,ds^{2}_{S^{d}} + \ldots, \quad u \rightarrow u_0\,,
\end{equation}
which is completely regular from the lower dimensional point of view (regular metric and a finite scalar field). The function \eqref{S} are given by
\begin{equation}
	\text{type III:}\quad
		S(\varphi) = S_0\sqrt{\varphi_0 - \varphi} +\mathcal{O}(\varphi_{0} - \varphi)\,, \quad
	\varphi\rightarrow \varphi_0^{-}\,,
	\label{functionstypeAreduced}
\end{equation}
where the coefficient $ S_0 $ is determined by the subleading terms in the expansions \eqref{typeAreduced} that are fixed by the equations of motion (see section \ref{subsec:IRsolutions}). Therefore type III geometries are characterized by an $S(\varphi)$ that vanishes at the IR endpoint $\varphi = \varphi_0$.

Dimensionally reducing type II geometries \eqref{typeB} using \eqref{lowerdimensionalfieldstext},  we obtain  (see Appendix \ref{app:highersolutionsandreduction})
\begin{equation}
	\text{type II:}\quad\begin{dcases}
		A(u) = \frac{b_{\text{c}}^{2}}{b^{2}}\,\log{\biggl(\frac{u_0-u}{\alpha}\biggr)} + \mathcal{O}(1)\\
		\varphi(u) =  -\frac{1}{b}\log{\biggl(\frac{u_0-u}{\bar{\alpha}}\biggr)} + \ldots
	\end{dcases}, \quad
	u\rightarrow u_0^{-}\,.
	\label{typeBreduced}
\end{equation}
The scale factor here corresponds to the $d+1$-dimensional metric
\begin{equation}
	ds^{2} = du^{2} + (u-u_0)^{2b_{\text{c}}^{2}\slash b^{2}}\alpha_{\scriptscriptstyle\text{IR}}^{2}\,ds^{2}_{S^{d}} + \ldots, \quad u \rightarrow u_0^{-},
\end{equation}
which is singular at $ u = u_0 $. This singularity in the metric is supported by a divergence $\varphi\rightarrow \infty$ in the scalar field at $\tilde{u} = \tilde{u}_0$. Therefore, from a lower-dimensional point of view type II geometries are  singular even though they uplift to regular metrics. Notice however that the {\em generic} singular solution of type 0 (discussed in the introduction and in Appendix \ref{app:reducedsolutions}) uplift to a singular geometry in $d+N+1$-dimensions.

At the endpoint of the geometry where $ \varphi \rightarrow \infty $, the $S(\varphi)$ function \eqref{S} diverges exponentially as (see Appendix \ref{app:highersolutionsandreduction})
\begin{equation}
	\text{type II:}\quad
	S(\varphi) = S_{\infty}^{\text{II}}\,e^{b\varphi} + \bigl(S_{\infty,1}^{\text{II}}+S_{*}^{\text{II}}\bigr)\,e^{(2\gamma_{\text{c}}-b)\,\varphi} +  \ldots\,, \quad
	\varphi\rightarrow \infty\,,
	\label{functionstypeBreduced}
\end{equation}
where ellipsis denote subleading terms in the $ \varphi \rightarrow \infty $ limit and the coefficient $ S_{\infty}^{\text{II}} $ is explicitly
\begin{equation}
	S_{\infty}^{\text{II}} =\sqrt{\frac{2N(d+N-1)}{d-1}}\,\frac{1}{\bar{\alpha}}\,.
	\label{eq:SBinfty_reduction}
\end{equation}
The function $W(\varphi)$ has the same divergence structure. The exponent $ \gamma_{\text{c}} $ is defined in \eqref{eq:reduction_exponents} and it satisfies $ \gamma_{\text{c}} < b $ above the confinement bound $b > b_{\text{c}}$. Therefore the second exponential in \eqref{functionstypeBreduced} is subleading with respect to the first one and its coefficients are determined by higher-dimensional parameters as (see Appendix \ref{app:highersolutionsandreduction})
\begin{equation}
	S_{\infty,1}^{\text{II}} = -\frac{d}{4}\frac{N-1}{N+1}\left(\frac{\bar{\alpha}}{\alpha}\right)^{2}\left(\frac{\alpha}{\alpha_{\IR}}\right)^{2}S_{\infty}^{\text{II}}\,,\quad S_{*}^{\text{II}} = \frac{(N-1)(d-N-1)}{N\,(N+1)(d+N-1)}\frac{\widetilde{\Lambda}}{\bar{\kappa}}\,S_{\infty}^{\text{II}}\,,
	\label{eq:Sinfty1B_text}
\end{equation}
where $ S_*^{\text{II}} $ is determined by the subleading terms in the expansion \eqref{typeBreduced} and is fixed once the equations of motion are imposed. As a result, the free parameter $\alpha_{\IR}$ of the type II geometry appears in $S_{\infty,1}^{\text{II}}$ in the coefficient of the subleading exponential in the lower-dimensional theory.

Lastly, the singular type I geometry \eqref{typeC} reduces to (see Appendix \ref{app:highersolutionsandreduction})
\begin{equation}
	\text{type I:}\quad\begin{dcases}
		A(u) = \log{\biggl(\frac{u_0-u}{\alpha}\biggr)} +\mathcal{O}(1)\\
		\varphi(u) =  -\frac{1}{b}\log{\biggl(\frac{u_0-u}{\bar{\alpha}}\biggr)} + \mathcal{O}(1)
	\end{dcases}, \quad
	u\rightarrow u_0^{-}
	\label{typeCreduced}
\end{equation}
where the subleading terms are fixed by equations of motion and can be computed explicitly. Since the scalar field diverges $ \varphi \rightarrow \infty $ the type I geometry is also singular from the lower-dimensional point of view.\footnote{The metric also has a conical singularity since the $ \mathcal{O}(1) $ terms in $ A(u) $ turn out to be non-zero on-shell.} The $S(\varphi)$ function \eqref{S} has a similar exponential divergence structure as for type II geometries,
\begin{equation}
		\text{type I:}\quad S(\varphi) = S_{\infty}^{\text{I}}\,e^{b\varphi} + S_{*}^{\text{I}}\,e^{(2\gamma_{\text{c}}-b)\,\varphi} +  \ldots\,, \quad
	\varphi\rightarrow \infty\,,
	\label{functionstypeCreduced}
\end{equation}
but the coefficients are different
\begin{equation}
	S_{\infty}^{\text{I}} = \sqrt{\frac{2N(N-1)}{d-1}}\,\frac{1}{\bar{\alpha}}\,, \quad S_{*}^{\text{I}} = -\frac{1}{d+N}\frac{\widetilde{\Lambda}}{\bar{\kappa}}\,S_{\infty}^{\text{I}}\,.
	\label{eq:SinftyC}
\end{equation}
Notice that the coefficient $S_{*}^{\text{I}}$ of the subleading exponential does not contain any free parameters and it is completely fixed by the equations of motion.

\subsection{Solutions of the Einstein-scalar theory}\label{subsec:IRsolutions}

In the previous section, we classified higher-dimensional $\text{AdS}_{d+1}\times S^{N}$ geometries into three topologically distinct types (I, II and III) based on which cycles shrink to zero size in the interior. They are classified by specific IR behaviors for the scale factors $(A_1,A_2)$ near the endpoint of the geometries. We dimensionally reduced their IR asymptotics to $d+1$ dimensions where they correspond to certain asymptotics for $(A,\varphi)$ which translate to the functions $(S,W)$. We found that for type III the scalar field remains bounded while for type I, II it diverges:
\begin{equation}
	\begin{dcases}
		\varphi\rightarrow \varphi_0\,,\quad S\rightarrow 0\,,\quad &\text{type III}\\
		\varphi\rightarrow \infty\,,\quad S\sim e^{b\varphi}\,,\quad &\text{type I and II}
	\end{dcases}\,.
 \label{eq:lower_dimensional_classification}
\end{equation}
In this section, we search for solutions of these types directly in the $d+1$-dimensional Einstein-scalar theory by solving the equation of motion for $S(\varphi)$ given in \eqref{Sequation}. We consider potentials $V(\varphi)$ which have exponential asymptotics of the type \eqref{exponentialpotential} with leading and subleading exponents $b$ and $\gamma$ respectively. Dimensional reduction over an $S^{N}$ always produces a leading exponent $b>b_{\text{c}}$ and the subleading exponent $\gamma = \gamma_{\text{c}}$ without further subleading terms. Hence, our analysis is slightly more general, but the lower-dimensional solutions are still classified into three types by their IR asymptotics \eqref{eq:lower_dimensional_classification} which we continue calling type I, II and III as summarized in Table \ref{lowerdimensionaltypes}. When $b>b_{\text{c}}$ and $\gamma = \gamma_{\text{c}}$, the solutions match with their higher-dimensional counterparts in the IR region. Details of the calculations are relegated to Appendix \ref{app:reducedsolutions}.

\begin{figure}
	\centering
	\begin{center}
		\textbf{Classification of solutions of the Einstein-scalar theory}
	\end{center}
	{\tabulinesep=1.6mm
		\begin{tabu} {|c|c|c|c|}
			\hline
			Geometry & $ A(u) $ & $ \varphi(u) $ & $ S(\varphi) $\\
			\hline
			type III &  $\displaystyle{\log{\biggl(\frac{u_0-u}{\alpha}\biggr)}} $& $ \displaystyle{\varphi_0} $ & $\displaystyle{\sqrt{-\frac{2V'(\varphi_{0})}{d+1}}\,\sqrt{\varphi_0 - \varphi} }$\\
			\hline
			type II & $\displaystyle{ \frac{b_{\text{c}}^{2}}{b^{2}}\log{\biggl(\frac{u_0-u}{\alpha}\biggr)}}$&$ \displaystyle{ -\frac{1}{b}\log{\biggl(\frac{u_0-u}{\bar{\alpha}}\biggr)}} $& $\displaystyle{ 2b\,\sqrt{\frac{-V_{\infty}}{1-2\,(b^{2}-b_{\text{c}}^{2})}}\,e^{b\varphi}}$ \\
			\hline
			type I & $\displaystyle{\log{\biggl(\frac{u_0-u}{\alpha}\biggr)}}$&$ \displaystyle{ -\frac{1}{b}\log{\biggl(\frac{u_0-u}{\bar{\alpha}}\biggr)}} $& $\displaystyle{ 2b_{\text{c}}\sqrt{-V_{\infty}}\,e^{b\varphi}}$\\
			\hline
	\end{tabu}}
	\caption{IR asymptotics of solutions of the $ d+1 $-dimensional Einstein-scalar theory above the confinement bound $ b > b_{\text{c}} $ that uplift to solutions of Einstein gravity in $ d+N+1 $ dimensions with an internal sphere $ S^{N} $. Type II and III solutions uplift to regular geometries while the type I solution uplifts to a singular geometry where $ S^{d} $ and $ S^{N} $ shrink at the same point.}
	\label{lowerdimensionaltypes}
\end{figure}

\paragraph{Type III solutions.} We first look for type III solutions for which $ \varphi \rightarrow \varphi_{0}^{-} $ such that $S(\varphi_0) = 0$. By expanding in a series around $\varphi_0$, we find that equation \eqref{Sequation} has the solution (see Appendix \ref{reduced:typeA})
\begin{equation}
S(\varphi) = S_0\,\sqrt{\varphi_0 - \varphi} + \mathcal{O}(\varphi_{0}-\varphi)^{3\slash 2}, \quad \varphi \rightarrow \varphi_0^{-},
\end{equation}
where the coefficient is explicitly
\begin{equation}
S_0 = \sqrt{-\frac{2V'(\varphi_{0})}{d+1}}\,.
\label{eq:S0_lower_dimensional}
\end{equation}
The validity of the solution requires that $ V'(\varphi_{0}) < 0 $ which is the case for the potentials studied in this paper. We can compare the coefficient \eqref{eq:S0_lower_dimensional} with the one obtained from dimensional reduction of a higher-dimensional type III solution (in which case $V(\varphi)$ has only the two exponentials) and it matches as shown in Appendix \ref{app:highersolutionsandreduction}. Therefore, we have recovered the type III solutions directly in the lower-dimensional theory.

\paragraph{Type II solutions.} In this case, we keep track of the  subleading exponential $\sim e^{2\gamma \varphi}$ in the large-$\varphi$ asymptotics \eqref{exponentialpotential}. We  search for solutions of \eqref{Sequation} whose  leading asympttotics is $ S \sim e^{b\varphi} $ when $\varphi\rightarrow \infty$. The solution including the first two subleading terms is given by (see Appendix \ref{app:solutionsunboundedscalar})
\begin{equation}
S(\varphi) = S_{\infty}^{\text{II}}\,e^{b\varphi} + S_{\infty,1}^{\text{II}}\,e^{(2\gamma_{\text{c}} - b)\,\varphi} + S_{*}^{\text{II}}\,e^{(2\gamma - b)\,\varphi} + \mathcal{O}(\varphi^{2}e^{-b\varphi})\,, \quad \varphi\rightarrow +\infty\,,
\label{eq:typeB_IR_expansion_lower}
\end{equation}
where $\gamma_{\text{c}} = \frac{1}{2\,(d-1)\,b}$, $ S_{\infty,1}^{\text{II}} $ is a free integration constant and the two other coefficients are given by
\begin{equation}
S_{\infty}^{\text{II}} = 2b\,\sqrt{\frac{-V_{\infty}}{1-2\,(b^{2}-b_{\text{c}}^{2})}}\,, \quad S_{*}^{\text{II}} = -\frac{(2b^{2}(d-1)-d)(2\gamma-b)}{2b(d-2b(d-1)(2\gamma-b))}\frac{V_{\infty,1}}{V_{\infty}}\,S_{\infty}^{\text{II}}\,.
\label{SinfB}
\end{equation}
Which of the two subleading terms in \eqref{eq:typeB_IR_expansion_lower} is dominant depends on the value of $\gamma$ (which appears  in the potetial at subleading order) relative to  $\gamma_{\text{c}}$ (which depends only on $b$, and satisfies $\gamma_{\text{c}} < b$ as long as $b_{\text{G}}>b>b_{\text{c}}$).

Notice that, using \eqref{gubsB} and \eqref{confinementbound}, $S_{\infty}^{\text{II}} $ is real and finite for $ b_{\text{c}} < b < b_{\text{G}} $, but diverges at the Gubser bound (for which $ b^{2} - b_{\text{c}}^{2} =1\slash 2 $) and becomes complex for $b>b_{\text{G}}$. This solution therefore exists only for $b<b_{\text{G}}$.

For any $b>b_{\text{c}}$ and any $\gamma <b$,  the leading  asymototics  of $T$ for type II solutions is given by (see Appendix \ref{app:solutionsunboundedscalar})
\begin{equation}
    T(\varphi) = T_{\infty,1}^{\text{II}}\,e^{2\gamma_{\text{c}}\varphi} + \ldots\,,\quad \varphi\rightarrow \infty\,,
   \label{eq:T_typeII_expansion}
\end{equation}
where
\begin{equation}
	T_{\infty,1}^{\text{II}} = -\biggl( 1+\frac{d-2}{2\,(d-1)\,b^{2}} \biggr)\,S_{\infty}^{\text{II}}\,S_{\infty,1}^{\text{II}}\,.
 \label{eq:typeB_T_coefficients_text}
\end{equation}
Notice that 1) the  (would-be) leading term $e^{2b\varphi}$ of $T$ is absent; 2)  any contribution from  subleading terms in the potential drops out from   \eqref{eq:T_typeII_expansion}, whether $\gamma < \gamma_{\text{c}}$ or not. The leading contribution to $T$ (which controls the curvature of the slices) therefore is controlled entirely by the free integration constant $ S_{\infty,1}^{\text{II}}$.  Positivity of the curvature of the slicing $T(\varphi) = d\kappa\,e^{-2A_{\text{II}}}>0$ at leading order imposes the constraint $T_{\infty,1}^{\text{II}}>0$ which implies that the integration constant must be negative $ S_{\infty,1}^{\text{II}}< 0 $.

When the lower-dimensional potential comes from dimensional reduction with $ \gamma = \gamma_{\text{c}} $ and $b$ fixed as \eqref{eq:reduction_exponents}, the expansion \eqref{eq:typeB_IR_expansion_lower} matches with the expansion of a dimensional reduction of a higher-dimensional type II solution \eqref{functionstypeBreduced}. In particular, the coefficients $S_\infty^{\text{II}}$, $S_*^{\text{II}}$ \eqref{SinfB} match exactly with \eqref{eq:SBinfty_reduction} and \eqref{eq:Sinfty1B_text} respectively as shown in Appendix \ref{app:solutionsunboundedscalar}. Via the formula \eqref{eq:Sinfty1B_text}, the integration constant $S_{\infty,1}^{\text{II}}$ in \eqref{eq:typeB_IR_expansion_lower} can also be identified with the ratio $\alpha\slash \alpha_{\IR}$ that parametrizes the higher-dimensional type II solutions. From \eqref{eq:typeB_IR_expansion_lower} we observe that the integration constant being negative $S_{\infty,1}^{\text{II}} < 0$ corresponds to $\alpha\slash \alpha_{\IR}>0$.

\paragraph{The type I solution.} Looking for solutions of \eqref{Sequation} of the type $ S \sim e^{b\varphi} $, when $\varphi\rightarrow \infty$, we also find the solution
\begin{equation}
S(\varphi) = S_{\infty}^{\text{I}}\,e^{b\varphi} + S_{*}^{\text{I}}\,e^{(2\gamma-b)\,\varphi} + \mathcal{O}(\varphi^{2}e^{-b\varphi})\,, \quad \varphi\rightarrow \infty\,,
\end{equation}
where  we have again kept the   contributions from subleading terms in the scalar potential. The coefficients are
\begin{equation}
S_{\infty}^{\text{I}} = 2b_{\text{c}}\sqrt{-V_{\infty}}\,, \quad S^{\text{I}}_* = -\frac{(d-1)\gamma(2\gamma-3b)+1\slash 2}{4\gamma^{2}-2(d+3)b\gamma+4db^{2}-1}\frac{V_{\infty,1}}{V_{\infty}}\,S_{\infty}^{\text{I}}\,.
\label{SinfC}
\end{equation}
This solution  contains neither  the subleading exponential $e^{(2\gamma_{\text{c}}-b)\,\varphi}$ found in type II solutions,  nor any  free integration constant. For $\gamma = \gamma_{\text{c}} = \frac{1}{2\,(d-1)\,b}$, these coefficients match with the ones \eqref{eq:SinftyC} obtained from dimensional reduction of the singular type I solution of the higher-dimensional theory (see Appendix \ref{app:solutionsunboundedscalar}).

In the higher-dimensional theory, it is clear that the type I solution is obtained in the $\bar{\alpha}_{\IR}\slash \bar{\alpha}\rightarrow 0$ and $ \alpha_{\IR}\slash \alpha\rightarrow 0$ limits of type II and III solutions respectively. Using the relations \eqref{varphi0kappaIR} and \eqref{eq:Sinfty1B_text} obtained from dimensional reduction, these translate to $\varphi_0\rightarrow \infty$ and $S_{\infty,1}^{\text{II}}\rightarrow -\infty$ limits in the Einstein-scalar theory. That the lower-dimensional type I solution is recovered in these limits, cannot be seen directly from the IR expansions of $S(\varphi)$, but we shall confirm it numerically in Section \ref{sec:numerics}. This shows the utility of using the higher-dimensional uplift in characterizing the space of solutions of Einstein-scalar theory (see Table \ref{fig:integrationconstant_types} for a summary).

We have observed that at the confinement bound $ b = b_{\text{c}} $, the type II coefficients \eqref{SinfB} reduce to the type I ones \eqref{SinfC}: $S_{\infty}^{\text{II}}\big\vert_{b = b_{\text{c}}} = S_{\infty}^{\text{I}} $ and $S_{*}^{\text{II}}\big\vert_{b = b_{\text{c}}} = S_{*}^{\text{I}}$. Therefore, at least in the near IR region, this suggests that the type II branch merges with the type I geometry at the confinement bound. The integration constant $S_{\infty,1}^{\text{II}}$ of the type II solution also disappears in the $ b \rightarrow b_{\text{c}} $ limit, because the subleading exponent $2\gamma_{\text{c}}-b = (2b_{\text{c}}^2-b^2)\slash b \rightarrow b_{\text{c}} $ coincides with the leading exponent, whose coefficient is fixed uniquely.

\begin{figure}
	\centering
	\begin{center}
	\textbf{Integration constants of the solutions}
	\end{center}
	{\tabulinesep=1.6mm
		\begin{tabu} {|c|c|c|c|}
			\hline
			Geometry & Free parameter & $ d+N+1$-dimensions & $d+1 $-dimensions \\
			\hline
			type III &  Yes & $ \tfrac{\bar{\alpha}_{\IR}}{\bar{\alpha}} > 0 $ & $\varphi_0>0$\\
			\hline
			type II &Yes &$\tfrac{\alpha_{\IR}}{\alpha} > 0 $& $S_{\infty,1}^{\text{II}}< 0$\\
			\hline
            type I &No &$\tfrac{\alpha_{\IR}}{\alpha} =\tfrac{\bar{\alpha}_{\IR}}{\bar{\alpha}} = 0 $& $\varphi_0 = \infty\,,\;S_{\infty,1}^{\text{II}} = -\infty$\\
            \hline
	\end{tabu}}
	\caption{Free parameters of the solutions. In the higher-dimensional theory, $\alpha_{\IR}$ and $\bar{\alpha}_{\IR}$ are the radii of the non-shrinking sphere in the IR ($S^d$ for type II and $S^N$ for type III). In the lower-dimensional theory, $\varphi_0$ is the endpoint value of the scalar field in type III solutions while $S_{\infty,1}^{\text{II}}$ is the coefficient of the subleading exponential $e^{(2\gamma_{\text{c}}-b)\,\varphi}$ in the expansion of the type II solution as $\varphi\rightarrow \infty$.}
	\label{fig:integrationconstant_types}
\end{figure}

\subsection{Solutions in the vicinity of the critical solution}\label{subsec:neartypeC_solutions}

We now study the behavior of type II and III solutions in the vicinity of the type I solution which lies at their interface. We  perform the analysis both in the higher-dimensional Einstein theory and in the lower-dimensional Einstein-scalar theory. Higher-dimensional perturbations have already been studied in previous works \cite{bohm1998inhomogeneous,bohm1999non,Kol:2002xz,Asnin:2006ip,Kalisch:2017bin,Aharony:2019vgs,Kiritsis:2020bds} both with and without a negative cosmological constant and they exhibit Efimov oscillations when $d+ N < 9$. In the Einstein-scalar theory with a generic exponential potential \eqref{exponentialpotential}, perturbations around the type I solution also exhibit analogous oscillations when $b> b_{\text{E}}$. They are a priori independent from the oscillations in the higher-dimensional theory, but we show that they are the same when the uplift exists.

\paragraph{Higher-dimensional theory.} The higher-dimensional Einstein's equations admit an exact type I solution $(A_1^{\text{I}}(\tilde{u}),A_{2}^{\text{I}}(\tilde{u}))$ given in \eqref{eq:typeChigherdimensional_text}. The equations can then be linearized around this solution which is most conveniently performed by defining
\begin{equation}
	A_{\pm} \equiv d\,A_1 \pm N\,A_2
	\label{eq:Apmdeftext}
\end{equation}
and by considering perturbations $\delta A_{\pm}$ around $ A_{\pm}^{\text{I}}$. As shown in Appendix \ref{subapp:higher_Efimov}, $\delta A_+$ amounts to an infinitesimal translation in $\tilde{u}$ in the type I solution and is, hence, unphysical (a diffeomorphism). Henceforth, we  set $\delta A_+ = 0$ and consider the physical perturbation $\delta A_-$. Solving the linearized equations, we obtain the infrared expansion (see Appendix \ref{subapp:higher_Efimov})
\begin{equation}
	\delta A_- =  \sum_{\pm} C_{\pm}\,\biggl(\frac{\tilde{u}_0-\tilde{u}}{\tilde{\ell}}\biggr)^{\tilde{\beta}_{\pm}} + \ldots\,,\quad \tilde{u}\rightarrow \tilde{u}_0^{-}\,,
	\label{eq:deltaAm_int_text}
\end{equation}
where $C_{\pm}$ are two integration constants and the exponents are given by (see also \cite{Aharony:2019vgs})
\begin{equation}
	\tilde{\beta}_{\pm} = -\frac{1}{2}\left(D-1\pm\sqrt{(D-1)(D-9)}\right)
	\label{eq:tildebeta_text}
\end{equation}
with $D = d+N$. Since these exponents have a negative real part $\text{Re}\,\tilde{\beta}_{\pm} < 0$ for $D>1$, the perturbative solution \eqref{eq:deltaAm_int_text} diverges as $\tilde{u}\rightarrow \tilde{u}_0^{-}$. Therefore perturbation theory breaks down in the IR and \eqref{eq:deltaAm_int_text} is only a valid solution in an intermediate regime between the IR and the UV.

When the type II and III solutions are near the type I solution, we can define the perturbations $\delta A_{-}^{\text{II},\text{III}} \equiv A_{-}^{\text{II},\text{III}} - A_{-}^{\text{I}}$ away from the type I solution. Then it follows that $\delta A_{-}^{\text{II},\text{III}} $ have an expansion of the form \eqref{eq:deltaAm_int_text} with two sets of integration constants $C_{\pm}^{\text{II},\text{III}}$. The constants $C_{+}$ and $C_{-}$ are not independent from each other, because type II and III solutions contain only a single parameter $\alpha_{\IR}$ or $\bar{\alpha}_{\IR}$ respectively (the IR radius of the sphere that does not shrink). Therefore they are related as $C^{\text{II}}_{\pm} = C^{\text{II}}_{\pm}(\alpha_{\IR}) $ and $C^{\text{III}}_{\pm} = C^{\text{III}}_{\pm}(\bar{\alpha}_{\IR}) $ where the functional dependence on the respective IR parameter can be fixed by imposing type II and III boundary conditions in the IR. As shown in detail in Appendix \ref{subapp:higher_Efimov}, this argument gives at leading order (see also \cite{Aharony:2019vgs,Kiritsis:2020bds})
\begin{equation}
	 C_{\pm}^{\text{II}} \propto\biggl(\frac{\tilde{\ell}}{\alpha_{\text{IR}}}\biggr)^{\tilde{\beta}_{\pm}}\,,\quad C_{\pm}^{\text{III}} \propto\biggl(\frac{\tilde{\ell}}{\bar{\alpha}_{\text{IR}}}\biggr)^{\tilde{\beta}_{\pm}}\,,\quad\alpha_{\IR},\bar{\alpha}_{\IR}\rightarrow 0\,.
	\label{eq:Cpm_alphaIR_text}
\end{equation}
Since $ \text{Re}\,\tilde{\beta}_{\pm} < 0$, they satisfy
\begin{equation}
	C_{\pm}^{\text{II}}\big\lvert_{\alpha_{\text{IR}} = 0}=C_{\pm}^{\text{III}} \big\lvert_{\bar{\alpha}_{\text{IR}} = 0}= 0
\end{equation}
as required by the fact that type II and III solutions reduce to type I when $\bar{\alpha}_{\text{IR}} = \alpha_{\text{IR}} = 0$. Unfortunately, we are not able to compute the proportionality constants analytically, because the perturbative solution \eqref{eq:deltaAm_int_text} is not valid all the way to $\tilde{u} = \tilde{u}_0$ where it could be exactly matched with the IR asymptotics. However, the functional dependence in \eqref{eq:Cpm_alphaIR_text} will be enough for our purposes below.

As mentioned in Section \ref{subsec:classification_higher}, the higher-dimensional Einstein's equations are invariant under a $\mathbb{Z}_2$-transformation $ A_1\leftrightarrow A_2 $, $ d\leftrightarrow N $, $\alpha\leftrightarrow \bar{\alpha}$ and $\alpha_{\IR}\leftrightarrow \bar{\alpha}_{\IR}$ mapping type II and III solutions to each other. This amounts to $A_1^{\text{II}} = A_{2}^{\text{III}}\big\vert_{E} $, $A_2^{\text{II}} = A_{1}^{\text{III}}\big\vert_{E}$, where the notation $ \big\vert_{E}$ with $E = \{\alpha\leftrightarrow \bar{\alpha}\,, \alpha_{\IR}\leftrightarrow \bar{\alpha}_{\IR}\,,d\leftrightarrow N\}$ indicates that we perform the exchanges inside the curly brackets. At the level of the perturbations, we obtain $\delta A_{-}^{\text{II}} =  -\delta  A_{-}^{\text{III}}\big\vert_{E}$ and that the integration constants are related as $  C_{\pm}^{\text{II}} = - C_{\pm}^{\text{III}}\big\vert_{E}$. See Appendix \ref{subapp:higher_Efimov} for details.

The integration constants \eqref{eq:Cpm_alphaIR_text} are oscillatory functions of $\tilde{\ell}\slash \alpha_{\IR},\tilde{\ell}\slash \bar{\alpha}_{\IR}$ when $d+N <9$, and the threshold value $d+ N = 9$ at which they become oscillatory, is the Efimov bound \cite{Aharony:2019vgs,Kiritsis:2020bds}. The oscillations of the integration constants translate to Efimov oscillations of the UV parameters of the dual field theory as functions of the IR parameters \cite{Aharony:2019vgs,Kiritsis:2020bds} (see also \cite{Karch:2009ph}). In this case, the dual theory lives on the $d+N$-dimensional conformal boundary of $\text{AdS}_{d+1}\times S^{N}$ which is $S^{d}\times S^{N}$ in this case. We shall see below, that Efimov oscillations also arise in the Einstein-scalar theory independently of the higher-dimensional description.\footnote{Note that the UV of the higher-dimensional theory $\tilde{u}\rightarrow -\infty$ is not the same as the UV of the Einstein-scalar theory located at $\varphi = 0$.}

\paragraph{Lower-dimensional theory.} We now look for perturbative solutions around the type I solution directly in the Einstein-scalar theory. The type I solution $S_{\text{I}}(\varphi)$ of the Einstein-scalar theory has the IR asymptotics
\begin{equation}
	S_{\text{I}}(\varphi) = \sqrt{\frac{-2V_{\infty}}{d-1}}\,e^{b\varphi} + \ldots\,,\quad \varphi\rightarrow \infty\,.
\end{equation}
As we have explained, this solution lies at the interface of type II and III families of solutions. We search for solutions in the form
\begin{equation}
	S(\varphi) = S_{\text{I}}(\varphi) + \delta S(\varphi)\,,
	\label{eq:perturbedtypeC}
\end{equation}
where we require the ratio $ \delta S(\varphi)\slash S_{\text{I}}(\varphi)\ll 1 $ to be small for all $0 < \varphi < \infty$. Because there are only type I, II and III solutions, \eqref{eq:perturbedtypeC} must be equal either to a type II or III solution in the range of parameters where they are close to the type I solution. As shown in Section \ref{subsec:IRsolutions}, this corresponds to $\varphi_0\rightarrow \infty$ for type III and $ S_{\infty,1}^{\text{II}}\rightarrow -\infty$ for type II respectively.

The equation for $S(\varphi)$ we want to solve is given by \eqref{Sequation}. By linearizing the equation, we find the general perturbative solution (see Appendix \ref{app:solutionsunboundedscalar})
\begin{equation}
	\delta S(\varphi) = S_+\,e^{\beta_+ \varphi} + S_-\,e^{\beta_- \varphi} + \ldots\,,\quad \varphi \rightarrow \infty\,,
	\label{eq:gendeltaStext}
\end{equation}
where the coefficients $ S_{\pm} $ are two free integration constants and the exponents $ \beta_{\pm} $ are given by
\begin{equation}
	\beta_{\pm} = \frac{b}{2}\left(d+1\pm\sqrt{d^{2}-10d+9+4b^{-2}}\right).
	\label{eq:betapmtext}
\end{equation}
It is useful to notice that these exponents can be written as
\begin{equation}
	\beta_{\pm} = \frac{(d+1)\,b}{2}\pm \sqrt{1-\frac{b^{2}}{b_{\text{E}}^{2}}}\,,
	\label{eq:typeCbetapmtext}
\end{equation}
where we have defined the Efimov bound
\begin{equation}
	b_{\text{E}} \equiv \frac{2}{\sqrt{(d-1)(9-d)}}\,.
\end{equation}
For dimensions $1 < d < 8$, the Efimov bound satisfies $b_{\text{c}} <b_{\text{E}} < b_{\text{G}} $, while for $d\geq 8$, it is equal or above the Gubser bound $b_{\text{E}} \geq b_{\text{G}}$. Both above and below the Efimov bound $b_{\text{E}}$, the solution \eqref{eq:gendeltaStext} diverges faster than $S_{\text{I}}(\varphi)$ in the IR region $\varphi\rightarrow \infty$. This means that the perturbative approximation $ \delta S(\varphi)\slash S_{\text{I}}(\varphi)\ll 1 $ breaks down in the IR region where $\varphi$ is large. Therefore, \eqref{eq:gendeltaStext} is only valid in an intermediate range of $\varphi$ between the IR and the UV.

We observe that when $b \leq b_{\text{E}}$, the exponents $\beta_{\pm}$ are real, while for $b>b_{\text{E}}$, they acquire an imaginary part. When the potential $V(\varphi)$ comes from dimensional reduction, the region $b > b_{\text{E}}$ coincides exactly with the Efimov regime $d+N<9$ of the higher-dimensional theory. This can be deduced from the identity
\begin{equation}
    \tilde{\beta}_{\pm} = \frac{b}{b^{2}-b_{\text{c}}^{2}}\,(b-\beta_{\pm})\,,
\label{eq:betatilde_beta}
\end{equation}
which implies that the higher-dimensional exponents $\tilde{\beta}_{\pm}$ \eqref{eq:tildebeta_text} are imaginary exactly when $\beta_{\pm}$ are. Notice that the Efimov regime $b_{\text{E}}< b <b_{\text{G}}$ of the Einstein-scalar theory exists independently of the higher-dimensional uplift and relies only on the presence of the exponential asymptotics \eqref{exponentialpotential} of the potential.

The relation \eqref{eq:betatilde_beta} implies that the solution \eqref{eq:gendeltaStext} is the dimensional reduction of the higher-dimensional solution \eqref{eq:deltaAm_int_text} which is shown in detail in Appendix \ref{subapp:Efimov_reduction}. The relation between lower- and higher-dimensional integration constants turns out to be
\begin{equation}
	S_{\pm} \propto \frac{1}{\bar{\alpha}}\biggl(\frac{\bar{\alpha}}{\tilde{\ell}}\biggr)^{\tilde{\beta}_{\pm}}C_{\pm}\,.
	\label{eq:SpmCpmrelation_text}
\end{equation}
We now define $\delta S_{\text{II}} \equiv S_{\text{II}} - S_{\text{I}}$ and $\delta S_{\text{III}} \equiv S_{\text{III}} - S_{\text{I}}$ which are small when the type II and III solutions $S_{\text{II},\text{III}}$ are close to the type I solution. In particular, for type III this amounts to $\varphi_0 \rightarrow \infty$, and for type II, to $S_{\infty,1}^{\text{B}}\rightarrow -\infty$. This is shown in Table \ref{fig:integrationconstant_types} and explained in the discussion around it. Then $\delta S_{\text{II}}$ and $\delta S_{\text{III}}$ are given by \eqref{eq:gendeltaStext} with two sets of integration constants $S_{\pm}^{\text{II,III}}$. Since type II and III solutions involve only a single free parameter, $S_-$ and $S_+$ are not independent, but functions of the IR parameters as $S_{\pm}^{\text{II}} =S_{\pm}^{\text{II}}(S_{\infty,1}^{\text{II}}) $ and $S_{\pm}^{\text{III}} =S_{\pm}^{\text{III}}(\varphi_0) $. As shown in Section \ref{subsec:classification_higher}, the IR parameters are given by
\begin{equation}
    \varphi_0\propto \log{\frac{\bar{\alpha}}{\bar{\alpha}_{\IR}}} \,,\quad S_{\infty,1}^{\text{II}} \propto -\left(\frac{\alpha}{\alpha_{\IR}}\right)^{2}
\label{eq:pIR_uplift}
\end{equation}
so that $S_{\pm}^{\text{II}} $ and $S_{\pm}^{\text{III}} $ are functions of $\alpha_{\IR}$ and $\bar{\alpha}_{\IR}$ respectively. By substituting \eqref{eq:Cpm_alphaIR_text} to \eqref{eq:SpmCpmrelation_text} gives
\begin{equation}
	S_{\pm}^{\text{III}} \propto\biggl(\frac{\bar{\alpha}}{\bar{\alpha}_{\IR}}\biggr)^{\tilde{\beta}_{\pm}} \,,\quad  S_{\pm}^{\text{II}}  \propto \left(\frac{\alpha}{\alpha_{\IR}}\right)^{\tilde{\beta}_{\pm}}\,,\quad \alpha_{\IR},\bar{\alpha}_{\IR}\rightarrow 0\,.
	\label{eq:SpmIRparameters_text}
\end{equation}
By combining with \eqref{eq:pIR_uplift} it follows that (see also Appendix \ref{subapp:Efimov_reduction})
\begin{equation}
	S_{\pm}^{\text{III}} \propto  e^{(b-\beta_{\pm})\,\varphi_0}\,,\quad S_{\pm}^{\text{II}}\propto (-S_{\infty,1}^{\text{II}})^{P_{\pm}}\,,\quad \varphi_0,\vert S_{\infty,1}^{\text{II}}\vert \rightarrow \infty\,,
	\label{eq:SpmAB_text}
\end{equation}
where the exponents are given by
\begin{equation}
	P_{\pm} = \frac{1}{2b}\frac{b-\beta_{\pm}}{1-(b_{\text{c}}\slash b)^{2}}\,.
	\label{eq:Ppm_text}
\end{equation}
One may check that the real parts of the exponents are negative $\text{Re}\,(b-\beta_{\pm}),\,\text{Re}\,P_{\pm} < 0$ when $b > b_{\text{c}}$ so that the integration constants \eqref{eq:SpmAB_text} go to zero $S_{\pm}^{\text{II,III}}\rightarrow 0$ when $\varphi_0\rightarrow \infty$ and $S_{\infty,1}^{\text{II}}\rightarrow -\infty$ as expected from type II and III solutions reducing to the type I solution in these limits.

\section{Free energy of the dual field theory}\label{sec:free_energy}

In this section, we  compute the renormalized on-shell action of the Einstein-scalar theory which is identified with the free energy of the dual field theory. We also analyze its behavior analytically near the critical point dual to the type I solution. We assume that the potential $V(\varphi)$ of the Einstein-scalar theory has a maximum at $\varphi = 0$ with AdS radius $\ell$ and exponential asymptotics \eqref{exponentialpotential} in the deep IR $\varphi\rightarrow \infty$.

\subsection{On-shell action of the Einstein-scalar theory}

We now compute the free energy of the dual field theory using Einstein-scalar theory. The regularized free energy is defined as
\begin{equation}
	F_{\text{reg}} = -I_{\text{on-shell}}^{\text{reg}}\,,
\end{equation}
where the regularized Euclidean on-shell Einstein-scalar action \eqref{einsteinscalar} is given by \cite{Ghosh:2017big,Ghosh:2020qsx}
\begin{equation}
	I_{\text{on-shell}}^{\text{reg}} = M_{\text{p}}^{d-1}\int d^{d}x\sqrt{\zeta}\,\biggl( \frac{2\kappa}{d}\int_{\log{\epsilon}}^{u_0}du\,e^{(d-2)A(u)}+2\,(d-1)\bigl[e^{dA(u)}\,\dot{A}(u)\bigr]\bigg\lvert^{u = u_0}_{u = \log{\epsilon}}\,\biggr)\,,
	\label{onshellactionstart_text}
\end{equation}
where $ \epsilon \rightarrow 0 $ is the UV cut-off and we use the notation $f(u)\lvert_{u=u_1}^{u=u_2}\, \equiv f(u_2)-f(u_1)$. This action also includes the Gibbons--Hawking--York term at the cut-off surface $ u = \log{\epsilon} $ which makes the variational principle for the metric (in this case $ A(u) $) well defined. As shown in \cite{Ghosh:2017big,Ghosh:2018qtg} and reviewed in Appendix \ref{app:onshelllowerdimensional}, the renormalized free energy may be written as a pure boundary term calculated in terms of its values in the IR and the UV as
\begin{equation}
	F_{\text{reg}} = M_{\text{p}}^{d-1}\,\widetilde{\Omega}_d\,\biggl(W(\varphi)\,T(\varphi)^{-\frac{d}{2}}+ U(\varphi)\,T(\varphi)^{-\frac{d}{2}+1}\biggr)\bigg\lvert^{\varphi = \varphi_{\scriptscriptstyle\text{IR}}}_{\varphi = \varphi_{\UV}}\, ,
	\label{Freg1text}
\end{equation}
where $U(\varphi)$ solves the first-order differential equation
\begin{equation}
	SU'-\frac{d-2}{2(d-1)}\,WU +\frac{2}{d} = 0\,,
	\label{Uequationtext}
\end{equation}
and we have defined $\varphi_{\IR} = \varphi(u_0)$, $ \varphi_{\UV} \equiv \varphi(\log{\epsilon})\rightarrow 0 $ and $\widetilde{\Omega}_d$ is a constant given in \eqref{eq:Omega_tilde_d}.
Notice that the function $U$ is determined by equation \eqref{Uequationtext}  up to an integration constant, but different solutions lead to the same result when inserted in  the free energy \eqref{Freg1text}, as the contributions which contain the integration constant from the UV and the IR cancel each other.\footnote{This can be checked explicitly, but it is manifest by the fact that the original form \eqref{onshellactionstart_text} of the on-shell action does not contain this extra freedom.}

The form of the IR boundary term in \eqref{Freg1text} is sensitive to the on-shell IR behavior of the functions $W,T,U$ and has to be analyzed separately for type I, II and III solutions which is done in Appendix \ref{app:onshelllowerdimensional}. It turns out that the IR boundary term can always be made to vanish by fixing the integration constant $\mathcal{U}$ appearing in $U(\varphi)$ in a particular way. The integration constant $\mathcal{U}$ arises due to the first-order nature of the equation \eqref{Uequationtext} and it appears as the coefficient of the divergent term in the IR expansion of $U$ as
\begin{equation}
  U(\varphi) = \mathcal{U}\times\left\{\begin{alignedat}{3}
    &(\varphi_{0}-\varphi)^{-\frac{d}{2}+1}  + \ldots, \quad &&\varphi \rightarrow \varphi^{-}_{0}\,,\quad  &&\text{type III}\\
		&e^{\frac{d-2}{2b\,(d-1)}\,\varphi}  + \ldots\,, \quad &&\varphi \rightarrow \infty\,,\quad  &&\text{type II}\\
		&e^{(d-2)\,b\varphi} + \ldots\,, \quad &&\varphi \rightarrow \infty\,,\quad &&\text{type I}
  \end{alignedat}\right.
  \label{eq:U_IR_expansion_text}
\end{equation}
As shown in Appendix \ref{app:onshelllowerdimensional}, the first term $W(\varphi_{\IR})\,T(\varphi_{\IR})^{-\frac{d}{2}}$ vanishes below the Gubser bound $b_{\text{c}} < b < b_{\text{G}}$, while the second $U(\varphi_{\IR})\,T(\varphi_{\IR})^{-\frac{d}{2}+1}$ vanishes if we set $\mathcal{U} = 0$, which may be interpreted as a regularity condition on $U$. Given this choice of the integration constant, the regularized free energy is a pure boundary term in the UV of the form
\begin{equation}
	F_{\text{reg}} = -M_{\text{p}}^{d-1}\,\widetilde{\Omega}_d\,\biggl( U(\varphi_{\UV})\,T(\varphi_{\UV})^{-\frac{d}{2}+1}+W(\varphi_{\UV})\,T(\varphi_{\UV})^{-\frac{d}{2}}\biggr)\,.
	\label{Freg2text}
\end{equation}
This expression is UV divergent when $\varphi_{\UV}\rightarrow 0$, corresponding to $ \epsilon \rightarrow 0 $, which can be seen using the on-shell UV expansions of $ W $, $ T $ and $ U $ derived in Appendix \ref{app:UVasymptotics}. The free energy must be renormalized by the addition of counterterms that cancel these divergences. Therefore, we define the renormalized free energy as
\begin{equation}
	F \equiv \lim_{\epsilon \rightarrow 0}\bigl(F_{\text{reg}} + F_{\text{ct}}\bigr)\,,
	\label{eq:Fren}
\end{equation}
where $ F_{\text{ct}} $ is a local and diffeomorphism invariant counterterm supported on the cut-off surface. Its form was derived in full generality in \cite{Balasubramanian:1999re,Henningson:1998gx,Papadimitriou:2011qb} (see also \cite{Papadimitriou:2007sj,Kiritsis:2014kua,Ghosh:2017big,Ghosh:2020qsx}). In  the notation of \cite{Ghosh:2020qsx}, it takes the form
\begin{equation}
	F_{\text{ct}} = M_{\text{p}}^{3}\int d^{4}x\sqrt{h}\,\Bigl[W_{\text{ct}}(\varphi_{\UV}) +R^{(h)}(\varphi_{\UV})\,U_{\text{ct}}(\varphi_{\UV})+[R^{(h)}(\varphi_{\UV})]^{2}\,Y_{\text{ct}}(\varphi_{\UV}) \Bigr]\, ,
	\label{eq:Fcttext}
\end{equation}
where $h_{\mu\nu}$ is the induced metric on the cut-off surface $ u = \log{\epsilon} $, $R^{(h)}$ is its Ricci scalar and the functions $ W_{\text{ct}} $, $ U_{\text{ct}} $, $ Y_{\text{ct}} $ satisfy a set of three first-order differential equations given in \eqref{eq:Wct} - \eqref{eq:Yct}.\footnote{The differential equations are chosen such that that $ W_{\text{ct}} $, $ U_{\text{ct}} $, $ Y_{\text{ct}} $ have the correct divergence structure to cancel all divergences of the regularized free energy in any dimension $d$ and $W_{\text{ct}}$, $ U_{\text{ct}} $ satisfy the same equations as $W$, $U$ in flat slicing in which $T = 0$, $S= W'$.} The three integration constants $ \mathcal{A}_{\text{ct}} $, $ \mathcal{B}_{\text{ct}} $ and $ \mathcal{C}_{\text{ct}} $ of these equations parametrize the renormalization scheme. Using the fact that $R^{(h)} = T(\varphi)$ and $\sqrt{h} = \kappa^{d\slash 2}\,T(\varphi)^{-\frac{d}{2}}\,\sqrt{\zeta}$, the renormalized free energy in $ d=4 $ becomes explicitly
\begin{align}
	&F_{\text{ren}} = -M_{\text{p}}^{3}\,\widetilde{\Omega}_{4}\\
	&\times\lim_{\epsilon\rightarrow 0}\,\Bigl[ \bigl(W(\varphi_{\UV}) - W_{\text{ct}}(\varphi_{\UV})\bigr)\,T(\varphi_{\UV})^{-2} + \bigl(U(\varphi_{\UV}) - U_{\text{ct}}(\varphi_{\UV})\bigr)\,T(\varphi_{\UV})^{-1}-Y_{\text{ct}}(\varphi_{\UV})  \Bigr]\,.\nonumber
\end{align}
The on-shell UV asymptotics of $ W, T $ and $U$ derived in Appendix \ref{app:UVasymptotics} follow from the UV expansion of $S$ given in equation \eqref{eq:S_UV_expansion_text}. Therefore they involve the two integration constants $\mathcal{R}$ and $\mathcal{C}$ appearing in $S$. There is an additional third integration constant $\mathcal{B}$ which in $d = 4$ appears in the UV expansion of $U$ as (here we show only the relevant terms)
\begin{equation}
    U(\varphi) =  \ell\,\biggl(\frac{1}{4}+ \mathcal{B}\,\varphi^{2\slash \Delta_-}+ \frac{\mathcal{R}}{48\Delta_-}\,\varphi^{2\slash \Delta_-}\log{\varphi}+ \ldots \biggr) \,.
    \label{eq:U_uv_expansion_text}
\end{equation}
On-shell, $\mathcal{B} = \mathcal{B}(p,\mathcal{U})$ is a function of the IR parameter $p \equiv \varphi_0,\vert S_{\infty,1}^{\text{II}}\vert$ (of type III and II solutions respectively) and the integration constant $\mathcal{U}$ appearing in the IR expansion \eqref{eq:U_IR_expansion_text}. To remove the IR boundary term from the on-shell action \eqref{Freg1text}, we have made the convenient choice $\mathcal{U} = 0$, and we  simply denote $\mathcal{B}(p)\equiv \mathcal{B}(p,0)$ from this point on.\footnote{As mentioned before, the value of the renormalized energy is independent of the choice of $\mathcal{U}$, because the change in the IR boundary value due to $\mathcal{U}$ is cancelled by a change in the UV boundary value due to $\mathcal{B}(\mathcal{U})$.} By inverting the equation $\mathcal{R} = \mathcal{R}(p)$, it follows that $\mathcal{B} = \mathcal{B}(p,0) = \mathcal{B}(\mathcal{R})$ and is a function of the dimensionless curvature only. The same applies to the dimensionless vev $ \mathcal{C} = \mathcal{C}(\mathcal{R}) $. Using the on-shell UV asymptotics $ W_{\text{ct}},U_{\text{ct}},Y_{\text{ct}} $ given in Appendix \ref{app:UVasymptotics},  we thus obtain  in $d = 4$ \cite{Ghosh:2020qsx}
\begin{equation}
	F(\mathcal{R}) = -\mathcal{N}\,\left[ \mathcal{R}^{-2}\,(\mathcal{C}(\mathcal{R}) - \mathcal{C}_{\text{ct}})+\mathcal{R}^{-1}\,(\mathcal{B}(\mathcal{R}) - \mathcal{B}_{\text{ct}})-\frac{1}{192}-\mathcal{A}_{\text{ct}}  \right]\,,
\end{equation}
where the normalization factor
\begin{equation}
    \mathcal{N} \equiv (M_{\text{p}}\ell)^{3}\,\widetilde{\Omega}_{4} = 384\pi^2\,(M_{\text{p}}\ell)^{3}\,.
\label{eq:curlyN}
\end{equation}
The constant $ 1\slash 192 $ comes from an order $\mathcal{R}^2\,\varphi^{4\slash \Delta_-}$ term in the expansion of $W$ whose coefficient is fixed by matching with the conformal limit $\varphi_- =\mathcal{C} = 0$ \cite{Ghosh:2020qsx}. From this point on, we  work in the renormalization scheme
\begin{equation}
	\mathcal{A}_{\text{ct}} = -\frac{1}{192}\,,\quad \mathcal{B}_{\text{ct}} = \mathcal{C}_{\text{ct}} = 0\,,
\end{equation}
so that the renormalized free energy is simply
\begin{equation}
	F(\mathcal{R}) = -\mathcal{N}\,[ \mathcal{R}^{-2}\,\mathcal{C}(\mathcal{R})+\mathcal{R}^{-1}\,\mathcal{B}(\mathcal{R})]\,.
	\label{eq:renormalizedFtext}
\end{equation}
One can prove the identity \cite{Ghosh:2020qsx}
\begin{equation} \label{identity}
	\mathcal{C}'(\mathcal{R})-\mathcal{B}(\mathcal{R}) + \mathcal{R}\,\mathcal{B}'(\mathcal{R}) = \frac{1}{96}\,\mathcal{R}\,,
\end{equation}
which is dimension-dependent. The right-hand side here is appropriate to four dimensions and it is due to the conformal anomaly. The three-dimensional analog can be found in \cite{Ghosh:2018qtg}. By taking the derivative of \eqref{eq:renormalizedFtext} and using this relation, we obtain the equation
\begin{equation}
	F'(\mathcal{R}) = \mathcal{N}\,\biggl(\frac{2\,\mathcal{C}(\mathcal{R})}{\mathcal{R}^{3}}-\frac{1}{96}\,\frac{1}{\mathcal{R}}\biggr)\,,
 \label{eq:F_diff_eq}
\end{equation}
which proves to be useful below.

\subsection{Efimov oscillations}\label{subsec:Efimov_oscillations}

In this section, we  study analytically the behavior of the UV parameters $(\mathcal{R},\mathcal{C})$ and the free energy $F$ of type II and III solutions in the regime of IR parameters where they are close to the type I solution. By introducing the notation
\begin{equation}
    p \equiv
    \begin{dcases}
        \vert S_{\infty,1}^{\text{II}}\vert\,,\quad &\text{type II}\\
        \varphi_0\,,\quad &\text{type III}\
    \end{dcases}\,
\label{eq:p_notation}
\end{equation}
for the IR parameters, this regime amounts to $p\rightarrow \infty$.

\paragraph{UV parameters.}

The type I solution is unique and corresponds to a single point $(\mathcal{R},\mathcal{C}) = (\mathcal{R}_{\text{I}},\mathcal{C}_{\text{I}})$ in the plane of UV parameters.  $(\mathcal{R}_{\text{I}},\mathcal{C}_{\text{I}})$ specify the values of dimensionless curvature and vev of the dual theory for which the type I geometry is a valid bulk solution. We  consider small perturbations $\mathcal{R} = \mathcal{R}_{\text{I}}+\delta\mathcal{R}$ and $\mathcal{C} = \mathcal{C}_{\text{I}}+\delta\mathcal{C}$ around the type I value for which the corresponding bulk solution is either a type II or III geometry parametrized by the IR parameter $p= \varphi_0$ or $p = \vert S_{\infty,1}^{\text{II}}\vert$ respectively. Therefore $\delta\mathcal{R} = \delta\mathcal{R}(p)$ and $\delta\mathcal{C} = \delta\mathcal{C}(p)$ are not independent, but functions of the same IR parameter. The type I solution is recovered in the $p\rightarrow \infty$ limit so that we obtain the expansion
\begin{equation}
    \mathcal{R}(p) = \mathcal{R}_{\text{I}} + \delta \mathcal{R}(p)\,,\quad \mathcal{C}(p) = \mathcal{C}_{\text{I}} + \delta \mathcal{C}(p)\,,\quad p\rightarrow \infty\,,
    \label{eq:R_C_IR_expansion}
\end{equation}
where $\lim_{p\rightarrow \infty}\delta \mathcal{R}(p)=\lim_{p\rightarrow \infty}\delta \mathcal{C}(p)=0$ for both type II and III geometries.

\begin{figure}[t]
\centering
\begin{subfigure}{.32\textwidth}
  \centering
  \begin{tikzpicture}
		\node (img1)  {\includegraphics{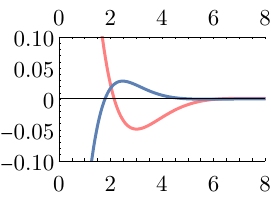}};
		\node[below=of img1, node distance=0cm, yshift=1.4cm, xshift=2cm] {$  \varphi_0 $};
		\node[above=of img1, node distance=0cm, yshift=-1.4cm, xshift=1.6cm] {$  \log{\vert S_{\infty,1}^{\text{II}}\vert} $};
	\end{tikzpicture}
  \caption{$b>b_{\text{E}}$}
  \label{plot:analytical_Efimov_above}
\end{subfigure}
\hfill
\begin{subfigure}{.32\textwidth}
  \centering
  \begin{tikzpicture}
		\node (img1)  {\includegraphics{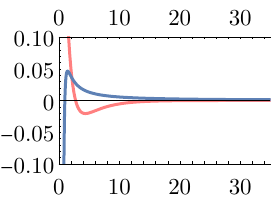}};
		\node[below=of img1, node distance=0cm, yshift=1.65cm, xshift=2.4cm] {$  \varphi_0 $};
		\node[above=of img1, node distance=0cm, yshift=-1.7cm, xshift=2.6cm] {$  \vert S_{\infty,1}^{\text{II}}\vert $};
	\end{tikzpicture}
  \caption{$b<b_{\text{E}}$}
 \label{plot:analytical_Efimov_below_peaks}
\end{subfigure}
\hfill
\begin{subfigure}{.32\textwidth}
  \centering
  \begin{tikzpicture}
		\node (img1)  {\includegraphics{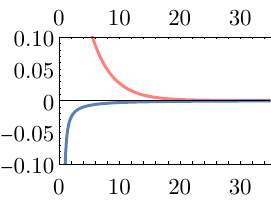}};
		\node[below=of img1, node distance=0cm, yshift=1.65cm, xshift=2.4cm] {$  \varphi_0 $};
		\node[above=of img1, node distance=0cm, yshift=-1.7cm, xshift=2.6cm] {$  \vert S_{\infty,1}^{\text{II}}\vert $};
	\end{tikzpicture}
 \caption{$b<b_{\text{E}}$}
\label{plot:analytical_Efimov_below}
\end{subfigure}
\caption{Plots of analytical results for $\mathcal{R}-\mathcal{R}_{\text{I}}$ (or equivalently $\mathcal{C}-\mathcal{C}_{\text{I}}$) on the vertical axis for type II (blue) and III (pink) solutions as functions of IR parameters $p =\varphi_0,\vert S_{\infty,1}^{\text{II}}\vert $ in the limit $p\rightarrow \infty$. We have plotted three qualitatively different possibilities depending on the value of $b$ and details of the potential for $d = 4$ where $b_{\text{E}} \approx 0.52$. In (a) for $b=0.65$ above the Efimov bound $b>b_{\text{E}}$, there is an infinite number of oscillations around the type I value at zero. Figures (b) and (c) are both below the Efimov bound $b<b_{\text{E}}$ with $b=0.47$, but in (b) type II and III branches have a single peak, while in (c) there are no peaks. In both cases, the type II and III branches approach the same type I value when $\varphi_0,\vert S_{\infty,1}^{\text{II}}\vert\rightarrow \infty$.}
\end{figure}
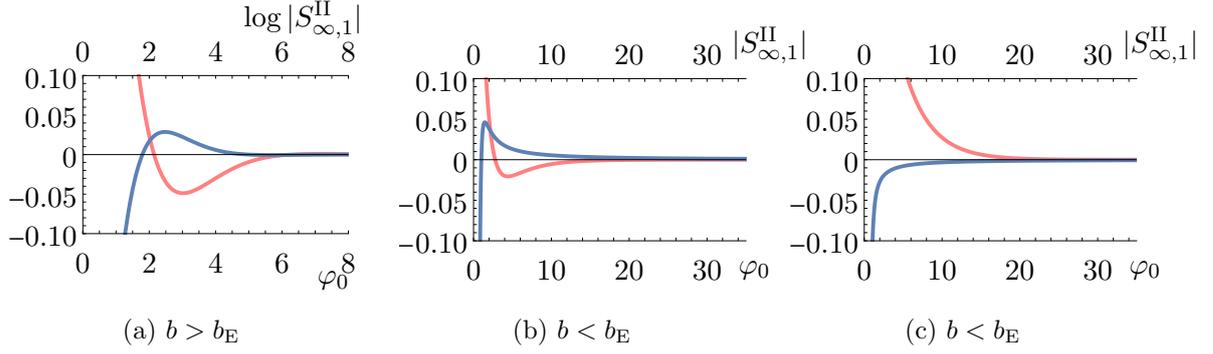

In Section \ref{subsec:neartypeC_solutions}, we solved the equation \eqref{Sequation} for $S(\varphi)$ to linear order around the type I solution $S_{\text{I}}(\varphi)$ in the IR region $\varphi\rightarrow \infty$. The perturbative solution $\delta S(\varphi)$ when $\varphi\rightarrow \infty$ is given in terms of two integration constants $S_{\pm} = S_{\pm}(p) $ which are fixed in terms of the IR parameter $p\rightarrow \infty$ by matching to a type II or III solution respectively. The equation for $\delta S(\varphi)$ relates the IR integration constants $S_{\pm}$ to the UV integration constants $(\mathcal{R},\mathcal{C})$ appearing in the UV expansion \eqref{eq:S_UV_expansion_text} of $S(\varphi)$ when $\varphi\rightarrow 0$. We assume the type I solution is an attractor so that any small perturbation around it in the IR remains small all they way to the UV.\footnote{We see this numerically in Section \ref{subsec:num_solutions} where the type II and III solutions remain close to the type I solution in the UV.} Since the equation for the perturbation $\delta S(\varphi)$ is linear, there must exist a linear relation
\begin{equation}
	\begin{pmatrix}
		\delta \mathcal{R}(p)\\
		\delta \mathcal{C}(p)
	\end{pmatrix}=M\cdot \begin{pmatrix}
		S_-(p)\\
		S_+(p)
	\end{pmatrix}\equiv  \begin{pmatrix}
		m_{\mathcal{R}}^{-} &m_{\mathcal{R}}^{+}\\
		m_{\mathcal{C}}^{-} &m_{\mathcal{C}}^{+}
	\end{pmatrix}\begin{pmatrix}
		S_-(p)\\
		S_+(p)
	\end{pmatrix}\,,\quad p\rightarrow \infty\,,
 \label{eq:M_matrix}
\end{equation}
with some unknown invertible two by two matrix $M$ that may in principle be determined by solving the linearized equation for $\delta S(\varphi)$ for all $0 \leq \varphi<\infty$. This we are not able to do however, because we do not know the type I solution $S_{\text{I}}(\varphi)$ except in the IR region $\varphi \rightarrow \infty$. By using \eqref{eq:SpmAB_text}, we obtain for type II solutions as $ S_{\infty,1}^{\text{II}}\rightarrow -\infty$ and for type III solutions as $\varphi_0\rightarrow \infty$,
\begin{equation}
	\begin{pmatrix}
		\delta \mathcal{R}_{\text{III}}\\
		\delta \mathcal{C}_{\text{III}}
	\end{pmatrix}= M_{\text{III}}\cdot
	\begin{pmatrix}
		e^{(b-\beta_{-})\,\varphi_0}\\
		e^{(b-\beta_{+})\,\varphi_0}
	\end{pmatrix}\,,\quad
	\begin{pmatrix}
		\delta \mathcal{R}_{\text{II}}\\
		\delta \mathcal{C}_{\text{II}}
	\end{pmatrix}= M_{\text{II}}\cdot
	\begin{pmatrix}
		(-S_{\infty,1}^{\text{II}})^{P_-}\\
		(-S_{\infty,1}^{\text{II}})^{P_+}
	\end{pmatrix}\,,
\label{eq:deltaRC_type_AB}
\end{equation}
with distinct two by two matrices $M_{\text{II},\text{III}}$ that relate the UV integration constants to the IR ones. Note that the type III matrix $M_{\text{III}}$ is not independent from the type II matrix $M_{\text{II}}$, but they are related via the $\mathbb{Z}_2$-transformation of Einstein's equations mapping type II and III solutions to each other in the higher-dimensional description. The uplift is only valid in the IR region $\varphi\rightarrow \infty$ where the potential is exponential, but this is enough for our purposes, because we are interested in the regime $p\rightarrow \infty$, where the endpoint $\varphi_0\rightarrow \infty$ of the type III solution also probes the exponential region. Since $  C_{\pm}^{\text{II}} = - C_{\pm}^{\text{III}}\big\vert_{E}$ under the $\mathbb{Z}_2$-transformation and equation \eqref{eq:SpmCpmrelation_text}, we have the relation $ S_{\pm}^{\text{II}}\propto - S_{\pm}^{\text{III}}\big\vert_{E}$ which means that $\delta \mathcal{R}_{\text{III}}$ and $\delta \mathcal{R}_{\text{II}}$ must have opposite signs.

\begin{figure}[t]
\centering
\begin{subfigure}{.32\textwidth}
  \centering
  \begin{tikzpicture}
		\node (img1)  {\includegraphics{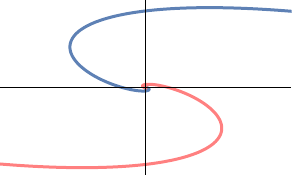}};
	\end{tikzpicture}
  \caption{$b>b_{\text{E}}$}
  \label{plot:analytical_Efimov_above_spiral}
\end{subfigure}
\hfill
\begin{subfigure}{.32\textwidth}
  \centering
  \begin{tikzpicture}
		\node (img1)  {\includegraphics{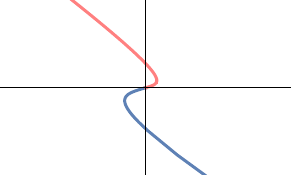}};
	\end{tikzpicture}
  \caption{$b<b_{\text{E}}$}
 \label{plot:analytical_peaks}
\end{subfigure}
\hfill
\begin{subfigure}{.32\textwidth}
  \centering
  \begin{tikzpicture}
		\node (img1)  {\includegraphics{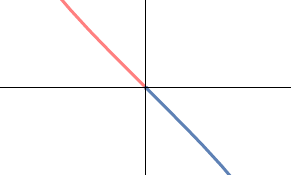}};
	\end{tikzpicture}
 \caption{$b<b_{\text{E}}$}
 \label{plot:analytical_no_peaks}
\end{subfigure}
\caption{Analytically derived behaviors of $\mathcal{C}-\mathcal{C}_{\text{I}}$ (vertical axis) as a function of $ \mathcal{R}-\mathcal{R}_{\text{I}} $ (horizontal axis) near the type I value $(\mathcal{C}_{\text{I}},\mathcal{R}_{\text{I}})$ (at the origin) for type II (blue) and III (pink) solutions. We have plotted three qualitatively different possibilities depending on the value of $b$ and details of the potential in $d = 4$ where $b_{\text{E}} \approx 0.52$. In (a) for $b=0.65$ above the Efimov bound $b>b_{\text{E}}$, $\mathcal{C}(\mathcal{R})$ is always multi-valued and exhibits an Efimov spiral circling around the type I value in the center an infinite number of times. For $b=0.47$ below the Efimov bound $b<b_{\text{E}}$ in figure (b), $\mathcal{C}(\mathcal{R})$ has a multi-valued region due to $\mathcal{C}(p)$ and $\mathcal{R}(p)$ exhibiting peaks as functions of IR parameters. If there are no peaks, $\mathcal{C}(\mathcal{R})$ is a single-valued function as in (c) with $b=0.47$.}
\end{figure}

It is useful to write \eqref{eq:deltaRC_type_AB} in a slightly different form as (focusing on the $\delta \mathcal{R}$ component)
\begin{equation}
    \delta \mathcal{R}_{\text{III}}
	= n_{\mathcal{R},\text{III}}^{-}\exp{\biggl(-\frac{b^{2}}{4b_{\text{c}}^{2}}\,\varphi_0\biggr)}\sin{\left(\sqrt{(b\slash b_{\text{E}})^{2}-1}\,\varphi_0 + n_{\mathcal{R},\text{III}}^+\right)}
	\label{eq:deltaRA}
\end{equation}
for type III and
\begin{equation}
	\delta \mathcal{R}_{\text{II}}=  n_{\mathcal{R},\text{II}}^-\exp{\biggl(-\frac{1}{8b_{\text{c}}^{2}}\frac{\log{(-S_{\infty,1}^{\text{II}})}}{1-(b_{\text{c}}\slash b)^{2}}\biggr)}\sin{\biggl(\frac{1}{2b}\frac{\sqrt{(b\slash b_{\text{E}})^{2}-1}}{1-(b_{\text{c}}\slash b)^{2}}\log{(-S_{\infty,1}^{\text{II}})} +  n_{\mathcal{R},\text{II}}^+\biggr)}
	\label{eq:deltaRB}
\end{equation}
for type II respectively. The constants $n_{\mathcal{R}}^{\pm}$ are given in terms of the components $m_{\mathcal{R}}^{\pm}$ of the matrix $M$ and $ \delta \mathcal{C}_{\text{II},\text{III}} $ have the exact same expressions in terms of $n_{\mathcal{C}}^{\pm}$. We observe that above the Efimov bound $b > b_{\text{E}}$, $(\mathcal{R}(p),\mathcal{C}(p))$ exhibit Efimov oscillations with a decreasing amplitude around their type I values as functions of IR parameters. They lead to an infinite number of peaks (turning points) in $\mathcal{R}(p)$ and $\mathcal{C}(p)$, see  Figure \ref{plot:analytical_Efimov_above}, and to the Efimov spiral \cite{Jarvinen:2011qe,Jarvinen:2015ofa,Kiritsis:2020bds} in $\mathcal{C}(\mathcal{R})$, see Figure \ref{plot:analytical_Efimov_above_spiral}.

Below the Efimov bound $b < b_{\text{E}}$, the sine function in \eqref{eq:deltaRA} - \eqref{eq:deltaRB} becomes a hyperbolic sine so that
\begin{align}
    \delta \mathcal{R}_{\text{III}} &= m_{\mathcal{R},\text{III}}^-\,e^{(b-\beta_-)\,\varphi_0}+m_{\mathcal{R},\text{III}}^+\,e^{(b-\beta_+)\,\varphi_0}\,,\nonumber\\
    \delta \mathcal{R}_{\text{II}} &= m_{\mathcal{R},\text{II}}^-\,(-S_{\infty,1}^{\text{II}})^{P_-}+m_{\mathcal{R},\text{II}}^+\,(-S_{\infty,1}^{\text{II}})^{P_+}
\label{eq:deltaR_II_III}
\end{align}
and similarly for $\delta\mathcal{C}_{\text{II,III}}$. This decays to zero without Efimov oscillations, however, if the coefficients have opposite signs, there may be at most a single peak in both type II and III branches in as in Figure \ref{plot:analytical_Efimov_below_peaks}. Figure \ref{plot:analytical_Efimov_below} shows the case when there are no peaks. The corresponding behaviors of $\mathcal{C}(\mathcal{R})$ are plotted in Figures \ref{plot:analytical_peaks} and \ref{plot:analytical_no_peaks}.

\subsection{The quantum phase transition}\label{subsec:phasetransition}

In the previous section we have shown that the curvature parameter $\mathcal{R}(p)$ of type II and III solutions approach the type I value $\mathcal{R}_{\text{I}}$ from different directions as functions of large IR parameters $p\rightarrow \infty$ defined in \eqref{eq:p_notation}. The situation changes when $p\rightarrow 0$ and $\mathcal{R}$ is far away from $\mathcal{R}_{\text{I}}$: in that case only either a type II or III solution exist. One can show analytically \cite{Ghosh:2017big,Ghosh:2018qtg,Ghosh:2020qsx} that for type III solutions $\mathcal{R} \sim \varphi_0^{-2\slash \Delta_-}\rightarrow \infty$ when $\varphi_0\rightarrow 0$. This means that type III solutions extend to infinite values of the curvature\footnote{An intuitive way of understanding this limit  is that  if $\varphi_0$ asymptotes the UV maximum, then the solution approaches the AdS solution with constant $\varphi$, i.e.  the source $\varphi_-$  goes to zero and $\mathcal{R} \to\infty$.}, but have no zero-curvature limit (if this existed  it would be obtained for $\varphi_0\to \infty$, since in the flat space solution the scalar field runs to infinity). In particular, the curvature of type III solutions must be bounded from below by a positive value.

If there are solutions  which connect  to  $\mathcal{R} \to 0$ they must be of  type II.  Indeed, this is the case because taking $p = \vert S_{\infty,1}^{\text{II}}\vert =0$ in type II solutions leads,  by equation \eqref{eq:typeB_T_coefficients_text}, to $T=0$, i.e. the flat space solution. Therefore, we have type III solutions at large curvature and type II solutions at small curvature, with the two branches  approaching each other at the value $\mathcal{R}_{\text{I}}$. This suggests  that there may be a phase transition in the vicinity of $\mathcal{R}_{\text{I}}$ and the nature of this transition depends on the value of $b$ and the matrices $M_{\text{II},\text{III}}$ in equation \eqref{eq:deltaRC_type_AB}. We confirm these statements in a numerical example in Section \ref{sec:numerics}.

For a fixed $\mathcal{R}$ near the critical point $\vert\mathcal{R}-\mathcal{R}_{\text{I}}\vert\ll 1$, both type II and III can be valid solutions for the same $\mathcal{R}$ which happens when $ \mathcal{R}(p) $ exhibits peaks. This is always guaranteed when $b$ is above the Efimov bound $b>b_{\text{E}}$ due to Efimov oscillations producing an infinite number of peaks as shown in Figure \ref{plot:analytical_Efimov_above}. Below the Efimov bound, $b_{\text{c}}<b<b_{\text{E}}$, there may also be a single peak as in Figure \ref{plot:analytical_Efimov_below_peaks}. In both of these cases, type II and III solutions compete when $\vert\mathcal{R}-\mathcal{R}_{\text{I}}\vert\ll 1$ and there is a phase transition at the point $\mathcal{R} = \mathcal{R}_{\text{c}} $ where the free energy of the type II and III branches cross. As a function of the IR parameter, the crossing will happen before the two branches reach the type I solution so that $\mathcal{R}_{\text{c}}\neq \mathcal{R}_{\text{I}}$. Therefore, there is no reason to expect the derivative of the free energy to be continuous at the crossing point, so that the transition must be of first-order. The free energy exhibits a swallow-tail, and the type I solution is thermodynamically unstable.

The situation changes if the exchange in dominance between type II and III happens exactly at the type I solution. This happens if $\mathcal{R}(p)$ does not have peaks which may only be the case below the Efimov bound (Figure \ref{plot:analytical_Efimov_below}). It follows that the type I solution is thermodynamically favored and the only available solution at $\mathcal{R}=\mathcal{R}_{\text{I}}$, because neither type II or III solution reach $\mathcal{R}=\mathcal{R}_{\text{I}}$. We now prove that in this case the transition must be at least second-order, and more generally, that it has to be of finite-order.

Let $F_{\text{II,III}}(\mathcal{R}_{\text{II,III}})$ denote the free energy of type II and III branches by respectively which become equal when they reach the type I solution. The renormalized free energy satisfies the differential equation
\begin{equation}
	F'(\mathcal{R}) = \mathcal{N}\,\biggl(\frac{2\,\mathcal{C}(\mathcal{R})}{\mathcal{R}^{3}}-\frac{1}{96}\,\frac{1}{\mathcal{R}}\biggr)\,,\quad F(\mathcal{R}_{\text{I}}) = F_{\text{I}}\,,
\end{equation}
where $ \mathcal{N}$ is defined in \eqref{eq:curlyN}. From this we obtain that the difference of the free energies of the type II and III solutions is given by
\begin{equation}
	\frac{F_{\text{III}}'(\mathcal{R})-F_{\text{II}}'(\mathcal{R})}{2\,\mathcal{N}} =\frac{\mathcal{C}_{\text{III}}(\mathcal{R})-\mathcal{C}_{\text{II}}(\mathcal{R})}{\mathcal{R}^{3}}\,.
\label{eq:deltaF_prime_text}
\end{equation}
Due to the fact that $\mathcal{C}_{\text{III}}(\mathcal{R}_{\text{I}}) = \mathcal{C}_{\text{II}}(\mathcal{R}_{\text{I}}) $ at the transition point, we obtain
\begin{equation}
	\frac{F'_{\text{III}}(\mathcal{R}_{\text{I}})-F'_{\text{II}}(\mathcal{R}_{\text{I}})}{2\,\mathcal{N}} = \frac{\mathcal{C}_{\text{III}}(\mathcal{R}_{\text{I}})-\mathcal{C}_{\text{II}}(\mathcal{R}_{\text{I}})}{\mathcal{R}^{3}_{\text{I}}} = 0\,,
	\label{eq:continuous_first_derivative_F}
\end{equation}
so that the first derivative of the free energy is continuous by continuity of $\mathcal{C}(\mathcal{R})$.\footnote{The continuity of $\mathcal{C}$ is explained in Section \ref{subsec:Efimov_oscillations} and can be seen from equation \eqref{eq:R_C_IR_expansion}.} Hence, the transition is at least second-order below the Efimov bound $b_{\text{c}}<b<b_{\text{E}}$ when $\mathcal{C}$ is a single-valued function of $\mathcal{R}$ (as in Figure \ref{plot:analytical_no_peaks}).

As shown in Section \ref{subsec:Efimov_oscillations}, the vev has the expansion
\begin{equation}
	\mathcal{C}(\mathcal{R}) = \mathcal{C}_{\text{I}} + \delta \mathcal{C}(p(\mathcal{R}))\,,\quad p\rightarrow \infty\,,
\end{equation}
which when substituted to \eqref{eq:deltaF_prime_text} gives
\begin{equation}
	\frac{F_{\text{III}}'(\mathcal{R})-F_{\text{II}}'(\mathcal{R})}{2\,\mathcal{N}} =\frac{\delta \mathcal{C}_{\text{III}}(p(\mathcal{R}))-\delta \mathcal{C}_{\text{II}}(p(\mathcal{R}))}{\mathcal{R}^{3}}\,,\quad \mathcal{R}\rightarrow \mathcal{R}_{\text{I}}\,.
	\label{eq:deltaF_prime_deltaC_text}
\end{equation}
The IR parameters $p = p(\mathcal{R})$ as functions of the dimensionless curvature may be obtained from the expansion
\begin{equation}
	\mathcal{R}(p) = \mathcal{R}_{\text{I}} + \delta \mathcal{R}(p)\,,
\end{equation}
where $\delta \mathcal{R}(p)$ for type II and III are given explicitly in \eqref{eq:deltaR_II_III}. The results at leading order in the limit $\mathcal{R}\rightarrow \mathcal{R}_{\text{I}} $ are given by
\begin{align}
	\varphi_0 &= \frac{1}{b-\beta_-}\biggl[\log{\biggl(\frac{\mathcal{R} - \mathcal{R}_{\text{I}}}{m^-_{\mathcal{R},\text{III}}}\biggr)} -\frac{m_{\mathcal{R},\text{III}}^+}{m_{\mathcal{R},\text{III}}^-}\biggl(\frac{\mathcal{R} - \mathcal{R}_{\text{I}}}{m^-_{\mathcal{R},\text{III}}}\biggr)^{(\beta_--\beta_+)\slash(b-\beta_-)} + \ldots\biggr]\,,\\
	S_{\infty,1}^{\text{II}} &= -\biggl(\frac{\mathcal{R} - \mathcal{R}_{\text{I}}}{m^-_{\mathcal{R},\text{II}}}\biggr)^{1\slash P_-}+\frac{1}{P_-}\frac{m_{\mathcal{R},\text{II}}^+}{m_{\mathcal{R},\text{II}}^-}\biggl(\frac{\mathcal{R} - \mathcal{R}_{\text{I}}}{m^-_{\mathcal{R},\text{II}}}\biggr)^{(P_+-P_-+1)\slash P_-} + \ldots\,.
\end{align}
Substituting to \eqref{eq:deltaF_prime_text},  we obtain  for both type II and type III solutions the expansion
\begin{equation}
	\delta \mathcal{C}(p) = \frac{m_{\mathcal{C}}^{-}}{m_{\mathcal{R}}^{-}}\,(\mathcal{R}-\mathcal{R}_{\text{I}}) + \frac{\det{M}}{(m_{\mathcal{R}}^{-})^{\delta+1}}\,(\mathcal{R}-\mathcal{R}_{\text{I}})^{\delta}+\ldots\,,\quad \mathcal{R}\rightarrow \mathcal{R}_{\text{I}}\,,
\end{equation}
where the determinant $\det{M} =m_{\mathcal{R}}^{-}m_{\mathcal{C}}^{+}-m_{\mathcal{R}}^{+}m_{\mathcal{C}}^{-} $ (non-zero by invertibility), the exponent\footnote{The reason why we obtain the same exponent for both type II and III solutions is because in terms of the higher-dimensional integration constants $\alpha_{\IR}$ and $\bar{\alpha}_{\IR}$, $S_{\pm}$ are controlled by the exponents $\tilde{\beta}_{\pm}$ as given in \eqref{eq:SpmIRparameters_text} which are the same for type II and III. It follows that $\delta = \tilde{\beta}_{+}\slash \tilde{\beta}_{-}  = (b-\beta_+)\slash (b-\beta_-)$ via \eqref{eq:betatilde_beta}.}
\begin{equation}
	\delta = \frac{b-\beta_+}{b-\beta_-} = \frac{(d-1)\,b+2\sqrt{1-(b\slash b_{\text{E}})^2}}{(d-1)\,b-2\sqrt{1-(b\slash b_{\text{E}})^2}}
\label{eq:delta_exponent}
\end{equation}
and the ellipsis denotes terms that vanish faster than $(\mathcal{R}-\mathcal{R}_{\text{I}})^{\delta}$. Substituting to \eqref{eq:deltaF_prime_deltaC_text}, we obtain
\begin{equation}
	\frac{F_{\text{III}}'(\mathcal{R})-F_{\text{II}}'(\mathcal{R})}{2\,\mathcal{N}} =\mathcal{F}_1\,\frac{\mathcal{R}-\mathcal{R}_{\text{I}}}{\mathcal{R}^{3}} + \mathcal{F}_\delta\,\frac{(\mathcal{R}-\mathcal{R}_{\text{I}})^{\delta}}{\mathcal{R}^{3}}+\ldots\,,\quad \mathcal{R}\rightarrow \mathcal{R}_{\text{I}}\,,
\label{eq:Fprime_diff_expansion}
\end{equation}
where we have defined
\begin{equation}
	\mathcal{F}_1\equiv \frac{m_{\mathcal{C},\text{III}}^{-}}{m_{\mathcal{R},\text{III}}^{-}}-\frac{m_{\mathcal{C},\text{II}}^{-}}{m_{\mathcal{R},\text{II}}^{-}}\,,\quad \mathcal{F}_\delta \equiv \frac{\det{M_{\text{III}}}}{(m_{\mathcal{R},\text{III}}^{-})^{\delta+1}}-\frac{\det{M_{\text{II}}}}{(m_{\mathcal{R},\text{II}}^{-})^{\delta+1}}\,.
\end{equation}
For $b_{\text{c}}< b \leq b_{\text{E}}$, the exponent satisfies $1 \leq \delta < \infty$. Therefore below the Efimov bound $b< b_{\text{E}}$, we have  $\delta > 1$, and the second term in \eqref{eq:Fprime_diff_expansion} is always subleading with respect to the first term. It follows that the discontinuity in the second derivative of the free energy is given by
\begin{equation}
	\frac{F_{\text{III}}''(\mathcal{R}_{\text{I}})-F_{\text{II}}''(\mathcal{R}_{\text{I}}) }{2\,\mathcal{N}}=\mathcal{R}_{\text{I}}^{-3}\,\mathcal{F}_1 =\mathcal{R}_{\text{I}}^{-3}\left(\frac{m_{\mathcal{C},\text{III}}^{-}}{m_{\mathcal{R},\text{III}}^{-}}-\frac{m_{\mathcal{C},\text{II}}^{-}}{m_{\mathcal{R},\text{II}}^{-}}\right) \,.
\label{eq:second_order_transition_condition}
\end{equation}
If the right-hand side vanishes, $\mathcal{F}_{1} = 0$, then the second derivative is continuous and the transition is at least third-order. We now calculate discontinuities in higher-order derivatives under the assumption that $\mathcal{F}_1 = 0$. In this case, the derivative of the free energy has the expansion
\begin{equation}
	\frac{F_{\text{III}}'(\mathcal{R})-F_{\text{II}}'(\mathcal{R})}{2\,\mathcal{N}} = \mathcal{F}_\delta\,\frac{(\mathcal{R}-\mathcal{R}_{\text{I}})^{\delta}}{\mathcal{R}^{3}_{\text{I}}}+\ldots\,,\quad \mathcal{R}\rightarrow \mathcal{R}_{\text{I}}\,,
	\label{eq:Fprime_diff_expansion_2}
\end{equation}
where we assume the generic situation in which $\delta$ is not an integer and $\mathcal{F}_\delta\neq 0$. How many derivatives vanish is dictated by $\lceil \delta\rceil\geq 2$, which gives assuming the transition is at least third-order, that
\begin{equation}
	\text{order of the transition} = 1+\lceil \delta\rceil\geq 3 \,.
\end{equation}
Therefore, we observe that below the Efimov bound $b_{\text{c}} < b < b_{\text{E}}$, the transition cannot be infinite-order in the generic situation in which $\mathcal{F}_\delta\neq 0$.

To summarize, above the Efimov bound $b>b_{\text{E}}$ the transition is always first-order, while below the Efimov bound $b_{\text{c}}<b<b_{\text{E}}$, there are two distinct situations depending on whether $\delta\mathcal{R}(p)$ and $\delta\mathcal{C}(p)$ exhibit a single peak (case (i)) or no peaks (case (ii)). In the case (i), the transition is first-order, but in the case (ii), it is either second-order or of order $1+\lceil \delta\rceil$ with $\delta$ given by \eqref{eq:delta_exponent}.

\section{Numerical RG flows in the Einstein-scalar theory}\label{sec:numerics}

We  now consider Einstein-scalar theory with the potential
\begin{equation}
	V(\varphi) = -\frac{d(d-1)}{\ell^{2}}+\biggl(\frac{\Delta_-(\Delta_--d)}{2\ell^{2}}-4V_{\infty}\,b^{2}\biggr)\,\varphi^{2}+4V_{\infty}\sinh^{2}{(b\varphi)}\,,
	\label{potentialwithfixedpoint}
\end{equation}
which is analogous to the potentials in \cite{Gubser:2008ny,Gubser:2008yx,Gursoy:2016ggq,Kiritsis:2016kog}. This potential has a UV fixed point at $ \varphi = 0 $ where
\begin{equation}
	V(\varphi) = -\frac{d(d-1)}{\ell^{2}}-\frac{1}{2}m^{2}\varphi^{2} + \mathcal{O}(\varphi^{3})\,,\quad \varphi \rightarrow 0\,,
\end{equation}
where $m^2\ell^2 = \Delta_-(d-\Delta_-) $. When $ \varphi \gg 1 $, the potential has the exponential behavior
\begin{equation}
	V(\varphi) = V_\infty\,e^{2b\varphi} +\mathcal{O}(\varphi^{2})\,, \quad \varphi\rightarrow \infty\,,
\end{equation}
which does not contain any subleading exponential of the type $e^{2\gamma \varphi}$ with $\gamma < b$ as in \eqref{exponentialpotential}. Therefore $V_{\infty,1} = 0$, and in the IR, this potential has an uplift to a higher-dimensional Einstein theory with a vanishing cosmological constant $ \widetilde{\Lambda} = 0 $.

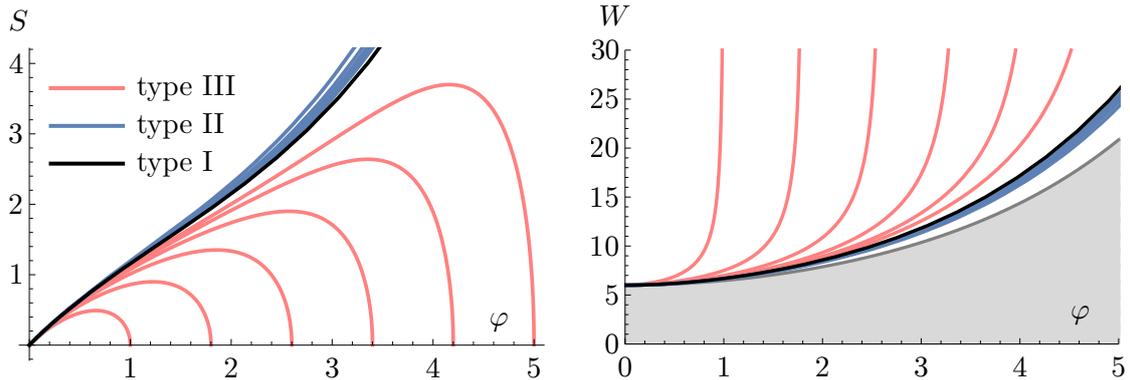
\begin{figure}[t]
	\begin{subfigure}[t]{0.5\textwidth}
		\centering
		\begin{tikzpicture}
			\node (img1)  {\includegraphics[scale=1]{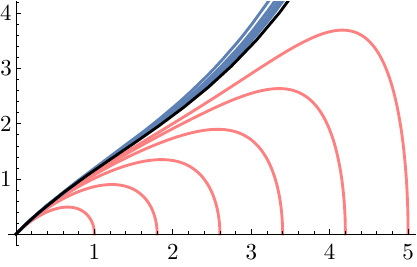}};
			\node[right=of img1, node distance=0cm, yshift=-1.4cm, xshift=-2cm] {$  \varphi $};
			\node[above=of img1, node distance=0cm, anchor=center,yshift=-0.8cm, xshift=-3.4cm] {$ S $};
			\draw[line width=1.5pt, color=pink] (-3,1.7) -- (-2,1.7) node[right, color=black] {type III};
			\draw[line width=1.5pt, color=mathematica1] (-3,1.2) -- (-2,1.2) node[right, color=black] {type II};
			\draw[line width=1.5pt, color=black] (-3,0.7) -- (-2,0.7) node[right, color=black] {type I};
		\end{tikzpicture}
		\label{plot:Ssolutions}
	\end{subfigure}
	\begin{subfigure}[t]{0.5\textwidth}
		\centering
		\begin{tikzpicture}
			\node (img1)  {\includegraphics[scale=1]{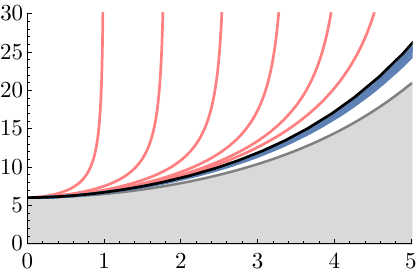}};
			\node[right=of img1, node distance=0cm, yshift=-1.4cm, xshift=-2cm] {$ \varphi $};
			\node[above=of img1, node distance=0cm, anchor=center, yshift=-0.9cm, xshift=-3.2cm] {$ W $};
		\end{tikzpicture}
		\label{plot:Wsolutions}
	\end{subfigure}
	\caption{Numerical solutions for $ S(\varphi) $ and $ W(\varphi) $ with $ b=0.47 $ (below the Efimov bound). The grey shaded area in the right figure is the region $W \leq \sqrt{-3V(\varphi)}$ which is a lower bound for $W(\varphi)$ on-shell.}
	\label{plot:SWsolutions}
\end{figure}

\subsection{Numerical solutions}\label{subsec:num_solutions}

We choose to work with four-dimensional theories $ d = 4 $ for which the bounds are
\begin{equation}
	b_{\text{c}} \approx 0.408\,, \quad b_{\text{E}} \approx 0.516\,, \quad b_{\text{G}} \approx 0.816\,.
\label{Ef}\end{equation}
We choose our parameters as
\begin{equation}
	\alpha = \ell =  1\,, \quad V_{\infty} = -1\,, \quad \Delta_- = \frac{3}{2}\,,
\end{equation}
where $\alpha$ is the UV radius of the $S^d$, and solve the second-order equation \eqref{Sequation} for $S(\varphi)$ numerically with boundary conditions imposed in the IR. Since the equation is second-order, we need to impose two boundary conditions: the first one is the value of $S(\varphi)$ in the IR (which can be infinite) while the second one is a regularity condition on the first derivative of $S(\varphi)$. To obtain type III solutions, we require that the solution behaves as \eqref{functionstypeAreduced} near $\varphi = \varphi_0$ which amounts to the two IR boundary conditions
\begin{equation}
    S(\varphi_0) = 0\,,\quad \lim_{\varphi\rightarrow \varphi_0}\sqrt{\varphi-\varphi_0}\,S'(\varphi) = \frac{S_0}{2} = \sqrt{-\frac{V'(\varphi_{0})}{2\,(d+1)}}\,.
\end{equation}
For the potential \eqref{potentialwithfixedpoint} used for our numerical calculations, type II solutions have the IR expansion
\begin{equation}
	S(\varphi) = 2b\,\sqrt{\frac{-V_{\infty}}{1-2\,(b^{2}-b_{\text{c}}^{2})}}\,e^{b\varphi} + S_{\infty,1}^{\text{II}}\,e^{(2\gamma_{\text{c}} - b)\,\varphi} +\ldots\,, \quad \varphi\rightarrow \infty\,,
 \label{eq:typeB_asymptotics_numerics}
\end{equation}
where the free parameter is the coefficient  $S_{\infty,1}^{\text{II}}$ of the subleading exponential and $S_{*}^{\text{II}} = 0$ since $V_{\infty,1} = 0$. Therefore to obtain type II solutions, we impose \eqref{eq:typeB_asymptotics_numerics} as an IR boundary condition on $S(\varphi)$ and its derivative at a fixed but large value of $\varphi$. The type I solution is obtained in the same manner by imposing the boundary condition
\begin{equation}
    S(\varphi) = \sqrt{\frac{-2V_{\infty}}{d-1}}\,e^{b\varphi} +\ldots\,, \quad \varphi\rightarrow \infty\,.
\end{equation}
As a result, we obtain a single type I solution and two one-parameter families of type II and III solutions parametrized by discrete (required to solve numerically the equation of motion) values of $\varphi_0>0$ and $S_{\infty,1}< 0$ respectively.

\begin{figure}[t]
	\begin{subfigure}[t]{0.5\textwidth}
		\centering
		\begin{tikzpicture}
			\node (img1)  {\includegraphics[scale=1]{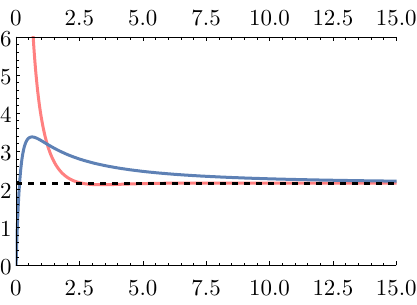}};
			\node[below=of img1, node distance=0cm, yshift=1.2cm, xshift=3.2cm] {$  \varphi_0 $};
			\node[above=of img1, node distance=0cm, yshift=-1.3cm, xshift=3cm] {$  \vert S_{\infty,1}^{\text{II}}\vert $};
			\node[left=of img1, node distance=0cm, anchor=center,yshift=0cm, xshift=0.8cm] {$ \mathcal{R} $};
			\draw[line width=1.5pt, color=pink] (-0.5,1.5) -- (1,1.5) node[right, color=black] {type III};
			\draw[line width=1.5pt, color=mathematica1] (-0.5,1) -- (1,1) node[right, color=black] {type II};
			\draw[line width=1.5pt, color=black,dashed] (-0.5,0.5) -- (1,0.5) node[right, color=black] {type I};
		\end{tikzpicture}
	\end{subfigure}
	\begin{subfigure}[t]{0.5\textwidth}
		\centering
		\begin{tikzpicture}
			\node (img1)  {\includegraphics[scale=1]{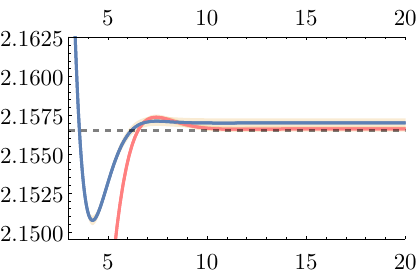}};
			\node[below=of img1, node distance=0cm, yshift=1.2cm, xshift=3.2cm] {$  \varphi_0 $};
			\node[above=of img1, node distance=0cm, yshift=-1.4cm, xshift=2.7cm] {$  \log{\vert S_{\infty,1}^{\text{II}}\vert} $};
			\node[left=of img1, node distance=0cm, anchor=center,yshift=0cm, xshift=0.8cm] {$ \mathcal{R} $};
		\end{tikzpicture}
		\label{plot:RIRparamscloseEfimov}
	\end{subfigure}
	\caption{The dimensionless curvature $ \mathcal{R} $ as a function of the IR parameters $ \varphi_0 $ and $ S_{\infty,1}^{\text{II}} $ of the type III and II solutions respectively for $ b=0.65 $ in $ d=4 $ (above the Efimov bound $ b_{\text{E}} \approx 0.52 $). The right plot is the same as the left plot, but with a longer range for the IR parameters (logarithmic scale for type II) and zoomed into a smaller interval of $\mathcal{R}$. The black dashed line is the type I value, while the shaded yellow regions enclose numerical noise.}
	\label{plot:RIRparamsEfimov}
\end{figure}
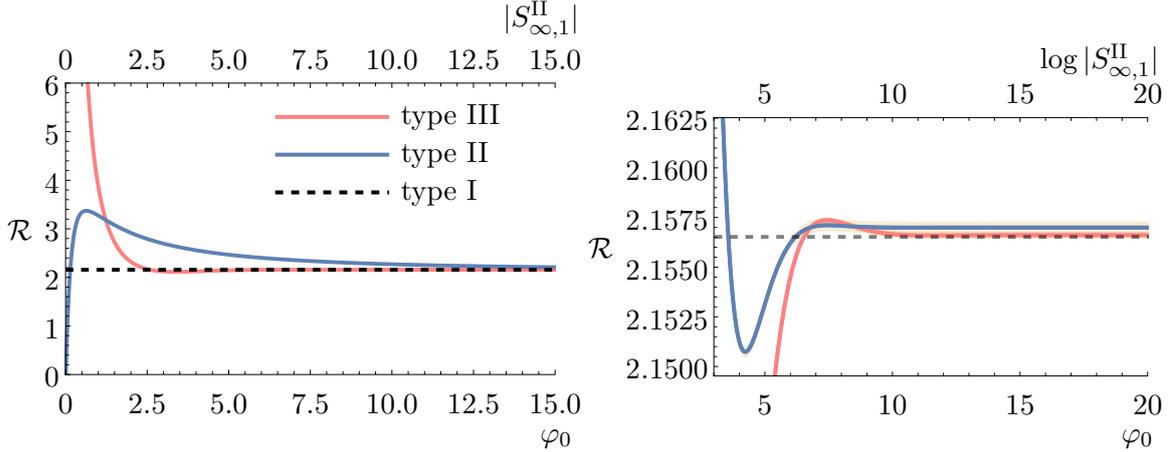

To visualize the solutions, we consider $b = 0.47$ which is below the Efimov bound. The resulting solutions $S(\varphi)$ are given in the left plot of Figure \ref{plot:SWsolutions}. The type I solution appears as an interface between type II and III solutions. We observe that type III solutions approach type I from below as $\varphi_0 \rightarrow \infty$, while type II solutions approach it from above as $S_{\infty,1}^{\text{II}}\rightarrow -\infty$. In all cases, we observe the expected linear behavior $S\sim \frac{\Delta_-}{\ell}\,\varphi$ in the UV $\varphi \rightarrow 0^+$. From the $S(\varphi)$ solutions we obtain $W(\varphi)$ using the formula \eqref{Wtext}. The results are given in the right plot of Figure \ref{plot:SWsolutions} where we again observe  that the type I solution is an interface between type II and III solutions. The solutions go to $W(0) = 2\,(d-1) = 6$ in the UV and are bounded from below by $ \sqrt{-\frac{4\,(d-1)}{d}\,V(\varphi)}=\sqrt{-3V(\varphi)}$ as required \cite{Ghosh:2017big}.

\begin{figure}[t]
	\begin{subfigure}[t]{0.5\textwidth}
		\centering
		\begin{tikzpicture}
			\node (img1)  {\includegraphics[scale=1]{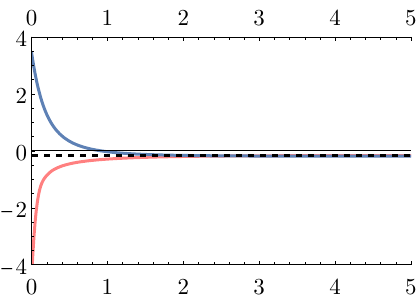}};
			\node[below=of img1, node distance=0cm, yshift=1.2cm, xshift=3.2cm] {$  \varphi_0 $};
			\node[above=of img1, node distance=0cm, yshift=-1.3cm, xshift=3cm] {$  \vert S_{\infty,1}^{\text{II}}\vert $};
			\node[left=of img1, node distance=0cm, anchor=center,yshift=0cm, xshift=1.1cm] {$ \mathcal{C} $};
			\draw[line width=1.5pt, color=pink] (-0.5,1.5) -- (1,1.5) node[right, color=black] {type III};
			\draw[line width=1.5pt, color=mathematica1] (-0.5,1) -- (1,1) node[right, color=black] {type II};
			\draw[line width=1.5pt, color=black,dashed] (-0.5,0.5) -- (1,0.5) node[right, color=black] {type I};
		\end{tikzpicture}
	\end{subfigure}
	\begin{subfigure}[t]{0.5\textwidth}
		\centering
		\begin{tikzpicture}
			\node (img1)  {\includegraphics[scale=1]{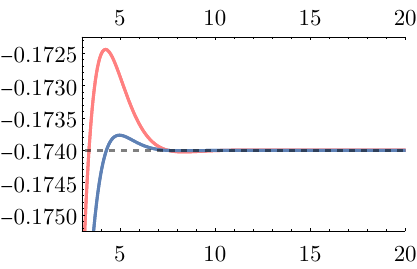}};
			\node[below=of img1, node distance=0cm, yshift=1.2cm, xshift=3.2cm] {$  \varphi_0 $};
			\node[above=of img1, node distance=0cm, yshift=-1.4cm, xshift=2.7cm] {$  \log{\vert S_{\infty,1}^{\text{II}}\vert} $};
			\node[left=of img1, node distance=0cm, anchor=center,yshift=0cm, xshift=0.8cm] {$ \mathcal{C} $};
		\end{tikzpicture}
		\label{plot:vevIRparamscloseefimov}
	\end{subfigure}
	\caption{The dimensionless vev $ \mathcal{C} $ as a function of the IR parameters $ \varphi_0 $ and $ S_{\infty,1}^{\text{II}} $ of the type III and II solutions respectively for $ b=0.65 $ in $ d=4 $ (above the Efimov bound $ b_{\text{E}} \approx 0.52  $). The right plot is the same as the left plot, but with a longer range for the IR parameters and zoomed into a smaller interval of $\mathcal{R}$. The black dashed line is the type I value. The shaded yellow regions enclosing numerical noise are so small that they are not visible.}
	\label{plot:vevIRparamsefimov}
\end{figure}
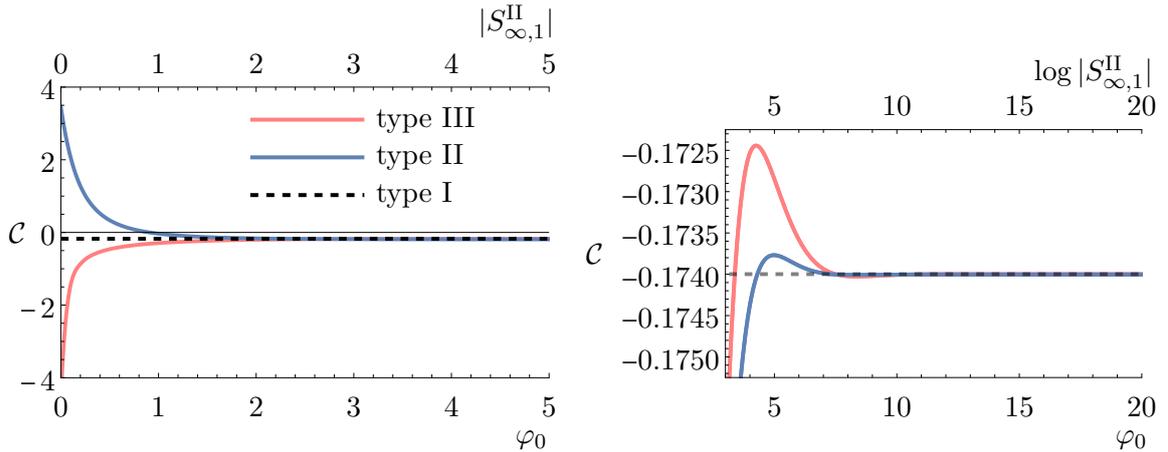

\subsection{Field theory parameters}

Given the families of numerical solutions $S(\varphi)$ and $W(\varphi)$, we can extract the two UV parameters $(\mathcal{R}(p),\mathcal{C}(p))$ as functions of the single IR parameter $p$ \eqref{eq:p_notation} (equal to $\varphi_0$ for type III and $\vert S_{\infty,1}^{\text{II}}\vert$ for type II solutions) from the UV behavior of the solutions in the limit $\varphi\rightarrow 0^+$ derived in Appendix \ref{app:UVasymptotics}. In the case of interest with $d = 4$ and $\Delta_- = 3\slash 2$, the leading terms are
\begin{equation}
	S(\varphi) = \frac{3}{2\ell}\,\varphi +\frac{8}{3}\frac{\mathcal{C}}{\ell}\,\varphi^{5\slash 3}  + \mathcal{O}(\varphi^2)\,,\quad W(\varphi) = \frac{6}{\ell}+\frac{\mathcal{R}}{4\ell}\,\varphi^{4\slash 3}+ \mathcal{O}(\varphi^2)\,,
\end{equation}
where we used $-1+d\slash \Delta_- = 5\slash 3$ and $2\slash \Delta_- = 4\slash 3$. Therefore $\mathcal{R}$ can be extracted from $W(\varphi)$ while $\mathcal{C}$ is obtained from $S(\varphi)$.

\paragraph{Above the Efimov bound.} We first consider the case $b \approx 0.65$ which is above the Efimov bound. First, we find that the type I solution corresponds to
\begin{equation}
    (\mathcal{R}_{\text{I}},\mathcal{C}_{\text{I}}) \approx  (2.15652, -0.173998)\,.
    \label{eq:typeC_value_Efimov}
\end{equation}
For type II and III solutions, numerical results for $\mathcal{R}$ are plotted in Figure \ref{plot:RIRparamsEfimov}. In Section \ref{subsec:Efimov_oscillations}, we have derived analytically the asymptotics ($p$ is the IR parameter \eqref{eq:p_notation})
\begin{equation}
    \mathcal{R}_{\text{II,III}}(p) = \mathcal{R}_{\text{I}}^{\text{II,III}} + \delta \mathcal{R}_{\text{II,III}}(p)\,,\quad p\rightarrow \infty\,,
\end{equation}
where $\delta \mathcal{R}_{\text{II,III}}(p)$ are given by \eqref{eq:deltaRA} and \eqref{eq:deltaRB} respectively containing two integration constants each. These constants together with the asymptotic values $\mathcal{R}_{\text{I}}^{\text{II,III}}$ can be fitted to numerics. We obtain a very good fit when the asymptotic value is given by
\begin{equation}
    \mathcal{R}_{\text{I}}^{\text{II,III}}  = \begin{dcases}
    2.15661\pm 0.00017\,,\quad &\text{type III}\\
    2.15699\pm 0.00031\,,\quad &\text{type II}\\
    \end{dcases}\,,
\end{equation}
where the error is due to numerical noise.\footnote{We suspect that this noise is caused by having to take a numerical derivative of $S(\varphi)$ to derive $W(\varphi)$. Indeed, we can observe that the noise in $\mathcal{C}$ is much smaller since it is extracted directly from $S(\varphi)$.} Analytically, we know that the asymptotic values for the type II and III branches must both coincide with the type I value $\mathcal{R}_{\text{I}}^{\text{III}} = \mathcal{R}_{\text{I}}^{\text{II}} = \mathcal{R}_{\text{I}}$ which is confirmed by the numerics within the error margin. Therefore the slightly different asymptotic limit for the type II and III branches is a numerical artifact.

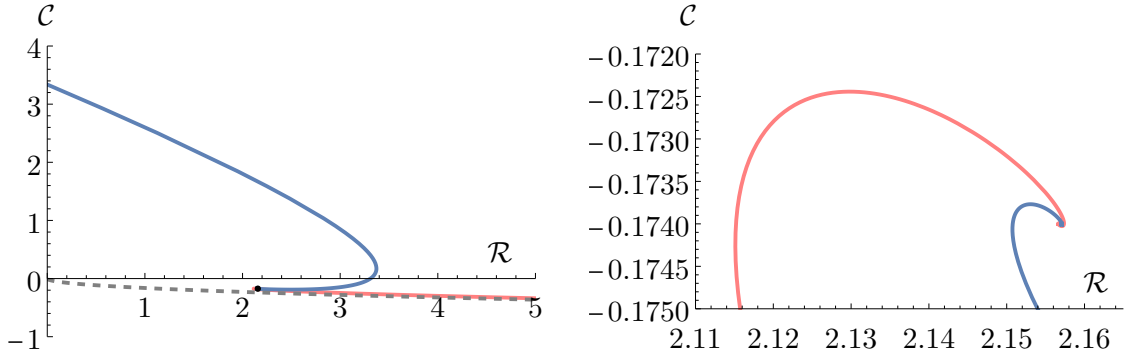
\begin{figure}[t]
	\begin{subfigure}[t]{0.5\textwidth}
		\centering
		\begin{tikzpicture}
			\node (img1)  {\includegraphics[scale=1]{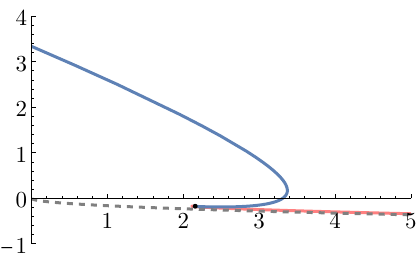}};
			\node[right=of img1, node distance=0cm, yshift=-0.8cm, xshift=-2cm] {$  \mathcal{R} $};
			\node[above=of img1, node distance=0cm, anchor=center,yshift=-1cm, xshift=-3cm] {$ \mathcal{C} $};
		\end{tikzpicture}
	\end{subfigure}
	\begin{subfigure}[t]{0.5\textwidth}
		\centering
		\begin{tikzpicture}
			\node (img1)  {\includegraphics[scale=1]{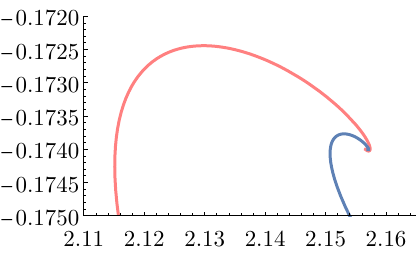}};
			\node[right=of img1, node distance=0cm, yshift=-1.15cm, xshift=-1.8cm] {$ \mathcal{R} $};
			\node[above=of img1, node distance=0cm, anchor=center, yshift=-0.9cm, xshift=-2.2cm] {$ \mathcal{C} $};
		\end{tikzpicture}
	\end{subfigure}
	\caption{The dimensionless vev $ \mathcal{C} $ as a function of $\mathcal{R}$ as obtained from numerics above the Efimov bound $b\approx 0.65$ in $d = 4$. The black dot is the type I value \eqref{eq:typeC_value_Efimov} while the grey dashed line is the analytical $\mathcal{R}\rightarrow \infty$ asymptotic $\mathcal{C}(\mathcal{R}) = -\frac{3\sqrt{3}}{32}\,\mathcal{R}^{1\slash 2}+\ldots$ (derived in \cite{Ghosh:2020qsx}) when $\Delta_- = 3\slash 2$. The right plot is the same as the left plot, but zoomed in around the type I value (the black dot in the left figure) where the Efimov spiral becomes visible.}
	\label{plot:vevcurlyRefimov}
\end{figure}

Similarly, for type II and III solutions, numerical results for $\mathcal{C}$ are plotted in Figure \ref{plot:vevIRparamsefimov}. Analytically, we know that $\mathcal{C}$ obeys analogous asymptotics
\begin{equation}
    \mathcal{C}_{\text{II,III}}(p) = \mathcal{C}_{\text{I}}^{\text{II,III}}  + \delta \mathcal{C}_{\text{II,III}}(p)\,,\quad p\rightarrow \infty\,,
\end{equation}
where $\delta \mathcal{C}_{\text{II,III}}(p)$ are given by the same formulae \eqref{eq:deltaRA} and \eqref{eq:deltaRB}, but with a set of integration constants independent from the ones in $\mathcal{R}$. A good fit to the numerics is obtained when the asymptotic limits are
\begin{equation}
    \mathcal{C}_{\text{I}}^{\text{II,III}} = \begin{dcases}
    -0.173999\pm 2\times 10^{-6}\,,\quad &\text{type III}\\
    -0.174003\pm 7\times 10^{-6}\,,\quad &\text{type II}\\
    \end{dcases}\,.
\end{equation}
Analytics require that $\mathcal{C}_{\text{I}}^{\text{III}} = \mathcal{C}_{\text{I}}^{\text{II}}  = \mathcal{C}_{\text{I}}$ which is confirmed by numerics to a very high accuracy. We observe that the error is orders of magnitude smaller compared to $\mathcal{R}$.

In Figure \ref{plot:vevcurlyRefimov}, the dimensionless vev $\mathcal{C} = \mathcal{C}(\mathcal{R}) $ is plotted as a function of the dimensionless curvature $\mathcal{R}$. As $\mathcal{R}\rightarrow \infty$, the type III branch approaches the dashed line in the left plot which is the analytical large-$\mathcal{R}$ asymptotic derived in \cite{Ghosh:2020qsx}. Because of the oscillatory behavior of $\mathcal{R}(p)$ and $\mathcal{C}(p)$ as functions of the IR parameters, $\mathcal{C} = \mathcal{C}(\mathcal{R}) $ exhibits an Efimov spiral which is visible in the right plot of Figure \ref{plot:vevcurlyRefimov}. Because of the spiral, $\mathcal{C}(\mathcal{R})$ is a multi-valued function of $\mathcal{R}$ in an intermediate regime suggesting the presence of a quantum phase transition from the type III branch to the type II branch as $\mathcal{R}$ becomes smaller.

\begin{figure}[t]
	\begin{subfigure}[t]{0.5\textwidth}
		\centering
		\begin{tikzpicture}
			\node (img1)  {\includegraphics[scale=1]{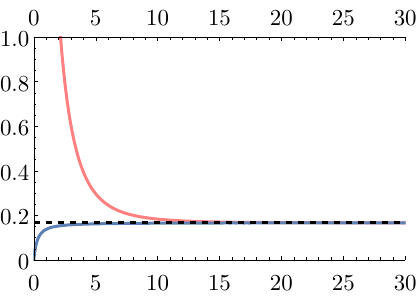}};
			\node[below=of img1, node distance=0cm, yshift=1.2cm, xshift=3.2cm] {$  \varphi_0 $};
			\node[above=of img1, node distance=0cm, yshift=-1.3cm, xshift=3cm] {$  \vert S_{\infty,1}^{\text{II}}\vert $};
			\node[left=of img1, node distance=0cm, anchor=center,yshift=0cm, xshift=0.8cm] {$ \mathcal{R} $};
			\draw[line width=1.5pt, color=pink] (-0.5,1) -- (1,1) node[right, color=black] {type III};
			\draw[line width=1.5pt, color=mathematica1] (-0.5,0.5) -- (1,0.5) node[right, color=black] {type II};
			\draw[line width=1.5pt, color=black,dashed] (-0.5,0) -- (1,0) node[right, color=black] {type I};
		\end{tikzpicture}
	\end{subfigure}
	\begin{subfigure}[t]{0.5\textwidth}
		\centering
		\begin{tikzpicture}
			\node (img1)  {\includegraphics[scale=1]{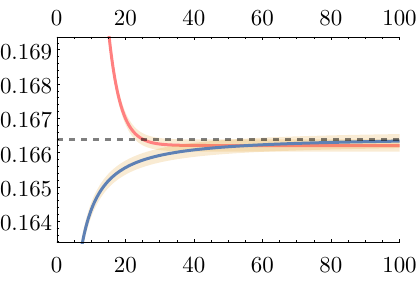}};
			\node[below=of img1, node distance=0cm, yshift=1.2cm, xshift=3.2cm] {$  \varphi_0 $};
			\node[above=of img1, node distance=0cm, yshift=-1.4cm, xshift=3cm] {$  \vert S_{\infty,1}^{\text{II}}\vert $};
			\node[left=of img1, node distance=0cm, anchor=center,yshift=0cm, xshift=0.8cm] {$ \mathcal{R} $};
		\end{tikzpicture}
		\label{plot:RIRparamsclose}
	\end{subfigure}
	\caption{The dimensionless curvature $ \mathcal{R} $ as a function of the IR parameters $ \varphi_0 $ and $ S_{\infty,1}^{\text{II}} $ of the type III and II solutions respectively for $ b=0.47 $ in $ d=4 $ below the Efimov bound $ b_{\text{E}} \approx 0.52  $. The right plot is the same as the left plot, but with a longer range for the IR parameters and zoomed into a smaller interval of $\mathcal{R}$. The black dashed line is the type I value, while the shaded yellow regions enclose numerical noise.}
	\label{plot:RIRparams}
\end{figure}
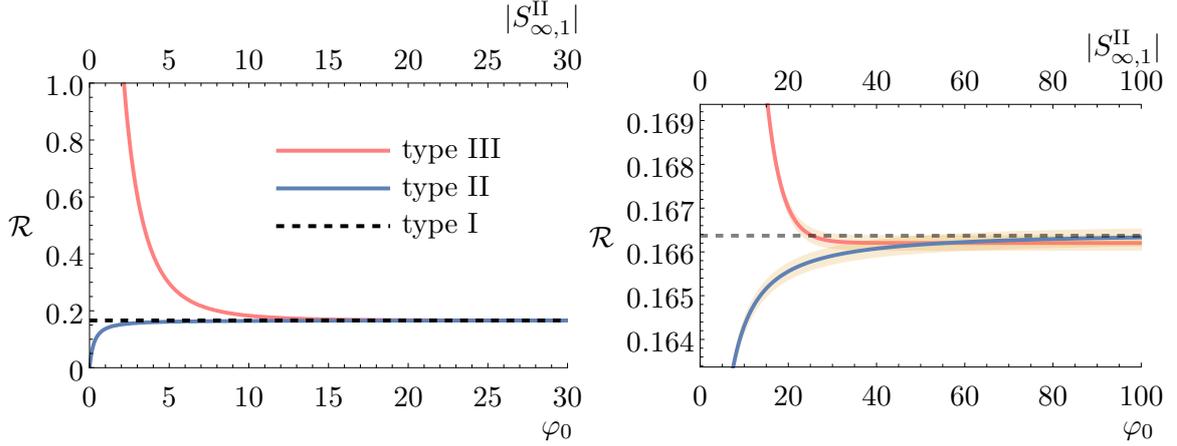

\paragraph{Below the Efimov bound.} We now consider the case when $b \approx 0.47$ below the Efimov bound, but above the confinement bound. First, we find that the type I solution corresponds to
\begin{equation}
    (\mathcal{R}_{\text{I}},\mathcal{C}_{\text{I}}) \approx  (0.1664,-0.12650)\,.
    \label{eq:typeC_value_below_Efimov}
\end{equation}
For type II and III solutions, numerical results for $\mathcal{R}$ are plotted in Figure \ref{plot:RIRparams}. In Section \ref{subsec:Efimov_oscillations}, we derived analytically the asymptotics
\begin{equation}
    \mathcal{R}_{\text{II,III}}(p) = \mathcal{R}_{\text{I}}^{\text{II,III}} + \delta \mathcal{R}_{\text{II,III}}(p)\,,\quad p\rightarrow \infty\,,
\end{equation}
where $\delta \mathcal{R}_{\text{II,III}}(p)$ are given by \eqref{eq:deltaR_II_III}. A good fit to the numerics is obtained when the asymptotic value is given by
\begin{equation}
    \mathcal{R}_{\text{I}}^{\text{II,III}}  = \begin{dcases}
    0.1662\pm 0.0002\,,\quad &\text{type III}\\
    0.1665\pm 0.0002\,,\quad &\text{type II}\\
    \end{dcases}\,,
\end{equation}
where the error is due to numerical noise. Asymptotically, the type II and III branches approach the same type I value within the numerical noise as expected based the on the analytical calculation. We emphasize, that the overlap region between the type II and III branches is a numerical artifact.

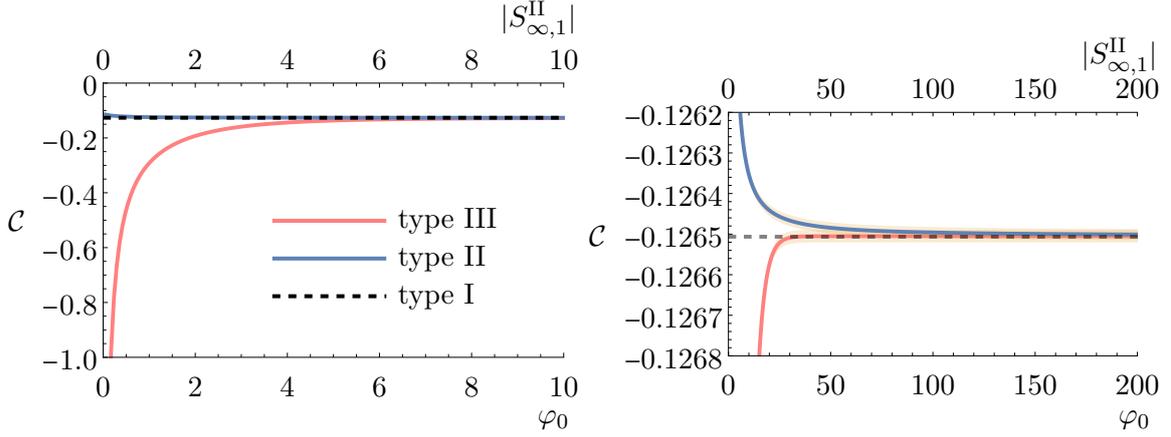
\begin{figure}[t]
	\begin{subfigure}[t]{0.5\textwidth}
		\centering
		\begin{tikzpicture}
			\node (img1)  {\includegraphics[scale=1]{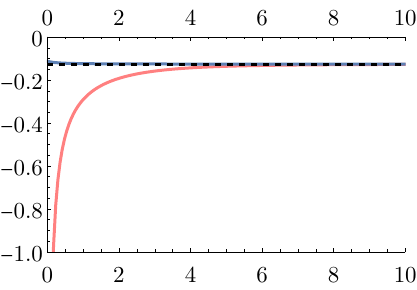}};
			\node[below=of img1, node distance=0cm, yshift=1.2cm, xshift=3.2cm] {$  \varphi_0 $};
			\node[above=of img1, node distance=0cm, yshift=-1.3cm, xshift=3cm] {$  \vert S_{\infty,1}^{\text{II}}\vert $};
			\node[left=of img1, node distance=0cm, anchor=center,yshift=0cm, xshift=0.8cm] {$ \mathcal{C} $};
			\draw[line width=1.5pt, color=pink] (-0.5,0) -- (1,0) node[right, color=black] {type III};
			\draw[line width=1.5pt, color=mathematica1] (-0.5,-0.5) -- (1,-0.5) node[right, color=black] {type II};
			\draw[line width=1.5pt, color=black,dashed] (-0.5,-1) -- (1,-1) node[right, color=black] {type I};
		\end{tikzpicture}
	\end{subfigure}
	\begin{subfigure}[t]{0.5\textwidth}
		\centering
		\begin{tikzpicture}
			\node (img1)  {\includegraphics[scale=1]{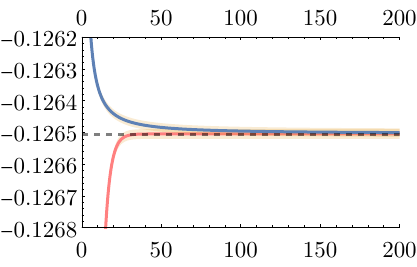}};
			\node[below=of img1, node distance=0cm, yshift=1.2cm, xshift=3.2cm] {$  \varphi_0 $};
			\node[above=of img1, node distance=0cm, yshift=-1.4cm, xshift=3cm] {$  \vert S_{\infty,1}^{\text{II}}\vert $};
			\node[left=of img1, node distance=0cm, anchor=center,yshift=0cm, xshift=0.8cm] {$ \mathcal{C} $};
		\end{tikzpicture}
		\label{plot:vevIRparamsclose}
	\end{subfigure}
	\caption{The dimensionless vev $ \mathcal{C} $ as a function of the IR parameters $ \varphi_0 $ and $ S_{\infty,1}^{\text{II}} $ of the type II and III solutions respectively for $ b=0.47 $ in $ d=4 $ below the Efimov bound $ b_{\text{E}} \approx 0.52  $. The right plot is the same as the left plot, but with a longer range for the IR parameters and zoomed into a smaller interval of $\mathcal{R}$. The black dashed line is the type I value, while the shaded yellow regions enclose numerical noise.}
	\label{plot:vevIRparams}
\end{figure}

For type II and III solutions, numerical results for $\mathcal{C}$ are plotted in Figure \ref{plot:vevIRparams}. Analogously to above, $\mathcal{C}$ obeys the asymptotics
\begin{equation}
    \mathcal{C}_{\text{II,III}}(p) = \mathcal{C}_{\text{I}}^{\text{II,III}}  + \delta \mathcal{C}_{\text{II,III}}(p)\,,\quad p\rightarrow \infty\,.
\end{equation}
A good fit is obtained when the asymptotic limits are given by
\begin{equation}
    \mathcal{C}_{\text{I}}^{\text{II,III}} = \begin{dcases}
    -0.12651\pm 0.00002\,,\quad &\text{type III}\\
    -0.12650\pm 0.00001\,,\quad &\text{type II}\\
    \end{dcases}\,,
\label{eq:below_Efimov_asymptotics_C}
\end{equation}
which are equal up to numerical error as expected by analytics. We observe that the error is an order of magnitude smaller compared to $\mathcal{R}$ and the overlap region between type II and III branches is much smaller.

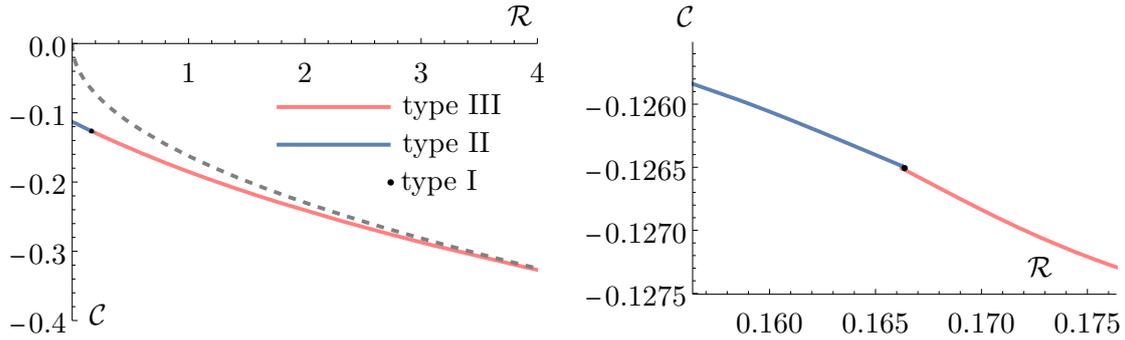
\begin{figure}[t]
	\begin{subfigure}[t]{0.5\textwidth}
		\centering
		\begin{tikzpicture}
			\node (img1)  {\includegraphics[scale=1]{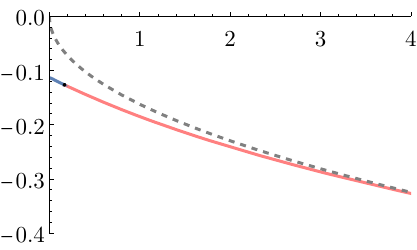}};
			\node[above=of img1, node distance=0cm, yshift=-1.3cm, xshift=3.2cm] {$  \mathcal{R} $};
			\node[below=of img1, node distance=0cm, anchor=center,yshift=1.5cm, xshift=-2.35cm] {$ \mathcal{C} $};
			\draw[line width=1.5pt, color=pink] (0,1) -- (1.5,1) node[right, color=black] {type III};
			\draw[line width=1.5pt, color=mathematica1] (0,0.5) -- (1.5,0.5) node[right, color=black] {type II};
            \fill[] (1.5,0) circle (1.2pt) node[right, color=black] {type I};
		\end{tikzpicture}
	\end{subfigure}
	\begin{subfigure}[t]{0.5\textwidth}
		\centering
		\begin{tikzpicture}
			\node (img1)  {\includegraphics[scale=1]{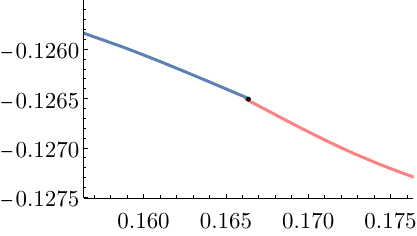}};
			\node[right=of img1, node distance=0cm, yshift=-1cm, xshift=-2.5cm] {$ \mathcal{R}$};
			\node[above=of img1, node distance=0cm, anchor=center, yshift=-0.8cm, xshift=-2.2cm] {$ \mathcal{C} $};
		\end{tikzpicture}
		\label{plot:vevRplotclose}
	\end{subfigure}
	\caption{The dimensionless vev $ \mathcal{C} $ as a function of the dimensionless curvature $ \mathcal{R} $ below the Efimov bound $ b=0.47 $ in $ d=4 $. The black dot is the type I value \eqref{eq:typeC_value_below_Efimov} while the grey dashed line is the analytical $\mathcal{R}\rightarrow \infty$ asymptotic $\mathcal{C}(\mathcal{R}) = -\frac{3\sqrt{3}}{32}\,\mathcal{R}^{1\slash 2}+\ldots$ (derived in \cite{Ghosh:2020qsx}) when $\Delta_- = 3\slash 2$. The right plot is the same as the left plot, but zoomed in around the type I value (the black dot in the left figure).}
	\label{plot:vevRplot}
\end{figure}

In Figure \ref{plot:vevRplot}, we plot the dimensionless vev as a function of the dimensionless curvature $\mathcal{C} = \mathcal{C}(\mathcal{R}) $.
As $\mathcal{R}\rightarrow \infty$, the type III branch approaches the dashed line in the left plot which is the analytical large-$\mathcal{R}$ asymptotic derived in \cite{Ghosh:2020qsx}. In the right plot, we observe that the type II and III branches meet at the type I value $(\mathcal{R}_{\text{I}},\mathcal{C}_{\text{I}})$ within numerical error.

From our numerical plots it is clear that neither $\mathcal{R}$ or $\mathcal{C}$ exhibit peaks as functions of the IR parameters $p = \varphi_0,\vert S_{\infty,1}^{\text{II}}\vert $. Therefore our numerics below the Efimov bound $b_{\text{c}}< b <b_{\text{E}}$ are in line with the case (ii) described at the end of Section \ref{subsec:phasetransition} in which $\mathcal{C}$ is a single valued function for all $\mathcal{R}$ and in which the type I solution is the only available solution at $\mathcal{R} = \mathcal{R}_{\text{I}}$. Therefore, if also the free energy is a single valued function of $\mathcal{R}$ there will be a phase transition at $\mathcal{R} = \mathcal{R}_{\text{I}} $, and as proven analytically in Section \ref{subsec:phasetransition}, it must be second- or higher-order. We confirm that this is what happens in the next section.

\subsection{Renormalized free energy}

We  now calculate the renormalized free energy $F(\mathcal{R})$ by solving the first-order differential equation \eqref{eq:F_diff_eq} numerically using the results for $(\mathcal{R},\mathcal{C})$ obtained in the previous section. To solve the equation, we need to know the value of the free energy at some value of $\mathcal{R}$. In the limit $\mathcal{R}\rightarrow \infty$, the corresponding bulk solution is a type III geometry with $\varphi_0\rightarrow 0$ and one can show analytically that \cite{Ghosh:2020qsx}
\begin{equation}
	\mathcal{R}^{-2}\,\mathcal{C}_{\text{III}}(\mathcal{R}) = \mathcal{O}(\mathcal{R}^{-\Delta_-})\,,\quad \mathcal{R}^{-1}\,\mathcal{B}_{\text{III}}(\mathcal{R}) = \frac{1}{96}\biggl(1+\log{\frac{\mathcal{R}}{48}}\biggr)+\mathcal{O}(\mathcal{R}^{-\Delta_-})\,,\quad \mathcal{R}\rightarrow \infty\,.
\end{equation}
By substituting into expression \eqref{eq:renormalizedFtext}, we obtain that the free energy of the type III solution has the asymptotics
\begin{equation}
	F_{\text{III}}(\mathcal{R}) = -\frac{\mathcal{N}}{96}\biggl(1+\log{\frac{\mathcal{R}}{48}}\biggr)+\mathcal{O}(\mathcal{R}^{-\Delta_-}) \,,\quad \mathcal{R}\rightarrow \infty\,,
 \label{eq:F_boundary_condition}
\end{equation}
which we may impose as a boundary condition to solve \eqref{eq:F_diff_eq} for $F_{\text{III}}(\mathcal{R})$. The result is valid for all $\mathcal{R}$ down to $\lim_{\varphi_0\rightarrow \infty}\mathcal{R}_{\text{III}}(\varphi_0) = \mathcal{R}_{\text{I}}$ where the free energy of the type III solution takes the type I value $F_{\text{III}}(\mathcal{R}_{\text{I}}) = F_{\text{I}}$. For smaller $\mathcal{R}$, the free energy is controlled by the type II solution $F_{\text{II}}(\mathcal{R})$. To solve for the type II free energy using \eqref{eq:F_diff_eq}, we impose $F_{\text{III}}(\mathcal{R}_{\text{I}})$ as a boundary condition for $F_{\text{II}}(\mathcal{R})$ at $\mathcal{R} = \mathcal{R}_{\text{I}}$. The result for $F_{\text{II}}(\mathcal{R})$ is valid down to $\mathcal{R} = 0$, so that together with $F_{\text{III}}(\mathcal{R})$, we obtain the free energy for all $\mathcal{R}>0$.

The method for the calculation of the free energy outlined here has the advantage that it only requires the knowledge of $\mathcal{C} = \mathcal{C}(\mathcal{R})$ as input, which is determined by $\mathcal{R} = \mathcal{R}(p)$ and $\mathcal{C} = \mathcal{C}(p)$ obtained from the UV expansions of $S(\varphi)$ and $W(\varphi)$. On the other hand, direct application of the formula \eqref{eq:renormalizedFtext} would require the knowledge of $\mathcal{B} = \mathcal{B}(\mathcal{R}) $ which could be obtained from the UV asymptotics of the function $U(\varphi)$ (see Appendix \ref{app:UVasymptotics}). However, solving $U(\varphi)$ from the first-order differential equation \eqref{Uequationtext} leads to additional numerical uncertainties which are avoided by the method above.

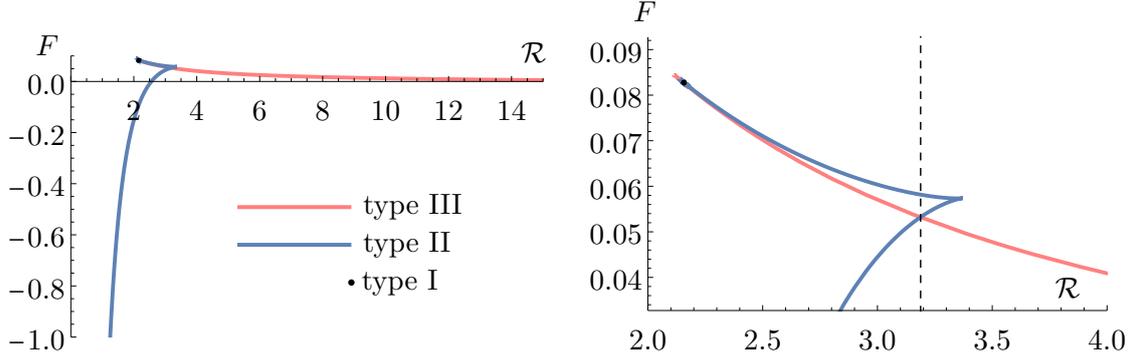
\begin{figure}[t]
	\begin{subfigure}[t]{0.5\textwidth}
		\centering
		\begin{tikzpicture}
		\node (img1)  {\includegraphics[scale=1]{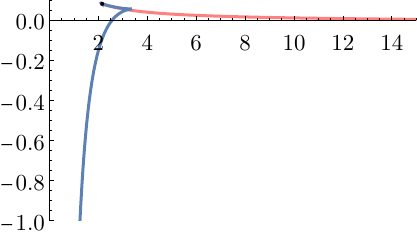}};
		\node[above=of img1, node distance=0cm, anchor=center,yshift=-1.1cm, xshift=3.4cm] {$ \mathcal{R} $};
		\node[above=of img1, node distance=0cm, anchor=center,yshift=-1cm, xshift=-3cm] {$ F $};
		\draw[line width=1.5pt, color=pink] (-0.5,0) -- (1,0) node[right, color=black] {type III};
		\draw[line width=1.5pt, color=mathematica1] (-0.5,-0.5) -- (1,-0.5) node[right, color=black] {type II};
		\fill[] (1,-1) circle (1.2pt) node[right, color=black] {type I};
		\end{tikzpicture}

	\end{subfigure}
	\begin{subfigure}[t]{0.5\textwidth}
		\centering
		\begin{tikzpicture}
		\node (img1)  {\includegraphics[scale=1]{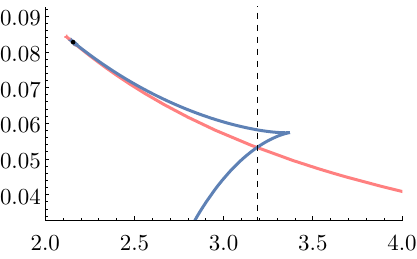}};
		\node[right=of img1, node distance=0cm, yshift=-1.25cm, xshift=-2.2cm] {$ \mathcal{R} $};
		\node[above=of img1, node distance=0cm, anchor=center, yshift=-0.9cm, xshift=-2.8cm] {$ F $};
		\end{tikzpicture}
	\end{subfigure}
	\caption{Renormalized free energy \eqref{eq:renormalizedFtext}  as a function of the dimensionless curvature $ \mathcal{R} $ above the Efimov bound $ b=0.65 $ in $ d=4 $ with the normalization set to unity $\mathcal{N}\equiv 1$. In the right plot, the swallow tail appearing in the left plot has been magnified and the vertical dashed line is the location $\mathcal{R} = \mathcal{R}_{\text{c}}$ of the first-order phase transition.}
	\label{FRabove}
\end{figure}

\paragraph{Numerics above the Efimov bound.} The numerically evaluated free energy above the Efimov bound is plotted in Figure \ref{FRabove}. The free energies $F_{\text{II}}(p)$ and $F_{\text{III}}(p)$ of the type II and III branches of solutions respectively are individually multi-valued functions of the IR parameters due to Efimov oscillations. This suggests the existence of a swallow-tail in the free energy $F$ as a function of $\mathcal{R}$,  which is visible in the right plot of Figure \ref{FRabove}. Because of the swallow-tail, the type I solution is thermodynamically unstable and there is a phase transition before the type II and III branches reach it. The transition occurs at $\mathcal{R} = \mathcal{R}_{\text{c}} \approx 3.19$ with the thermodynamically favored solution changing from the type II branch to the type III one as $\mathcal{R}$ is increased from zero to infinity. The transition is evidently first-order with a discontinuous first derivative confirming our analytical expectations.

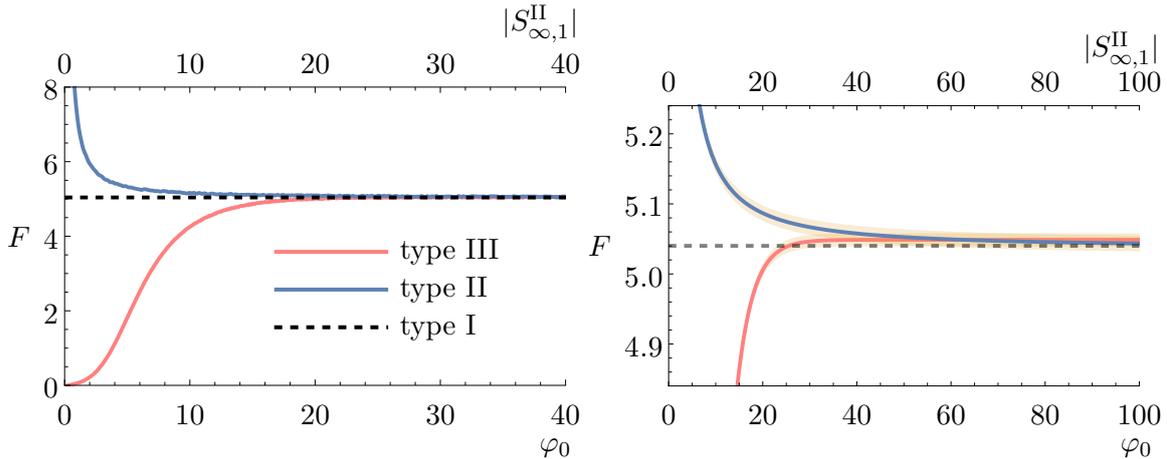
\begin{figure}[t]
	\begin{subfigure}[t]{0.5\textwidth}
		\centering
		\begin{tikzpicture}
			\node (img1)  {\includegraphics[scale=1]{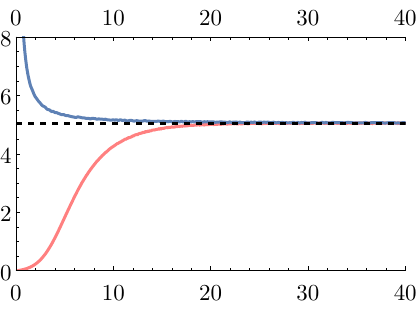}};
			\node[below=of img1, node distance=0cm, yshift=1.2cm, xshift=3.2cm] {$  \varphi_0 $};
			\node[above=of img1, node distance=0cm, yshift=-1.3cm, xshift=3cm] {$  \vert S_{\infty,1}^{\text{II}}\vert $};
			\node[left=of img1, node distance=0cm, anchor=center,yshift=0cm, xshift=0.8cm] {$ F $};
			\draw[line width=1.5pt, color=pink] (-0.5,-0.2) -- (1,-0.2) node[right, color=black] {type III};
			\draw[line width=1.5pt, color=mathematica1] (-0.5,-0.7) -- (1,-0.7) node[right, color=black] {type II};
			\draw[line width=1.5pt, color=black,dashed] (-0.5,-1.2) -- (1,-1.2) node[right, color=black] {type I};
		\end{tikzpicture}
	\end{subfigure}
	\begin{subfigure}[t]{0.5\textwidth}
		\centering
		\begin{tikzpicture}
			\node (img1)  {\includegraphics[scale=1]{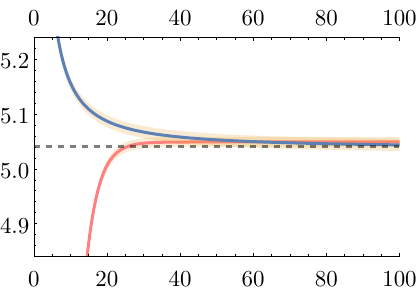}};
			\node[below=of img1, node distance=0cm, yshift=1.2cm, xshift=3.2cm] {$  \varphi_0 $};
			\node[above=of img1, node distance=0cm, yshift=-1.4cm, xshift=3cm] {$  \vert S_{\infty,1}^{\text{II}}\vert $};
			\node[left=of img1, node distance=0cm, anchor=center,yshift=0cm, xshift=0.8cm] {$ F $};
		\end{tikzpicture}
	\end{subfigure}
	\caption{Renormalized free energy of type II and III solutions as a function of the IR parameters for $ b=0.47 $ below the Efimov bound in $ d=4 $. The right plot is the same as the left plot, but with a longer range for the IR parameters and zoomed into a smaller interval of $\mathcal{R}$. The black dashed line is the type I value, while the shaded yellow regions enclose numerical noise.}
	\label{plot:FIRparameters}
\end{figure}

\paragraph{Numerics below the Efimov bound.} Below the Efimov bound there are no Efimov oscillations and we expect the transition to be at least second-order based on the previous section. The numerically evaluated free energy as a function of IR parameters of the type II and III solutions is plotted in Figure \ref{plot:FIRparameters}. We observe no Efimov oscillations and the analytical fit for the $p\rightarrow \infty$ asymptotic is given by
\begin{equation}
    F(p) = F_{\text{I}}^{\text{II,III}}+\delta F(p)\,,
\end{equation}
where by using \eqref{eq:F_diff_eq} the perturbation $\delta F(p)$ may be shown take the same form as $\delta\mathcal{R}(p)$ in equation \eqref{eq:deltaR_II_III}, but with a different set of coefficients. Since the type II and III branches meet at the type I solution, the asymptotic values should also be the same and equal to the type I value $F_{\text{I}}^{\text{II}} = F_{\text{I}}^{\text{III}} = F_{\text{I}}$. A good fit to the numerics is obtained with (with $\mathcal{N} \equiv 1$)
\begin{equation}
    F_{\text{I}}^{\text{II,III}} = \begin{dcases}
    5.04872 \pm 0.00928\,,\quad &\text{type III}\\
    5.03684 \pm 0.01111\,,\quad &\text{type II}\\
    \end{dcases}\,,
\label{eq:below_Efimov_asymptotic_F}
\end{equation}
which are equal up to error. The overlap region is therefore a numerical artifact and $F$ is a single valued function of IR parameters as required by analytical expectations. We observe no Efimov oscillations.

\begin{figure}[t]
	\begin{subfigure}[t]{0.5\textwidth}
		\centering
		\begin{tikzpicture}
		\node (img1)  {\includegraphics[scale=1]{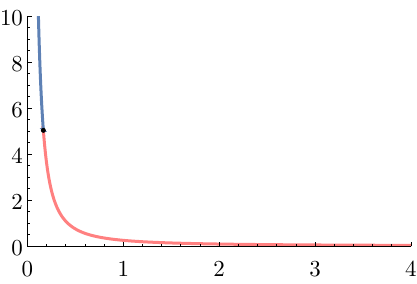}};
		\node[right=of img1, node distance=0cm, yshift=-1.4cm, xshift=-1.5cm] {$ \mathcal{R} $};
		\node[above=of img1, node distance=0cm, anchor=center,yshift=-1cm, xshift=-3cm] {$ F $};
		\draw[line width=1.5pt, color=pink] (-0.5,0) -- (1,0) node[right, color=black] {type III};
		\draw[line width=1.5pt, color=mathematica1] (-0.5,-0.5) -- (1,-0.5) node[right, color=black] {type II};
         \fill[] (1,-1) circle (1.2pt) node[right, color=black] {type I};
		\end{tikzpicture}
	\end{subfigure}
	\begin{subfigure}[t]{0.5\textwidth}
		\centering
		\begin{tikzpicture}
		\node (img1)  {\includegraphics[scale=1]{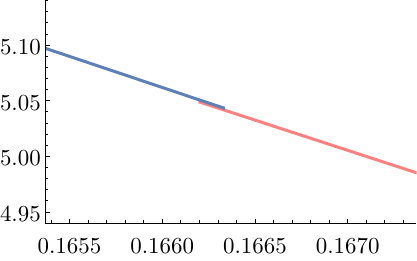}};
		\node[right=of img1, node distance=0cm, yshift=-1.2cm, xshift=-2cm] {$ \mathcal{R} $};
		\node[above=of img1, node distance=0cm, anchor=center, yshift=-0.8cm, xshift=-2.8cm] {$ F $};
		\end{tikzpicture}
	\end{subfigure}
	\caption{Renormalized free energy of type II and III solutions as a function of the dimensionless curvature $ \mathcal{R} $ below the Efimov bound $ b=0.47 $ in $ d=4 $. In the right plot, the region around the type I value of the left plot has been magnified. The overlap region between the type II and III branches in the right plot is a numerical artifact.}
	\label{plot:FR}
\end{figure}
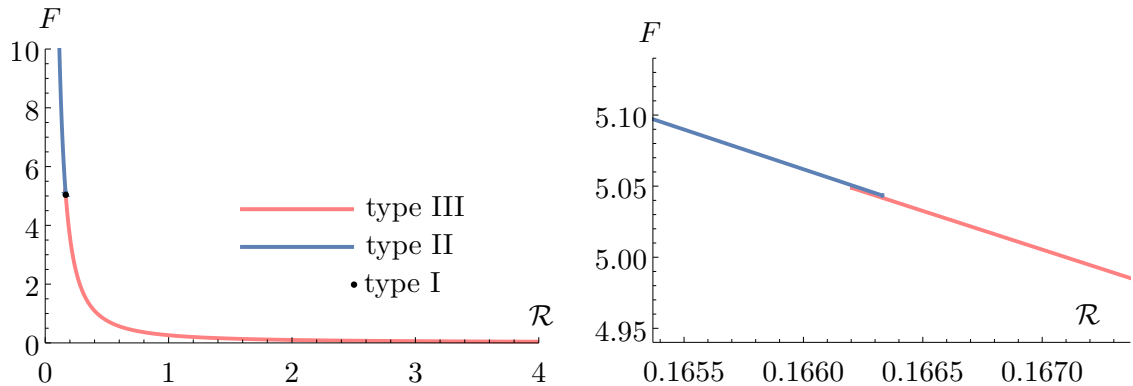

As a function of $\mathcal{R}$, the numerically evaluated free energy is plotted in Figure \ref{plot:FR}. It asymptotes to infinity along the type II branch as $\mathcal{R}\rightarrow 0$ and to zero along the type III branch as $\mathcal{R}\rightarrow \infty$. In the middle, there is a phase transition between the type II and III branches at the type I value $\mathcal{R} = \mathcal{R}_{\text{I}}\approx 0.166$ within numerical uncertainty. The plot in the region close to the transition is given on the right of Figure \ref{plot:FR}. There cannot be any swallow-tail, because there are no Efimov oscillations or peaks in $\mathcal{R}$ nor $F$ as functions of IR parameters. The overlap region on the right of Figure \ref{plot:FR} is, therefore, a numerical artifact due to the slightly different asymptotic values \eqref{eq:below_Efimov_asymptotic_F}. We emphasize that analytics require that the asymptotic values are the same.

The first derivative of the free energy must be continuous across the transition since $\mathcal{C}(\mathcal{R})$ is continuous to very high numerical accuracy \eqref{eq:below_Efimov_asymptotics_C}. We observe this in the right plot of Figure \ref{plot:FR} where the type II and III branches clearly have the same slope. This confirms that the transition is  second-order or higher. To determine  whether the transition is of third- or higher-order, we have to consider the ratio $m_{\mathcal{C}}^{-}\slash m_{\mathcal{R}}^{-}$ as indicated in \eqref{eq:second_order_transition_condition}: the second derivative of the free energy is continuous and the transition is of third-order if these ratios are equal for type II and III branches. Using the formula \eqref{eq:second_order_transition_condition} and our fits to the numerics, we obtain the estimate (with $\mathcal{N}=1$)
\begin{equation}
    F''_{\text{III}}(\mathcal{R}_{\text{I}})-F''_{\text{II}}(\mathcal{R}_{\text{I}}) \approx -7.1\,,
\end{equation}
which is clearly non-zero given our estimated errors. Therefore, the transition appears to be second-order and not higher-order.

We can analytically derive the maximum order of the transition in the current case. For $d = 4$ and $b = 0.47$, the exponent \eqref{eq:delta_exponent} is given by $\delta \approx 3.85$ so that the transition would be at most of order $1+\lceil \delta\rceil = 5$ if the second derivative of the free energy was not already discontinuous, as seen in equation \eqref{eq:Fprime_diff_expansion_2}.

\section{Discussion and outlook}

In this work, we have discussed  curvature-driven  phase transitions for field theories  on positively curved space-times, in terms of the bulk geometry of a gravity dual theory. We have studied this problem in a bottom-up holographic model with a scalar field subject to an exponential potential, chosen in such a way that the flat space-time QFT displays confinement behavior.

We have found that, when the boundary curvature is large enough, there is a phase transition between competing  geometries in the gravity dual. We have shown this to be the case for a large class of Einstein-scalar theories, for which the property of  confinement is related to the steepness of the potential at large field values, encoded in the parameter $b$ in equation \eqref{int2}.   For very steep potentials ($b>b_{\text{E}}$) the transition is first-order. It becomes at least second-order for $b_{\text{c}} < b < b_{\text{E}}$ and it disappears below the same critical value $b_{\text{c}}$ when one also loses flat-space confinement.

These results are qualitatively similar to  those obtained in the seemingly different setup of \cite{Marolf:2010tg,Blackman:2011in} in the case of the $S^2$ compactification.
In fact, we can interpret that setup in terms of the {\em higher}-dimensional picture discussed here, in the case $N=2$ in the notation of the present paper.  The phase transition found in \cite{Blackman:2011in} in the $S^2$ case  is a manifestation of the Hawking--Page type transition of  pure gravity solutions  featuring  warped products of Euclidean spheres the form \eqref{int4},  \cite{Aharony:2019vgs,Kiritsis:2020bds}. The low dimensionality ($d+2<9$) used in those examples places them in the Efimov regime, and the phase transition found for the $S^2\times S^d$ case is indeed first-order, in line with our general result in Section \ref{subsec:phasetransition}.
The other cases considered in \cite{Marolf:2010tg,Blackman:2011in}, in which the internal space is either  $S^1$ or $S^1\times S^1$ (with only one of the two circles shrinking to zero size) are special, as they corresponds to the case $N=1$ treated here. As discussed at the end of Section \ref{subsec:reduction}, this maps to the non-confining regime upon dimensional reduction on the internal $S^1$. This mismatch between the higher dimensional and lower-dimensional behavior requires further analysis.

The fact that that in a region of parameter space we find a second-order phase transition, leads to further questions. First,  the critical type I solution, contrary to the sub-critical ones, does not uplift to a regular geometry in $d+N+1$ dimensions, because in the uplift two spheres shrink at the same time. This seems to violate the general expectation that in such cases some vev turns  on and the singular solution is avoided (this is what indeed  happens when the transition is first-order, as the phase diagram jumps between two regular branches). In the case of a second-order transition, the singular solution is the only one available for ${\cal R} = {\cal R}_c$. It is possible that this is an artifact of these simplified models in which only gravity and a scalar field are present, and it would be interesting to have a top-down example of this phenomenon in string theory.\footnote{The fact that the transition is higher-order for values of $b$ for which the total dimension of the uplift is larger than ten, indicates that the corresponding scalar potentials cannot be obtained from critical string theory via dimensional reduction alone.}

Second, a second-order transition should be accompanied by some form of scale invariance at the critical point, but this is not manifest in the type I solution (either in the Einstein-scalar theory or in the uplift). Nevertheless, the type I solution does have special features which might be linked to criticality: in the Einstein-scalar theory, it is the point at which the mass spectrum of particle-like excitations of the boundary theory turns from discrete (in the type II solutions) to continuous (in the type III solutions), as it will be shown in forthcoming work \cite{spectra}; in the uplift, it is the solution in which the two spheres shrink to zero size at the same point. Whether this leads to an extra isometry corresponding to scale invariance is left as an open question.

In this work we have mostly focused on the leading term in the asymptotics of the potential at large $\varphi$,
\begin{equation}
V(\varphi) = V_{\infty}\, e^{2b\varphi} + \ldots\,.
\end{equation}
However, we have shown that  our results  are robust against possible subleading terms, as  these never modify the leading non-vanishing asymptotics  of the superpotentials $W,S,T$ for any correction to the potential of the form $e^{2\gamma\varphi}$ with $\gamma < b$.   Since this is true uniformly in the full range of $\gamma$, including when it is arbitrarily close to $b$, one can expect that our result  will still hold  in the more general case \eqref{genasympt}.

While on the gravity side there is a clear-cut description of the phase transition in terms of competing bulk geometries, one  important question, which deserves further analysis, is what exactly characterizes the phase transition on the field theory side. For example, does it share some similarities to the  deconfinement transition which takes place in flat space at finite temperature?  We have different definitions of confinement in flat space, but it is not straightforward to carry them over to positively curved space-time. For example, strictly speaking the Wilson loop test (which takes  the asymptotic limit of an infinite loop) on positive curvature Euclidean cannot be performed,  since on the sphere  there is a maximum distance between two probe quarks and one never reaches the asymptotic  regime where one should observe  the area law. All we know is that if the radius $\alpha$ is large enough there will be an approximate area law at intermediate distances $\Lambda^{-1} \ll L \ll \alpha$, but the Wilson loop does not give a clear-cut order parameter for confinement.

In Lorentzian signature i.e. in de Sitter space-time, on the other hand, the time dependence of the geometry makes it unclear what one should call an area law since geodesic observers grow further apart with time. Similarly, the existence of  a mass gap is not a distinctive signal of confinement, since  a positive curvature will also gap the theory. One could   use the free energy of the solution, which  according to the lore, for a large-$N$ gauge theory is supposed to be of $\mathcal{O}(1)$ in a confined phase and of $\mathcal{O}(N^2)$ in a deconfined phase.  However this is tricky: since the free energy is always defined up to an additive renormalization,  the difference between any two solutions in holography  will be, in the semiclassical limit, proportional to  $(M_{\text{p}}^3\ell^3) \sim N^2$ regardless of whether the solutions are confining or not. Indeed, we have seen  examples of this behavior in the phase transitions found  in  \cite{Ghosh:2017big}, where none of the competing phases is confining but the free energy difference away from the critical point was of order $N^2$.

One important feature which may be a good  distinguishing criterion for confinement/deconfinement is the spectrum: confining gauge theories have discrete spectra of gauge-invariant bound states (glueballs and mesons), while upon deconfinement the spectrum typically becomes continuous (mesons and glueballs ``melt''), or at least this is what happens in  the thermal case. There, one also finds  a discrete set of quasi-normal modes, corresponding to unstable quasi-particle excitations. All these are standard features found in various types of thermal phase transitions  in large-$N$ holographic theories, where  the particle spectrum is encoded  in the spectrum of normalizable  fluctuations around the dominant bulk solution, and can be obtained  by a relatively simple linear analysis.  For this reason, it would be very interesting to compute the generic features of the perturbations around type II and type III solutions discussed here,  and to work out the full spectrum of scalar and tensor perturbations in some specific examples. This would give us some physical handle on distinguishing the two phases from the point of view of the dual field theory. This will be addressed in the future \cite{spectra}.

More generally,  it is an interesting question whether one can find an order parameter in the field theory language which can clearly tell apart the two different types of geometries, type I-II and type III. One possibility is to turn to the topological charge sector of the gauge theory, which distinguishes the high and low temperature phases in the case of thermal phase transitions.  In finite-temperature Yang--Mills theory,  the topological susceptibility is exponentially small in $N$ above the critical temperature of the thermal deconfinement phase transition, while it is ${\cal O}(1)$  below \cite{Vicari:2008jw}. This is nicely reproduced in holographic theories of the kind we discuss here \cite{Gursoy:2008za,Gursoy:2009jd}: in the black hole geometries corresponding to the high temperature phase,  the susceptibility vanishes because  there are no non-trivial regular solutions for the bulk axion field dual to the $\tr{F\tilde{F}}$ operator. At low temperature instead, a non-trivial axion profile can be turned on, and this gives rise to a finite $\langle F\tilde F\rangle$. Given the similarities between the high-curvature solutions discussed here and the black-hole geometries relevant at finite temperature (in both cases the scalar field stops at a finite value), it is possible that a similar  behavior  of the bulk axion carries over to curvature-driven holographic phase transitions. We leave this possibility for future work.

The kind of physics we have studied here using holography has potentially important applications to the dynamics of confining gauge theories in a cosmological setting (e.g. during slow-roll inflation). The occurrence of a  curvature-induced deconfining transition in a strongly coupled sector may leave traces in the CMB and/or GW background. Moreover, understanding the details of these kinds of transitions is of   phenomenological importance if some of these details carry over to holographic theories close to  QCD.

On that note, it is important to stress that the class of theories we consider, however, fails to include precisely the phenomenological holographic models which are closest to real-world QCD:  Improved Holographic QCD model (IHQCD) for the glue sector \cite{Gursoy:2007cb,Gursoy:2007er} and V-QCD which includes flavor \cite{Jarvinen:2011qe}. The reason is that the scalar potential used in these models does not fit the simple exponential asymptotics \eqref{int2}, but rather behaves as
\be \label{int6}
V_{\text{IHQCD}} \sim \varphi^{1/2}\, e^{2b_{\text{c}} \varphi}\,, \qquad \varphi \to \infty\,.
\ee
These potentials produce linear glue-ball spectra $ m_n^{2}\sim n $ instead of the quadratic one $ m_n^{2}\sim n^2 $ coming from the potentials used in the current work. The crucial obstructions for  applying our analysis to  this case are 1) the power-law behavior which accompanies the exponential and 2) the fact that the exponent sits exactly at the critical point $b = b_{\text{c}}$ which separates confining from non-confining theories. Each of these two facts causes the IR expansion we use in this paper to break down: at the value  $b=b_{\text{c}}$ the type I and type II solutions become degenerate and the IR expansion is modified at leading and subleading order. At the same time, the power-law  in \eqref{int6} changes the way this expansion is organized, introducing power-like terms in addition to exponentials. Given the importance of this case, we are going to dedicate it to a separate work, where we introduce  tools  that are specific for these particular asymptotics \cite{JL}.

\section*{Acknowledgements}\label{ACKNOWL}
\addcontentsline{toc}{section}{Acknowledgements}

We thank C. Hoyos, E. Pr\'eau, J. L. Raymond, C. Rosen, J. Subils, for useful discussions. We also thank Melissa Faur, who contributed to  the initial stages of this work. J.~K. is supported by the Deutsche Forschungsgemeinschaft (DFG, German Research Foundation) through the German-Israeli Project Cooperation (DIP) grant ‘Holography and the Swampland’, as well as under Germany’s Excellence Strategy through the W\"{u}rzburg-Dresden Cluster of Excellence on Complexity and Topology in Quantum Matter - ct.qmat (EXC 2147, project-id 390858490). In addition, J.~K. thanks the Magnus Ehrnrooth foundation and the Osk. Huttunen foundation for support during earlier stages of this work.
This work is partially supported by the European MSCA grant HORIZON-MSCA-2022-PF-01-01 ``BlackHoleChaos" No.101105116 and by the H.F.R.I call ``Basic research Financing (Horizontal support of all Sciences)" under the National Recovery and Resilience Plan ``Greece 2.0" funded by the European Union -- NextGenerationEU (H.F.R.I. Project Number: 15384.),  by  ANR  grant ``XtremeHolo'' (ANR project n.284452) and by the In2p3 grant ``Extreme Dynamics" .

\newpage

\appendix
\renewcommand{\theequation}{\thesection.\arabic{equation}}
\addcontentsline{toc}{section}{Appendix\label{app}}
\section*{Appendix}

\section{Details of the dimensional reduction}\label{app:reductiondetails}

In this Appendix, we review the dimensional reduction of $d+N+1$-dimensional Einstein gravity on an $S^N$ to $d+1$-dimensional Einstein-scalar theory \cite{Gouteraux:2011ce,Gouteraux:2011qh}. To this end, we start with the $d+N+1$-dimensional Einstein action with a negative cosmological constant,
\begin{equation}
	I_{d+N+1} = M_{\text{p}}^{d+N-1}\int_{\widetilde{\mathcal{M}}} d^{d+N+1}\tilde{X} \sqrt{\tilde{g}}\,(\widetilde{R} -2\widetilde{\Lambda})\,,
 \label{eq:higher_einstein_action_app}
\end{equation}
in which:
\begin{equation}
    \widetilde{\Lambda} = -\frac{(d+N)(d+N-1)}{2\tilde{\ell}^2}\,.
\end{equation}
We take  the Euclidean  $ d+N+1 $-dimensional metric to be:
\begin{equation}
	ds^{2} = \tilde{g}_{ab}\,d\tilde{X}^{a}d\tilde{X}^{b} = d\tilde{u}^{2} + e^{2A_1(\tilde{u})}\,\zeta_{\mu\nu}\,dx^{\mu}dx^{\nu} + e^{2A_2(\tilde{u})}\,\bar{\zeta}_{mn}\,dy^{m}dy^{n}
	\label{highermetricapp}
\end{equation}
We consider
\begin{equation}
	\zeta_{\mu\nu}\,dx^{\mu}dx^{\nu} = \alpha^{2}ds^{2}_{S^{d}}, \quad \bar{\zeta}_{nm}\,dy^{n}dy^{m} = \bar{\alpha}^{2}ds^{2}_{S^{N}}\,,
\end{equation}
where the curvature radii $ \alpha $ and $ \bar{\alpha} $ are defined as
\begin{equation}
	\kappa = \frac{d-1}{\alpha^{2}}, \quad \bar{\kappa} = \frac{N-1}{\bar{\alpha}^{2}}.
\end{equation}
The components of the Ricci tensor  $\tilde{R}_{ab}$ of the metric \eqref{highermetricapp} are given by
\begin{align}
	\tilde{R}_{\tilde{u}\tilde{u}} &= -d\,A_1''-d\,(A_1')^{2} - N\,A_2''- N\,(A_2')^{2}\,,\nonumber\\
	\tilde{R}_{\mu\nu} &=\left[-A_1''-d\,(A_1')^{2}-N\,A_1'A_2'+\kappa e^{-2A_1}\right] e^{2A_1}\zeta_{\mu\nu}\,,\nonumber\\
	\tilde{R}_{mn} &=\left[-A_2''-N\,(A_2')^{2}-d\,A_1'A_2'+\bar{\kappa} e^{-2A_2}\right] e^{2A_2}\bar{\zeta}_{mn}\,,
	\label{eq:riccitensors}
\end{align}
where a prime denotes derivative with respect to $\tilde{u}$. The Ricci scalar  is
\begin{equation}
	\tilde{R} = -2d\,A_1''-d(d+1)\,(A_1')^{2}-2NA_2''-N(N+1)\,(A_2')^{2}-2dNA_1'A_2'+d\kappa e^{-2A_1}+N\bar{\kappa}e^{-2A_2}\,.
	\label{eq:ricciscalar}
\end{equation}
Einstein's equations are equivalent to $\tilde{R}_{ab} =\frac{2\widetilde{\Lambda}}{d+N-1}\,\tilde{g}_{ab}$ which using \eqref{eq:riccitensors} gives the three equations
\begin{align}
	A_1''+NA_1'A_2'+d\,(A_1')^{2}-\kappa\,e^{-2A_1}+\frac{2\widetilde{\Lambda}}{d+N-1} &= 0\,,\label{eq:appEinstein1}\\
	A_2''+dA_1'A_2'+N\,(A_2')^{2}-\bar{\kappa}\,e^{-2A_2}+\frac{2\widetilde{\Lambda}}{d+N-1} &= 0\,,\label{eq:appEinstein2}\\
	d\,A_1''+N\,A_2''+d\,(A_1')^{2}+N\,(A_2')^{2}+\frac{2\widetilde{\Lambda}}{d+N-1}&=0\,,\label{eq:appEinstein3}
\end{align}
where the first equation is the $(\mu,\nu)$ component, second the $(m,n)$ component and the last equation is the $(\tilde{u},\tilde{u})$ component. The third equation is not independent from the two others so that a priori there are a total of four integration constants corresponding to two second-order differential equations. However, by eliminating $dA_1'' + N A_2''$ in terms of $A_1'$ and $A_2'$ we obtain the first order equation
\begin{equation}
(dA_1'+NA_2')^{2}-d\,(A_1')^{2}-	N\,(A_2')^{2}-d\kappa\,e^{-2A_1}-N\bar{\kappa}\,e^{-2A_2}+2\widetilde{\Lambda}=0\,.
\end{equation}
Therefore, the actual number of integration constants of the system \eqref{eq:appEinstein1} -- \eqref{eq:appEinstein3} is three instead of four.

In what follows we  use an alternative form of the Einstein's equations \eqref{eq:appEinstein1} -- \eqref{eq:appEinstein3}. It can be obtained by starting with Einstein's equations in the form $ \tilde{R}_{ab} - \frac{1}{2}\tilde{R}\tilde{g}_{ab} + \widetilde{\Lambda} \tilde{g}_{ab} = 0 $ and substituting the Ricci tensor components \eqref{eq:riccitensors} and the Ricci scalar \eqref{eq:ricciscalar}. The $ (\tilde{u},\tilde{u}) $ component gives the equation
\begin{equation}
	d(d-1)\left(A_1'+\frac{N}{d-1}\,A_2'\right)^{2}-\frac{N(d+N-1)}{d-1}\,(A_2')^{2} + \Bigl(2\widetilde{\Lambda} - d\kappa e^{-2A_1}-N\bar{\kappa}e^{-2A_2}\Bigr) = 0\,,
	\label{einstein1}
\end{equation}
while the $ (\mu,\nu) $ components are equivalent to
\begin{align}
	&d(d-1)\left(A_1'+\frac{N}{d-1}\,A_2'\right)^{2}+2(d-1)\biggl(A_1''+\frac{N}{d-1}\,A_2''\biggr)+\frac{N(d+N-1)}{d-1}\,(A_2')^{2}\nonumber\\
	&-2NA_2'\left(A_1'+\frac{N}{d-1}\,A_2'\right)+ \Bigl(2\widetilde{\Lambda} - (d-2)\kappa\, e^{-2A_1}-N\bar{\kappa}\,e^{-2A_2}\Bigr) = 0\,,
	\label{einstein2}
\end{align}
and the $ (n,m) $ components give the same equation as the $ (\mu,\nu) $ components, but with the replacements $ A_1\leftrightarrow A_2 $, $ d \leftrightarrow N $ and $ \kappa \leftrightarrow \bar{\kappa} $.

We  now introduce the coordinate $ u = u(\tilde{u}) $, which will become the radial coordinate of the dimensionally reduced theory, and write \eqref{einstein1} as
\begin{equation}
	d(d-1)\biggl(\dot{A}_1+\frac{N}{d-1}\dot{A}_2\biggr)^{2}-\frac{N(d+N-1)}{d-1}\dot{A}_2^{2} + \biggl(\frac{d\tilde{u}}{du}\biggr)^{2}\Bigl(2\widetilde{\Lambda} - d\kappa\, e^{-2A_1}-N\bar{\kappa}\,e^{-2A_2}\Bigr) = 0\,,
	\label{einstein12}
\end{equation}
where the dot denotes derivative with respect to $ u $. If we define
\begin{gather}
	A(u) \equiv A_1(\tilde{u}) + \frac{N}{d-1}\,A_2(\tilde{u})\,, \quad \varphi(u) \equiv -\sqrt{\frac{2N(d+N-1)}{d-1}}\,A_2(\tilde{u})\,,\label{Avarphilower}\\
	V(\varphi) \equiv \left(\frac{d\tilde{u}}{du}\right)^{2}\left(2\widetilde{\Lambda}-N\bar{\kappa}\,e^{-2A_{2}(\tilde{u})}\right)\,, \quad \biggl(\frac{d\tilde{u}}{du}\biggr)^{2} \equiv e^{-\frac{2N}{d-1}A_{2}(\tilde{u})}\,,
	\label{lowerdimensionalfields}
\end{gather}
then the equation \eqref{einstein12} becomes
\begin{equation}
	d(d-1)\,\dot{A}^{2} - \frac{1}{2}\,\dot{\varphi}^{2} + V -d\kappa\, e^{-2A}= 0\,,
	\label{EOM2app}
\end{equation}
which we identify as the second equation of motion \eqref{EOM2} of the Einstein-scalar theory in $d+1$ dimensions. Substituting \eqref{Avarphilower} - \eqref{lowerdimensionalfields} to \eqref{einstein2} and using
\begin{equation}
	A_1''+\frac{N}{d-1}\,A_2'' = \biggl(\frac{du}{d\tilde{u}}\biggr)^{2}\ddot{A}+\frac{d^{2}u}{d\tilde{u}^{2}}\,\dot{A}\,,
\end{equation}
we obtain
\begin{align}
	d(d-1)\,\dot{A}^{2} +2 (d-1)\,\ddot{A}+\frac{1}{2}\,\dot{\varphi}^{2}+V-(d-2)\,\kappa\,e^{-2A}&\nonumber\\
	+2(d-1)\biggl(\frac{d\tilde{u}}{du}\biggr)^{2}\biggl[\frac{d^{2}u}{d\tilde{u}^{2}}-\frac{N}{d-1}\dot{A}_2\,\biggl(\frac{du}{d\tilde{u}}\biggr)^{2}\biggr]\,\dot{A} &= 0\,.
	\label{secondscalareq1}
\end{align}
From \eqref{lowerdimensionalfields} it follows that
\begin{equation}
	\frac{d^{2}u}{d\tilde{u}^{2}} = \frac{d}{d\tilde{u}}\Bigl(e^{\frac{N}{d-1}A_2(\tilde{u})}\Bigr) = \frac{N}{d-1}\dot{A}_2\,\biggl(\frac{du}{d\tilde{u}}\biggr)^{2}
\end{equation}
so that the second line in \eqref{secondscalareq1} vanishes leaving us with
\begin{equation}
	d(d-1)\,\dot{A}^{2} +2 (d-1)\,\ddot{A}+\frac{1}{2}\,\dot{\varphi}^{2}+V-(d-2)\,\kappa\,e^{-2A} = 0\,.
	\label{secondscalareq2}
\end{equation}
Subtracting \eqref{EOM2app} from this equation gives
\begin{equation}
	2 (d-1)\,\ddot{A}+\dot{\varphi}^{2}+2\kappa\,e^{-2A} = 0\,,
	\label{EOM1app}
\end{equation}
which we identify as the first equation of motion \eqref{EOM1} of the Einstein-scalar theory. Hence, we have shown that  Einstein's equations of the $ d+N+1 $-dimensional theory are equivalent to the equations of motion of the $ d+1 $-dimensional Einstein-scalar theory. The form of the potential $V = V(\varphi)$ is obtained from \eqref{Avarphilower} and \eqref{lowerdimensionalfields} to be
\begin{equation}
	V(\varphi) = 2\widetilde{\Lambda}\, e^{\sqrt{\frac{2N}{(d-1)(d+N-1)}}\,\varphi}-N\bar{\kappa}\,e^{\sqrt{\frac{2(d+N-1)}{N(d-1)}}\,\varphi}\,,
\end{equation}
which can be written as
\begin{equation}
	V(\varphi) = V_{\infty}\,e^{2b\varphi} + V_{\infty,1}\,e^{2\gamma_{\text{c}}\varphi}\,,
 \label{eq:potential_reduction_app}
\end{equation}
with the coefficients \eqref{eq:potential_coefficients} and exponents \eqref{eq:reduction_exponents}.

TFrom \eqref{Avarphilower} we also obtain that the functions \eqref{WST} of the first order formulation are given by
\begin{equation}
	S(\varphi) = -\sqrt{\frac{2N(d+N-1)}{d-1}}\,\dot{A}_2\,, \quad W(\varphi) = -2(d-1)\bigg(\dot{A}_1+\frac{N}{d-1}\dot{A}_2\biggr)\,,
	\label{SWA1A2app}
\end{equation}
where a dot denotes a derivative with respect to $ u $.

\section{Dimensional reductions of higher-dimensional solutions}\label{app:solution_reduction_details}

\subsection{Classification of solutions}\label{app:highersolutions}

The bulk metric is of the form
\begin{equation}
	ds^{2} = d\tilde{u}^{2} + e^{2A_1(\tilde{u})}\,\alpha^{2}ds^{2}_{S^{d}} + e^{2A_2(\tilde{u})}\,\bar{\alpha}^{2}ds^{2}_{S^{N}}.
	\label{highermetricapp2}
\end{equation}
We define
\begin{equation}
		\Omega(\tilde{u}) = \alpha e^{A_1(\tilde{u})}, \quad \bar{\Omega}(\tilde{u}) = \bar{\alpha} e^{A_2(\tilde{u})}.
\end{equation}
Let $ \tilde{u}_0 $ be finite and consider three types geometries that we call type I, II and III respectively:
\begin{equation}
	\begin{dcases}
		\Omega(\tilde{u}_0) = 0\,, \quad \bar{\Omega}(\tilde{u}_0)>0\,, \quad &\text{type III}\\
		\Omega(\tilde{u}_0) >0\,, \quad \bar{\Omega}(\tilde{u}_0) = 0\,, \quad &\text{type II}\\
		\Omega(\tilde{u}_0) = 0\,, \quad \bar{\Omega}(\tilde{u}_0) = 0\,, \quad &\text{type I}
	\end{dcases}\,.
\end{equation}
In type III geometries, the space-time sphere $ S^{d} $ shrinks to zero size at $ \tilde{u} = \tilde{u}_0 $, while for type II geometries, it is the internal sphere $ S^{N} $ that shrinks. In type I geometries, both spheres shrink at the same time.  The geometries, hence, exist in the range $ \tilde{u} \in (-\infty,\tilde{u}_0) $ and regularity restricts the behaviour of the metric when $ \tilde{u}\rightarrow \tilde{u}_0^{-} $ (approaching from below). Assuming $ \Omega(\tilde{u}) $ are smooth at $ \tilde{u}_0 $, we can Taylor expand:

\begin{equation}	
	\begin{dcases}
			\Omega(\tilde{u}) &= \alpha_{\scriptscriptstyle\text{IR}} + \omega_{1}\,(\tilde{u}_0-\tilde{u})^{p_{1}} + \omega_{2}\,(\tilde{u}_0-\tilde{u})^{p_{2}} +  \ldots\\
		\bar{\Omega}(\tilde{u}) &= \bar{\alpha}_{\scriptscriptstyle\text{IR}}+\bar{\omega}_1\,(\tilde{u}_0-\tilde{u})^{\bar{p}_{1}} + \bar{\omega}_2\,(\tilde{u}_0-\tilde{u})^{\bar{p}_2} + \ldots
	\end{dcases}, \quad \tilde{u}\rightarrow \tilde{u}_0^{-},
\end{equation}
where $ p_2 > p_1>0 $.

\paragraph{The type I geometry.} For type I geometries $ \alpha_{\IR} = \bar{\alpha}_{\IR} = 0 $ and the metric near $ \tilde{u} = \tilde{u}_0 $ is given by
\begin{equation}
	ds^{2} = d\tilde{u}^{2} + \omega_1^{2}\,(\tilde{u}-\tilde{u}_0)^{2p_1}\,ds^{2}_{S^{d}} + \bar{\omega}_1^{2}\,(\tilde{u}-\tilde{u}_0)^{2\bar{p_1}}\,ds^{2}_{S^{N}} +\ldots\,.
\end{equation}
This metric is singular even for the choice $ p_1 = \bar{p}_1 = \omega_1 = \bar{\omega}_1=1 $  (it is not a patch of $ \mathbb{R}^{d+N+1} $).  The ``regular'' type I geometry can be obtained as a limit of a regular type III geometry by taking $ \bar{\alpha}_{\scriptscriptstyle\text{IR}} \rightarrow 0 $ which we call a type $ \text{I}_{\text{a}} $ geometry. Similarly, taking $ \alpha_{\scriptscriptstyle\text{IR}}\rightarrow 0 $ limit of a type II geometry gives a type $ \text{I}_{\text{b}} $ geometry. These are characterized by the expansions
\begin{equation}
	\text{type }\text{I}_\text{a,b}:\quad\begin{dcases}
		A_1(\tilde{u}) =  p_1\log{\biggl(\frac{\tilde{u}_0-\tilde{u}}{\alpha}\biggr)}+A_{10} +A_{11}\,(\tilde{u}_0-\tilde{u})^{p_2-p_1} + \ldots\\
		A_2(\tilde{u}) =  \bar{p}_1\log{\biggl(\frac{\tilde{u}_0-\tilde{u}}{\bar{\alpha}}\biggr)}+A_{20} + A_{21}\,(\tilde{u}_0-\tilde{u})^{\bar{p}_2-\bar{p}_1} + \ldots
	\end{dcases}\,, \quad
	\tilde{u}\rightarrow \tilde{u}_0^{-}\,,
 \label{eq:typeC_IR_expansion_app}
\end{equation}
where $A_{10} = \log{(\omega_1 \alpha^{p_1-1})}$, $A_{11} = \frac{\omega_2}{\omega_1}$, $A_{20} = \log{(\bar{\omega}_1 \bar{\alpha}^{\bar{p}_1-1})}$ and $A_{21} = \frac{\bar{\omega}_2}{\bar{\omega}_1}$.

The Einstein's equations \eqref{eq:appEinstein1} - \eqref{eq:appEinstein3} admit a type I solution which can be found exactly for all values of the holographic radial coordinate. It is given by (this solution is also given in \cite{Aharony:2019vgs})
\begin{align}
	A_1(\tilde{u}) &= \log{\biggl[\frac{\tilde{\ell}}{\alpha}\sinh{\biggl(\frac{\tilde{u}_0-\tilde{u}}{\tilde{\ell}}\biggr)}\biggr]}+\frac{1}{2}\log{\biggl(\frac{d-1}{d+N-1}\biggr)}\,,\nonumber\\
	A_2(\tilde{u}) &= \log{\biggl[\frac{\tilde{\ell}}{\bar{\alpha}}\sinh{\biggl(\frac{\tilde{u}_0-\tilde{u}}{\tilde{\ell}}\biggr)}\biggr]}+\frac{1}{2}\log{\biggl(\frac{N-1}{d+N-1}\biggr)}\,.\label{eq:typeChigherdimensional}
\end{align}
This has an IR expansion of the form \eqref{eq:typeC_IR_expansion_app} with exponents $p_1 = \bar{p}_1 = 1$, $p_2 = \bar{p}_2 = 3$ and coefficients
\begin{equation}
    A_{10} = \frac{1}{2}\log{\biggl(\frac{d-1}{d+N-1}\biggr)}\,,\quad A_{20} = \frac{1}{2}\log{\biggl(\frac{N-1}{d+N-1}\biggr)}\,,\quad A_{11} = A_{21} = \frac{1}{6\tilde{\ell}^2}\,.
    \label{eq:typeC_IR_coefficients}
\end{equation}
The type I solution does not contain any free parameters and is, hence, unique. Therefore, the two limits considered above (type $I_a$ from the type III and  type $I_b$ from type II solutions) must be the same. In the flat space limit $\tilde{\ell}\rightarrow \infty$, it reduces to the solution given in \cite{Kol:2002xz}
\begin{align}
	A_1(\tilde{u}) &= \log{\biggl(\frac{\tilde{u}_0-\tilde{u}}{\alpha}\biggr)}+\frac{1}{2}\log{\biggl(\frac{d-1}{d+N-1}\biggr)}\,,\nonumber\\
	A_2(\tilde{u}) &= \log{\biggl(\frac{\tilde{u}_0-\tilde{u}}{\bar{\alpha}}\biggr)}+\frac{1}{2}\log{\biggl(\frac{N-1}{d+N-1}\biggr)}\,.\label{eq:typeChigherdimensionalflat}
\end{align}

\paragraph{Type II geometries.} Consider then a type II geometry with $ \bar{\alpha}_{\scriptscriptstyle\text{IR}} = 0 $. Absence of a conical singularity imposes that
\begin{equation}
	\bar{p}_{1} = 1\,, \quad \bar{\omega}_{1} = 1\,,
	\label{regularity}
\end{equation}
so that we obtain
\begin{equation}
	\text{type II:}\quad\begin{dcases}
		A_1(\tilde{u}) =  A_{10} +A_{11}\,(\tilde{u}_0-\tilde{u})^{p_1} + \ldots\\
		A_2(\tilde{u}) =  \log{\biggl(\frac{\tilde{u}_0-\tilde{u}}{\bar{\alpha}}\biggr)}+A_{20} + A_{21}\,(\tilde{u}_0-\tilde{u})^{\bar{p}_2-1} + \ldots
	\end{dcases}, \quad
	\tilde{u}\rightarrow \tilde{u}_0^{-}\,,
\end{equation}
where $ A_{10} = \log{\frac{\alpha_{\scriptscriptstyle\text{IR}}}{\alpha}} $, $ A_{11} = \frac{\omega_1}{\alpha_{\scriptscriptstyle\text{IR}}} $, $ A_{20} = \log{\bar{\omega}_1} $ and $ A_{21} = \frac{\bar{\omega}_2}{\bar{\omega}_1} $. The only free parameter of the type II geometry is $\alpha_{\scriptscriptstyle\text{IR}}$ which is the radius of $ S^{d} $ at the IR endpoint. Therefore, for type III solutions, we have near $ \tilde{u}_0 $ the expansions
\begin{align}
	A_1(\tilde{u}) &= A_{10} + A_{11}\,(\tilde{u}_0-\tilde{u})^{2} + \mathcal{O}(\tilde{u}_0-\tilde{u})^{4}\,,\label{typeBonshellexp1}\\
	A_2(\tilde{u}) &=\log{\biggl(\frac{\tilde{u}_0-\tilde{u}}{\bar{\alpha}}\biggr)}+ A_{20} + A_{21}\,(\tilde{u}_0-\tilde{u})^{2} + \mathcal{O}(\tilde{u}_0-\tilde{u})^{4}\,.
	\label{typeBonshellexp}
\end{align}
 Substituting into the three Einstein's equations \eqref{einstein1}, \eqref{einstein2} and the $ (n,m) $ component, we obtain
\begin{align}
	A_{20} &= 0,\\
	A_{11} &= -\frac{1}{2\,(N+1)}\biggl(\frac{2\widetilde{\Lambda}}{d+N-1}-\kappa_{\scriptscriptstyle\text{IR}}\biggr)\,,\label{onshellcoefficientstypeB0}\\
	A_{21} &= \frac{1}{6N(N+1)}\biggl(\frac{d-N-1}{d+N-1}\,2\widetilde{\Lambda}-d\kappa_{\scriptscriptstyle\text{IR}}\biggr)\,,
	\label{onshellcoefficientstypeB}
\end{align}
where we have defined $\kappa_{\IR} \equiv \frac{d-1}{\alpha_{\IR}^2} $ and used $ A_{10} = \frac{1}{2}\log{\frac{\kappa}{\kappa_{\scriptscriptstyle\text{IR}}}} $. Explicitly, the on-shell expansion becomes
\begin{align}
	A_1(\tilde{u}) &= \log{\frac{\alpha_{\scriptscriptstyle\text{IR}}}{\alpha}} +\frac{d-1}{2\,(N+1)}\biggl(\frac{\tilde{u}_0-\tilde{u}}{\alpha_{\scriptscriptstyle\text{IR}}}\biggr)^{2}+\frac{D}{2\,(N+1)}\biggl(\frac{\tilde{u}_0-\tilde{u}}{\tilde{\ell}}\biggr)^{2} + \ldots\,,\nonumber\\
	A_2(\tilde{u}) &=\log{\biggl(\frac{\tilde{u}_0-\tilde{u}}{\bar{\alpha}}\biggr)}-\frac{d(d-1)}{6N\,(N+1)}\biggl(\frac{\tilde{u}_0-\tilde{u}}{\alpha_{\scriptscriptstyle\text{IR}}}\biggr)^{2}-\frac{D\,(d-N-1)}{6N\,(N+1)}\biggl(\frac{\tilde{u}_0-\tilde{u}}{\tilde{\ell}}\biggr)^{2} + \ldots\,.
	\label{eq:typeBexplicitexp}
\end{align}

\paragraph{Type III geometries.} For a type III geometry $ \alpha_{\scriptscriptstyle\text{IR}} = 0 $. In this case, the absence of a conical singularity requires
\begin{equation}
	p_1 = 1\,, \quad \bar{\omega}_1 = 1
\end{equation}
so that near $ \tilde{u} = \tilde{u}_0 $ the metric \eqref{highermetricapp2} behaves as
\begin{equation}
	ds^{2} = d\tilde{u}^{2}  + (\tilde{u}-\tilde{u}_0)^{2}\,ds^{2}_{S^{d}}+ \bar{\alpha}_{\scriptscriptstyle\text{IR}}^{2}\,ds^{2}_{S^{N}} + \ldots\,, \quad \tilde{u}\rightarrow \tilde{u}_0^{-}\,.
\end{equation}
Hence, near $ \tilde{u} = \tilde{u}_0 $ the metric is locally $ \mathbb{R}^{d+1}\times S^{N} $ without a conical singularity at the origin of $\mathbb{R}^{d+1} $. Hence, regular type III geometries behave as
\begin{equation}
	\text{type III:}\quad\begin{dcases}
		A_1(\tilde{u}) = \log{\biggl(\frac{\tilde{u}_0-\tilde{u}}{\alpha}\biggr)} + A_{10} +A_{11}\,(\tilde{u}_0-\tilde{u})^{p_2 - 1} + \ldots\\
		A_2(\tilde{u}) =  A_{20} + A_{21}\,(\tilde{u}_0-\tilde{u})^{\bar{p}_1} + \ldots
	\end{dcases}, \quad
	\tilde{u}\rightarrow \tilde{u}_0^{-}\,,
\label{eq:general_typeA_expansion_app}
\end{equation}
where we have defined $ A_{10} = \log{\omega_1} $, $ A_{11} = \frac{\omega_2}{\omega_1} $, $ A_{20} = \log{\frac{\bar{\alpha}_{\scriptscriptstyle\text{IR}}}{\bar{\alpha}}} $ and $ A_{21} = \frac{\bar{\omega}_1}{\bar{\alpha}_{\scriptscriptstyle\text{IR}}} $. The only free parameter of the type III geometry is $\bar{\alpha}_{\scriptscriptstyle\text{IR}}$ which is the radius of $ S^{N} $ at the IR endpoint.

The Einstein's equations Einstein's equations \eqref{eq:appEinstein1} - \eqref{eq:appEinstein3} remain the same under the transformation $ A_1\leftrightarrow A_2 $ if one also transforms $ d\leftrightarrow N $, $\alpha\leftrightarrow \bar{\alpha}$ and $\alpha_{\IR}\leftrightarrow \bar{\alpha}_{\IR}$. Type II and III solutions are related by this transformation
\begin{equation}
	A_1^{\text{III}} = A_{2}^{\text{II}}\big\vert_{E}\,,\quad A_2^{\text{III}} = A_{1}^{\text{II}}\big\vert_{E}\,,
\label{eq:typeAB_exchange_relation}
\end{equation}
where $E = \{\alpha\leftrightarrow \bar{\alpha}\,, \alpha_{\IR}\leftrightarrow \bar{\alpha}_{\IR}\,,d\leftrightarrow N\}$ and the notation $ \big\vert_{E}$ indicates that we perform the transformation. From \eqref{typeBonshellexp1} - \eqref{onshellcoefficientstypeB} it follows that the type III solution is explicitly
\begin{align}
	A_1(\tilde{u}) &=\log{\biggl(\frac{\tilde{u}_0-\tilde{u}}{\alpha}\biggr)}+ A_{10}+ A_{11}\,(\tilde{u}_0-\tilde{u})^{2} + \mathcal{O}(\tilde{u}_0-\tilde{u})^{3}\,,\label{typeAsolutionA1}\\
	A_2(\tilde{u}) &= A_{20} + A_{21}\,(\tilde{u}_0-\tilde{u})^{2} + \mathcal{O}(\tilde{u}_0-\tilde{u})^{3}\,.
	\label{typeAsolutionA2}
\end{align}
with coefficients
\begin{align}
	A_{10} &= 0\,,\\
	A_{11} &= -\frac{1}{6d\,(d+1)}\biggl(\frac{d-N+1}{d+N-1}\,2\widetilde{\Lambda}+N\bar{\kappa}_{\scriptscriptstyle\text{IR}}\biggr)\,,\\
	A_{21} &= -\frac{1}{2\,(d+1)}\biggl(\frac{2\widetilde{\Lambda}}{d+N-1}-\bar{\kappa}_{\scriptscriptstyle\text{IR}}\biggr)\,,
	\label{onshellcoefficientstypeA}
\end{align}
where we have defined $\bar{\kappa}_{\IR} \equiv \frac{N-1}{\bar{\alpha}_{\IR}^2} $ and used $ A_{20} = \frac{1}{2}\log{\frac{\bar{\kappa}}{\bar{\kappa}_{\scriptscriptstyle\text{IR}}}} $. Explicitly, the on-shell expansion becomes ($D = d+N$)
\begin{align}
	A_1(\tilde{u}) &=\log{\biggl(\frac{\tilde{u}_0-\tilde{u}}{\alpha}\biggr)}-\frac{N(N-1)}{6d\,(d+1)}\biggl(\frac{\tilde{u}_0-\tilde{u}}{\bar{\alpha}_{\scriptscriptstyle\text{IR}}}\biggr)^{2}+\frac{D\,(d-N+1)}{6d\,(d+1)}\biggl(\frac{\tilde{u}_0-\tilde{u}}{\tilde{\ell}}\biggr)^{2} + \ldots\,,\nonumber\\
	A_2(\tilde{u}) &= \log{\frac{\bar{\alpha}_{\scriptscriptstyle\text{IR}}}{\bar{\alpha}}} +\frac{N-1}{2\,(d+1)}\biggl(\frac{\tilde{u}_0-\tilde{u}}{\bar{\alpha}_{\scriptscriptstyle\text{IR}}}\biggr)^{2}+\frac{D}{2\,(d+1)}\biggl(\frac{\tilde{u}_0-\tilde{u}}{\tilde{\ell}}\biggr)^{2} + \ldots\,.
	\label{eq:typeAexplicitexp}
\end{align}
In the flat space limit $\tilde{\ell}\rightarrow \infty$, it reduces to the solution given in \cite{Kol:2002xz}.

\subsection{Dimensional reduction}\label{app:highersolutionsandreduction}

We now dimensionally reduce the IR expansions of the type I, II and III geometries over the internal $S^{N}$ to $d+1$ dimensions. This will produce the IR expansions of type I, II and III geometries in the lower-dimensional theory.

\paragraph{Type III geometries.} Substituting the expansions \eqref{typeAsolutionA1} and \eqref{typeAsolutionA2} to \eqref{Avarphilower}, we obtain
\begin{equation}
A(u) = \log{\biggl(\frac{\tilde{u}_0-\tilde{u}}{\alpha}\biggr)}+\frac{N}{d-1}\log{\frac{\bar{\alpha}_{\IR}}{\bar{\alpha}}} + \mathcal{O}(\tilde{u}_0-\tilde{u})^{2}
\label{eq:typeAAutilde}
\end{equation}
and similarly
\begin{equation}
\varphi(u) = \varphi_{0} + \tilde{\varphi}_{1}\,(\tilde{u}_0-\tilde{u})^{2} + \mathcal{O}(\tilde{u}_0-\tilde{u})^{3}, \quad \tilde{u}\rightarrow \tilde{u}_0^{-}\,,
\label{typeAvarphi}
\end{equation}
where we have defined
\begin{equation}
	\varphi_{0} \equiv -\sqrt{\frac{2N(d+N-1)}{d-1}}\,\log{\frac{\bar{\alpha}_{\IR}}{\bar{\alpha}}}, \quad \tilde{\varphi}_1 = -\sqrt{\frac{2N(d+N-1)}{d-1}}\,A_{21}\,.
	\label{phi0upstairs}
\end{equation}
Here $ \tilde{\varphi}_1 < 0 $ so that $ \varphi(u) $ approaches $ \varphi_0 $ from below. The constant $ \varphi_{0} $ is the free parameter of the type III geometries in the Einstein-scalar theory. Using \eqref{typeAsolutionA2} the relation \eqref{lowerdimensionalfields} between $\tilde{u}$ and $u$ can be expanded as
\begin{equation}
\frac{du}{d\tilde{u}} =e^{\frac{N}{d-1}\,A_2(\tilde{u})} = e^{\frac{N}{d-1}\,A_{20}}\,\biggl(1 +\frac{N}{d-1}\, A_{21}\,(\tilde{u}_0-\tilde{u})^{2} + \mathcal{O}(\tilde{u}_0-\tilde{u})^{3}\biggr)\,,
\end{equation}
where $A_{20} = \log{\frac{\bar{\alpha}_{\IR}}{\bar{\alpha}}}$. Integrating both sides with the boundary condition $u(\tilde{u}_0) = u_0$ gives
\begin{equation}
u_0-u = e^{\frac{N}{d-1}\,A_{20}}\,(\tilde{u}_0-\tilde{u}) + \mathcal{O}(\tilde{u}_0-\tilde{u})^{3}\,.
\end{equation}
Inverting and substituting to \eqref{eq:typeAAutilde} and \eqref{typeAvarphi} gives
\begin{align}
A(u) &= \log{\biggl(\frac{u_0-u}{\alpha}\biggr)} + \mathcal{O}(u_0-u)^{2}\,,\nonumber\\
\varphi(u) &= \varphi_0 + \varphi_1\,(u_0-u)^{2} + \mathcal{O}(u_0-u)^{3}\,,
\label{eq:Aphi_typeA_app}
\end{align}
where we have defined
\begin{equation}
\varphi_1 \equiv \tilde{\varphi}_1\,e^{-\frac{2N}{d-1}A_{20}} = -\sqrt{\frac{2N(d+N-1)}{d-1}}\,A_{21}\,e^{-\frac{2N}{d-1}A_{20}}\,.
\end{equation}
By inverting \eqref{eq:Aphi_typeA_app} for $\varphi(u)$, we obtain
\begin{equation}
u_0-u = \frac{1}{\sqrt{-\varphi_1}}\sqrt{\varphi_0-\varphi} + \mathcal{O}(\varphi_0-\varphi)\,,
\label{eq:u_varphi_type_A}
\end{equation}
where the sign of the square root on the right-hand side is fixed by the requiring that $ u_0-u>0 $. The expansions of the functions \eqref{W} and \eqref{S} in $\varphi_0-\varphi$ are obtained by first taking derivatives of \eqref{eq:Aphi_typeA_app} with respect to $u$ and then substituting \eqref{eq:u_varphi_type_A}. The result is
\begin{equation}
S(\varphi)= \sqrt{\varphi_{0} - \varphi}\,[S_0 + \mathcal{O}(\sqrt{\varphi_{0} - \varphi})]\,,\quad W(\varphi) = \frac{1}{\sqrt{\varphi_{0} - \varphi}}\,[W_0 + \mathcal{O}(\varphi_{0} - \varphi)]\,,
\label{eq:typeA_dimensional_reduction}
\end{equation}
where the constants are given by
\begin{equation}
S_0 = 2\sqrt{-\varphi_1}\,, \quad W_0 = (d-1)\,S_0\,.
\label{eq:S0_W0_reductions}
\end{equation}
When the higher-dimensional equations of motion are satisfied $A_{21}$ is given by \eqref{onshellcoefficientstypeA}. Therefore on-shell
\begin{equation}
	\varphi_1 = \frac{Nb}{d+1}\,e^{2\gamma_{\text{c}}\varphi_0}\biggl(\frac{2\widetilde{\Lambda}}{d+N-1}-\bar{\kappa}_{\scriptscriptstyle\text{IR}}\biggr)\,,
	\label{eq:varphi1_midstep}
\end{equation}
where we have used that $-\frac{N}{d-1}A_{20} = \gamma_{\text{c}}\varphi_0 $ as follows from \eqref{phi0upstairs}. Here we can write
\begin{equation}
	\bar{\kappa}_{\IR} = \frac{\bar{\kappa}_{\IR}}{\bar{\kappa}}\,\bar{\kappa} = \frac{-V_{\infty}}{N}\,e^{2(b-\gamma_{\text{c}})\varphi_0}
\end{equation}
where have used $\bar{\kappa} = -\frac{1}{N}\,V_{\infty}$, $\frac{\bar{\kappa}_{\IR}}{\bar{\kappa}} = e^{-2A_{20}} = e^{2(b-\gamma_{\text{c}})\varphi_0} $ and
\begin{equation}
    b-\gamma_{\text{c}} = \sqrt{\frac{d-1}{2N(d+N-1)}}\,.
\label{bminusgammac}
\end{equation}
Substituting to \eqref{eq:varphi1_midstep} and by using $ 2\widetilde{\Lambda} = V_{\infty,1} $, we obtain
\begin{equation}
	\varphi_1 = \frac{1}{d+1}\,\biggl(\frac{Nb}{d+N-1}\,V_{\infty,1}\,e^{2\gamma_{\text{c}}\varphi_0}+b\,V_{\infty}\,e^{2b\varphi_0}\biggr) = \frac{1}{2}\frac{V'(\varphi_0)}{d+1}\,,
\end{equation}
where the potential $V(\varphi)$ is given by \eqref{eq:potential_reduction_app} and the last equality follows by $\frac{Nb}{d+N-1} = \gamma_{\text{c}}$. Hence, the coefficient \eqref{eq:S0_W0_reductions} can be written on-shell as
\begin{equation}
	S_0 = \sqrt{-\frac{2V'(\varphi_0)}{d+1}}\,.
 \label{eq:S0_reduction}
\end{equation}

\paragraph{Type II geometries.} Substituting the expansions \eqref{typeBonshellexp1} and \eqref{typeBonshellexp} to \eqref{Avarphilower}, we obtain
\begin{equation}
A(u) = \frac{N}{d-1}\,\log{\biggl(\frac{\tilde{u}_0-\tilde{u}}{\bar{\alpha}}\biggr)} + \mathcal{O}(1)\,, \quad \tilde{u}\rightarrow \tilde{u}_0^{-}\,,
\label{typeBAu}
\end{equation}
and similarly
\begin{equation}
(b-\gamma_{\text{c}})\,\varphi(u) =  -\log{\biggl(\frac{\tilde{u}_0-\tilde{u}}{\bar{\alpha}}\biggr)} -A_{21}\,(\tilde{u}_0-\tilde{u})^{2}+ \mathcal{O}(\tilde{u}_0-\tilde{u})^{4}\,,
\label{typeBphiasymptotics}
\end{equation}
where we have used \eqref{bminusgammac} and we observe that $ \varphi\rightarrow \infty $ as $ \tilde{u}\rightarrow \tilde{u}_0^{-} $. Using \eqref{typeBonshellexp} the relation \eqref{lowerdimensionalfields} between $\tilde{u}$ and $u$ can be expanded as
\begin{equation}
\frac{d\tilde{u}}{du} = e^{-\frac{N}{d-1}A_2(\tilde{u})} = \biggl(\frac{\tilde{u}_0-\tilde{u}}{\bar{\alpha}}\biggr)^{-\frac{N}{d-1}} + \ldots\,,
\end{equation}
which can be integrated to yield
\begin{equation}
\frac{\tilde{u}_0-\tilde{u}}{\bar{\alpha}} = \biggl(\frac{d+N-1}{d-1}\frac{u_0-u}{\bar{\alpha}}\biggr)^{\frac{d-1}{d+N-1}} + \ldots\,.
\end{equation}
Substituting to \eqref{typeBAu} and \eqref{typeBphiasymptotics} gives
\begin{align}
A(u) &= \frac{N}{d+N-1}\log{\biggl(\frac{u_0-u}{\bar{\alpha}}\biggr)} + \mathcal{O}(1)\,,\\
\varphi(u) &= -\sqrt{\frac{2N(d-1)}{d+N-1}}\,\log{\biggl(\frac{u_0-u}{\bar{\alpha}}\biggr)} + \ldots\,.
\end{align}
Written in terms of $ b $ and $ b_{\text{c}} $, this matches with equation \eqref{typeBreduced}.

We now compute the behavior of the function $ S(\varphi) $ using \eqref{SWA1A2app}. Taking $ u $-derivatives of \eqref{typeBonshellexp1} and \eqref{typeBonshellexp} gives
\begin{align}
\dot{A}_1 &= -2A_{11}\,(\tilde{u}_0-\tilde{u})\,\frac{d\tilde{u}}{du} + \mathcal{O}(\tilde{u}_0-\tilde{u})^{3}\,,\nonumber \\
\dot{A}_2 &= -\frac{1}{\tilde{u}_0-\tilde{u}}\frac{d\tilde{u}}{du} - 2A_{21}\,(\tilde{u}_0-\tilde{u})\,\frac{d\tilde{u}}{du} + \mathcal{O}(\tilde{u}_0-\tilde{u})^{3}\,.
\label{typeBdotAs}
\end{align}
By combining \eqref{Avarphilower} and \eqref{lowerdimensionalfields}, we can write
\begin{equation}
\frac{d\tilde{u}}{du} = e^{\gamma_{\text{c}}\varphi}\,,
\label{dutugammac}
\end{equation}
where we have again fixed the sign on the right-hand side to be positive as before. By inverting \eqref{typeBphiasymptotics}, we obtain
\begin{equation}
\frac{\tilde{u}_0-\tilde{u}}{\bar{\alpha}} = e^{-( b-\gamma_{\text{c}}) \,\varphi} -\bar{\alpha}^{2}A_{21}\,e^{-3(b-\gamma_{\text{c}}) \,\varphi}   + \mathcal{O}(e^{-4(b-\gamma_{\text{c}}) \,\varphi})\,, \quad \varphi \rightarrow \infty\,.
\label{utildevarphiB}
\end{equation}
Hence, we can expand
\begin{align}
\frac{1}{\tilde{u}_0-\tilde{u}}\frac{d\tilde{u}}{du} &=\frac{1}{\bar{\alpha}}\,e^{b\varphi} + \bar{\alpha}A_{21}\,e^{(2\gamma_{\text{c}}-b) \,\varphi} + \mathcal{O}(e^{(3\gamma_{\text{c}}-2b) \,\varphi})\,,\\
(\tilde{u}_0-\tilde{u})\,\frac{d\tilde{u}}{du} &= \bar{\alpha}\,e^{(2\gamma_{\text{c}}-b) \,\varphi}    + \mathcal{O}(e^{(3\gamma_{\text{c}}-2b) \,\varphi})\,.
\end{align}
Substituting to \eqref{typeBdotAs} gives
\begin{align}
\dot{A}_1 &= -2\bar{\alpha}A_{11}\,e^{(2\gamma_{\text{c}}-b) \,\varphi} + \mathcal{O}(e^{(3\gamma_{\text{c}}-2b) \,\varphi})\,,\\
\dot{A}_2 &= -\frac{1}{\bar{\alpha}}\, e^{b\varphi} -3\bar{\alpha}A_{21}\,e^{(2\gamma_{\text{c}}-b) \,\varphi}+ \mathcal{O}(e^{(3\gamma_{\text{c}}-2b) \,\varphi})\,,\label{eq:type_B_A_dot_2_exp}
\end{align}
where by \eqref{onshellcoefficientstypeB0} and \eqref{onshellcoefficientstypeB} the coefficients are on-shell explicitly
\begin{equation}
2\bar{\alpha}A_{11} = -\frac{\bar{\alpha}}{N+1}\biggl(\frac{2\widetilde{\Lambda}}{d+N-1}-\kappa_{\scriptscriptstyle\text{IR}}\biggr)\,, \quad 3\bar{\alpha}A_{21} = \frac{\bar{\alpha }}{2N(N+1)}\biggl(\frac{d-N-1}{d+N-1}\,2\widetilde{\Lambda}-d\kappa_{\scriptscriptstyle\text{IR}}\biggr)\,.
\end{equation}
Finally substituting \eqref{eq:type_B_A_dot_2_exp} to \eqref{SWA1A2app} gives the expansion
\begin{equation}
S(\varphi) =
S_{\infty}^{\text{II}}\,e^{b\varphi} + (S_{\infty,1}^{\text{II}}+S_{*}^{\text{II}})\,e^{(2\gamma_{\text{c}}-b)\,\varphi} + \mathcal{O}(e^{(3\gamma_{\text{c}}-2b) \,\varphi})\,,
\label{StypeBupstairsexpansion}
\end{equation}
where the coefficients are
\begin{gather}
S_{\infty}^{\text{II}} = \sqrt{\frac{2N(d+N-1)}{d-1}}\,\frac{1}{\bar{\alpha}}\,, \quad S_{\infty,1}^{\text{II}} = -\frac{d}{4}\frac{N-1}{N+1}\left(\frac{\bar{\alpha}}{\alpha}\right)^2\left(\frac{\alpha}{\alpha_{\IR}}\right)^{2}\,S_{\infty}^{\text{II}}\,,\nonumber\\
S_{*}^{\text{II}} = \frac{(N-1)(d-N-1)}{N\,(N+1)(d+N-1)}\frac{\widetilde{\Lambda}}{\bar{\kappa}}\,S_{\infty}^{\text{II}}\,.
\label{eq:typeB_reduction_coefficients}
\end{gather}
As a result, the free parameter $S_{\infty,1}^{\text{II}} $ of the type II solutions is given by $  \frac{\alpha}{\alpha_{\IR}} $. Using $ \bar{\alpha} = \sqrt{\frac{N(N-1)}{-V_{\infty}}} $ and $ 2\widetilde{\Lambda} = V_{\infty,1} $, we can write the coefficients as
\begin{align}
S_{\infty}^{\text{II}} &= \sqrt{\frac{(d+N-1)(-2V_{\infty})}{(N-1)(d-1)}}\,,\label{eq:SinftyB_higher}\\
S_{*}^{\text{II}} &= -\frac{1}{2}\frac{(N-1)(d-N-1)}{(N+1)(d+N-1)}\frac{V_{\infty,1}}{V_{\infty}}\,S_{\infty}^{\text{II}}\,.
\label{eq:Sstar_reduction}
\end{align}

\paragraph{The type I geometry.}

By substituting the IR expansion of the type I solution \eqref{eq:typeChigherdimensional} to \eqref{Avarphilower}, we obtain
\begin{equation}
A(u) = \log{\biggl(\frac{\tilde{u}_0-\tilde{u}}{\alpha}\biggr)}+\frac{N}{d-1}\,\log{\biggl(\frac{\tilde{u}_0-\tilde{u}}{\bar{\alpha}}\biggr)} + \mathcal{O}(1)\,, \quad \tilde{u}\rightarrow \tilde{u}_0^{-}\,,
\label{typeCAu}
\end{equation}
and similarly
\begin{equation}
(b-\gamma_{\text{c}})\,\varphi(u) =  -\log{\biggl(\frac{\tilde{u}_0-\tilde{u}}{\bar{\alpha}}\biggr)}-A_{20} -A_{21}\,(\tilde{u}_0-\tilde{u})^{2}+ \mathcal{O}(\tilde{u}_0-\tilde{u})^{4}\,,
\label{typeCphiasymptotics}
\end{equation}
where we have used \eqref{bminusgammac}. Hence, the scalar field diverges $ \varphi \rightarrow \infty $ when $\tilde{u}\rightarrow \tilde{u}_0^-$.

Using \eqref{SWA1A2app} together with $\frac{d\tilde{u}}{du} = e^{\gamma_{\text{c}}\varphi}$ gives
\begin{equation}
	S(\varphi) = -2N b\,e^{\gamma_{\text{c}}\varphi}\,A_2'(\tilde{u})\,,
	\label{eq:SWA1A2utildedertext}
\end{equation}
where the derivative is with respect to $\tilde{u}$. Substituting the type I solution \eqref{eq:typeChigherdimensional} gives
\begin{equation}
    S(\varphi) = \frac{2N b}{\tilde{\ell}}\,e^{\gamma_{\text{c}}\varphi}\,\sqrt{1+\csch^{2}{\biggl(\frac{\tilde{u}_0-\tilde{u}}{\tilde{\ell}}\biggr)}}\,.
    \label{eq:eq:SWA1A2utildedertext_phi}
\end{equation}
Solving the relation \eqref{lowerdimensionalfields} between $A_2$ and $\varphi$ given the type I solution \eqref{eq:typeChigherdimensional}, we obtain
\begin{equation}
	\sinh{\biggl(\frac{\tilde{u}_0-\tilde{u}}{\tilde{\ell}}\biggr)} = \frac{\bar{\alpha}}{\tilde{\ell}}\sqrt{\frac{d+N-1}{N-1}}\,e^{-(b-\gamma_{\text{c}})\varphi}\,,
	\label{eq:varphiutilderelation}
\end{equation}
where we have used $b-\gamma_{\text{c}} = \sqrt{\frac{d-1}{2N(d+N-1)}} $. Substituting to \eqref{eq:eq:SWA1A2utildedertext_phi} gives the type I solution
\begin{equation}
	S(\varphi) = \sqrt{\frac{2N(N-1)}{d-1}}\,\frac{1}{\bar{\alpha}}\,e^{b\varphi}\,\sqrt{ 1 - \frac{2}{d+N}\frac{\widetilde{\Lambda}}{\bar{\kappa}}\,e^{-2(b-\gamma_{\text{c}})\varphi}}\,.
\end{equation}
The expansion in $\varphi \rightarrow \infty$ is given by
\begin{equation}
S(\varphi) =
S_{\infty}^{\text{I}}\,e^{b\varphi} + S_{*}^{\text{I}}\,e^{(2\gamma_{\text{c}}-b)\,\varphi} + \ldots\,,\quad \varphi\rightarrow \infty\,,
\end{equation}
where the coefficients are
\begin{equation}
S_{\infty}^{\text{I}} = \sqrt{\frac{2N(N-1)}{d-1}}\,\frac{1}{\bar{\alpha}}\,, \quad S_{*}^{\text{I}} = -\frac{1}{d+N}\frac{\widetilde{\Lambda}}{\bar{\kappa}}\,S_{\infty}^{\text{I}}\,.
\end{equation}
Using $ \bar{\alpha} = \sqrt{\frac{N(N-1)}{-V_{\infty}}} $ and $ 2\widetilde{\Lambda} = V_{\infty,1} $, we can write these coefficients as
\begin{equation}
	S_{\infty}^{\text{I}} =\sqrt{\frac{-2V_{\infty}}{d-1}}\,, \quad S_{*}^{\text{I}} = \frac{1}{2}\frac{N}{d+N}\frac{V_{\infty,1}}{V_{\infty}}\,S_{\infty}^{\text{I}}\,.
	\label{eq:SinftyC_higher}
\end{equation}

\section{Solutions in the Einstein--scalar theory}\label{app:reducedsolutions}

In this Appendix, we  solve the equations of motion of the Einstein-scalar theory when the potential has exponential asymptotics
\begin{equation}
	V(\varphi) = V_{\infty}\,e^{2b\varphi} + V_{\infty,1}\,e^{2\gamma \varphi} + \ldots\,, \quad \varphi \rightarrow \infty\,,
	\label{potentialasymptoticsapp}
\end{equation}
where we assume that $b_{\text{c}} < b < b_{\text{G}}$, $ b > \gamma $ and $V_{\infty} < 0$. When the potential comes from dimensional reduction, it takes the exact form (valid for all $\varphi$)
\begin{equation}
    V(\varphi) = V_{\infty}\,e^{2b\varphi} + V_{\infty,1}\,e^{2\gamma_{\text{c}}\varphi}
    \label{reducedpotentialasymptoticsapp}
\end{equation}
with a subleading exponent $ \gamma = \gamma_{\text{c}} = \frac{1}{2\,(d-1)\,b} $ and coefficients \eqref{eq:potential_coefficients}. In this case, the condition $b > \gamma_{\text{c}}$ is equivalent to the confinement bound $b > b_{\text{c}}$. The solutions are classified into two types depending on whether the scalar field is finite $ \varphi \rightarrow \varphi_0 < \infty $ (type III solutions) or whether it diverges $ \varphi \rightarrow \infty $ (type 0, I and II solutions) at the IR endpoint.

\subsection{Bounded scalar field (type III solutions)}\label{reduced:typeA}

We search for solutions of the equation \eqref{Sequation} where $\varphi$ is bounded in the IR $\varphi(u_0) = \varphi_0 < \infty$. In this case $ \varphi \rightarrow \varphi_{0}^{-} $ as $u\rightarrow u_0^{-}$ and we can consider the ansatz \cite{Ghosh:2017big}
\begin{equation}
	S(\varphi) = \sqrt{\varphi_{0}-\varphi}\,\Bigl[S_0 + S_1\,\sqrt{\varphi_{0}-\varphi}+ S_2\,( \varphi_{0}-\varphi) + \mathcal{O}(\varphi_{0}-\varphi)^{3\slash 2}\Bigr]\,.
	\label{StypeA}
\end{equation}
By substituting to \eqref{Sequation}, we obtain\footnote{There is also a second solution $ S_0 = \sqrt{-2V'(\varphi_{0})} $ which gives a finite size $ A(u_0)\neq 0 $ for the $S^{d}$ at the IR endpoint (see equation \eqref{eq:Au_type_A_lower_app} below).  We do not consider this solution, but it is relevant for hyperbolically sliced flows, where it would correspond to a bounce (see \cite{Ghodsi:2024jxe}).}
\begin{equation}
	S_0 = \sqrt{-\frac{2V'(\varphi_{0})}{d+1}}\,, \quad S_1 = 0\,, \quad S_2 = \frac{1}{2(d+3)}\biggl(3V''(\varphi_{0})+\frac{2V(\varphi_0)}{d-1}\biggr)\frac{1}{S_0}\,.
 \label{eq:ScoefficientsA}
\end{equation}
When the potential comes from dimensional reduction \eqref{reducedpotentialasymptoticsapp}, we observe that the solution \eqref{StypeA} with the leading coefficient \eqref{eq:ScoefficientsA} matches with the dimensional reduction \eqref{eq:typeA_dimensional_reduction} of a type III solution with coefficient \eqref{eq:S0_reduction}.

Substituting \eqref{StypeA} to the equation \eqref{Wtext} for $W$ gives the expansion
\begin{equation}
	W(\varphi) = \frac{1}{\sqrt{\varphi_{0}-\varphi}}\Bigl[W_0 + W_1\,\sqrt{\varphi_{0}-\varphi}+ W_2\,( \varphi_{0}-\varphi)+ \mathcal{O}(\varphi_{0}-\varphi)^{3\slash 2}\Bigr]\,,
	\label{WexptypeA}
\end{equation}
where the coefficients are
\begin{equation}
	W_0 = (d-1)\,S_0\,, \quad W_1 = 0\,, \quad W_2 = \frac{d-1}{2(d+3)}\biggl[\frac{2(d+4)}{d(d-1)}\,V(\varphi_{0})-V''(\varphi_{0})\biggr]\,.
	\label{W0expressionapp}
\end{equation}
Since $ S(\varphi) = \dot{\varphi} $, the expansion \eqref{StypeA} implies
\begin{equation}
	\varphi(u) = \varphi_{0}+\varphi_1\,(u_0-u)^{2} + \varphi_2\,(u_0-u)^{4}  + \mathcal{O}(u_0-u)^{5}\,, \quad u\rightarrow u_0^{-}\,,
 \label{eq:varphiu_typeA_expansion}
\end{equation}
with coefficients
\begin{equation}
	\varphi_1 = -\frac{S_0^{2}}{4}\,, \quad \varphi_{2} = \frac{S_0^{3}S_2}{40}\,.
\end{equation}
Substituting \eqref{eq:varphiu_typeA_expansion} to \eqref{WexptypeA} gives an expansion for $W$ in powers of $u_0-u$. Using $ W(\varphi) = -2(d-1)\,\dot{A} $ then gives the expansion
\begin{equation}
	A(u) = \frac{W_0}{(d-1)\,S_0}\log{(u_0-u)} + A_0 + \frac{5S_0W_2+S_2W_0}{40(d-1)}\,(u_0-u)^{2} + \mathcal{O}(u_0-u)^{3}\,,
	\label{eq:Au_type_A_lower_app}
\end{equation}
where $ A_0 $ is an integration constant. Substituting the explicit values \eqref{W0expressionapp} for $W_0$ gives (focusing only on the leading term)
\begin{equation}
	A(u) = \log{\biggl(\frac{u_0-u}{\alpha}\biggr)} + \mathcal{O}(u_0-u)^{2}\,.
\end{equation}
where we have fixed $ A_0\equiv -\log{\alpha} $ by requiring the absence of a conical singularity in the metric \eqref{bulkansatz} at the IR endpoint.

For the $ T $ function \eqref{Wtext}, we obtain by substituting \eqref{StypeA} and \eqref{WexptypeA} the expansion
\begin{align}
	T(\varphi) = \frac{1}{\varphi_{0}-\varphi}\Bigl[T_0 + T_1\,\sqrt{\varphi_{0}-\varphi}+ \mathcal{O}( \varphi_{0}-\varphi)\Bigr]\,,\quad \varphi \rightarrow \varphi_{0}^{-}\,,
\end{align}
with the coefficients\footnote{The $ T_0 $ in \cite{Ghosh:2017big} has incorrectly an extra factor of $ (d+1)^{-1} $.}
\begin{equation}
	T_0 = \frac{d}{4}\, S_0 W_0 = \frac{d(d-1)}{4}\,S_0^{2}\,, \quad T_1 = 0\,.
	\label{T0U0expressionapp}
\end{equation}
We now consider the function $ U $ which satisfies the first-order differential equation \eqref{Uequationtext}. Using \eqref{StypeA} and \eqref{WexptypeA}, the equation becomes at leading order
\begin{equation}
	\bigl(S_0\sqrt{\varphi_0-\varphi}+\ldots\bigr)\,U'(\varphi) -\frac{d-2}{2(d-1)}\biggl(\frac{W_0}{\sqrt{\varphi_0-\varphi}}+\ldots\biggr)\,U(\varphi) + \ldots = 0\,.
 \label{eq:leadingorderUequation}
\end{equation}
This has the solution
\begin{equation}
	U(\varphi) = \mathcal{U}\,(\varphi_{0}-\varphi)^{-\frac{(d-2)W_0}{2(d-1)S_0}} + \ldots\,,
\end{equation}
where $ \mathcal{U} $ is an integration constant and the exponent is explicitly $ \frac{(d-2)W_0}{2(d-1)S_0} = -\frac{d}{2}+1 $. Including also the subleading terms in the expansions of $ S $ and $ W $ to the equation \eqref{eq:leadingorderUequation}, we then find
\begin{equation}
	U(\varphi) = \frac{\mathcal{U}}{(\varphi_{0}-\varphi)^{\frac{d}{2}-1}} +U_0\sqrt{ \varphi_{0}-\varphi} + \mathcal{O}( \varphi_{0}-\varphi)\,,
\end{equation}
where the coefficient
\begin{equation}
	U_0 = \frac{4}{d\,(d-1)\,S_0}\,.
 \label{eq:typeA_U0}
\end{equation}

\subsection{Unbounded scalar field (type 0, I and II solutions)}\label{app:solutionsunboundedscalar}

We  search for solutions of the equation \eqref{Sequation}, where $\varphi$ is unbounded in the IR $\varphi\rightarrow \infty$. Assuming first that $ S $ and its derivatives grow faster than $ V $ (and its derivatives) when $\varphi\rightarrow \infty$, we can truncate the equation \eqref{Sequation} into the first three terms which gives the (approximate) equation
\begin{equation}
	d\,SS''-\frac{d}{2}\,S^{2}-S'^{2} = 0\,.
\end{equation}
This equation has the type 0 solution
\begin{equation}
    S(\varphi) = s_1\biggl[\cosh{\biggl(\frac{\varphi-s_2}{2b_{\text{G}}}\biggr)}\biggr]^{2b_{\text{G}}^2}\,,
\end{equation}
where $ b_{\text{G}} $ is the Gubser bound \eqref{gubsB} and $s_{1,2}$ are integration constants. Therefore the type 0 solution has the expansion
\begin{equation}
	S(\varphi) = S_{\infty}^{0}\,e^{b_{\text{G}}\varphi} + S_{\infty,1}^{0}\,e^{(b_{\text{G}}-b_{\text{G}}^{-1})\,\varphi}+\ldots\,, \quad \varphi \rightarrow \infty\,,\quad \text{type 0}\,,
	\label{typeDsolution}
\end{equation}
where both $ S_{\infty}^{0} $ and $S_{\infty,1}^{0}$ are integration constants (functions of $s_{1,2}$). The type 0 solution exists as long as $ b < b_{\text{G}} $ and it has also been observed in flat slicing in \cite{Gursoy:2008za,Kiritsis:2016kog} and in hyperbolic slicing in \cite{Ghodsi:2024jxe}.

We  look for solutions to the full equation \eqref{Sequation} of the form
\begin{equation}
	S(\varphi) = S_{\infty}\,e^{b\varphi} + \ldots\,, \quad \varphi \rightarrow \infty\,,
	\label{leadingS}
\end{equation}
where $ b $ is the exponent in the leading term of the potential \eqref{potentialasymptoticsapp}. There are two branches of solutions
\begin{equation}
	S_{\infty} =
	\begin{dcases}
        S_{\infty}^{\text{I}} \equiv 2b_{\text{c}}\sqrt{-V_{\infty}}, \quad &\text{type I}\\
		S_{\infty}^{\text{II}} \equiv 2b\,\sqrt{\frac{-V_{\infty}}{1-2\,(b^{2}-b_{\text{c}}^{2})}}, \quad &\text{type II}
	\end{dcases}\,.
	\label{eq:Sinfty_lower}
\end{equation}
We observe that $ b_{\text{c}} \leq b < b_{\text{G}} $ amounts to $2b_{\text{c}}\sqrt{-V_{\infty}} \leq S_{\infty}^{\text{II}} < \infty$. At the confinement bound $ b = b_{\text{c}} $, the type I and type II solutions coincide to leading order in $ \varphi\rightarrow \infty $, while as $b$ approaches the Gubser bound $ b \rightarrow b_{\text{G}}^{-} $, the type II coefficient diverges $ S_{\infty}^{\text{II}} \rightarrow \infty $.

Consider for a moment the special case of a potential $V(\varphi)$ coming from dimensional reduction \eqref{reducedpotentialasymptoticsapp} with $b = \sqrt{\frac{d+N-1}{2N(d-1)}}$. Substituting to \eqref{eq:Sinfty_lower} we  obtain
\begin{equation}
	S_{\infty}^{\text{II}} = \sqrt{\frac{(d+N-1)(-2V_{\infty})}{(N-1)(d-1)}}\,,\quad S_{\infty}^{\text{I}} = \sqrt{\frac{-2V_{\infty}}{d-1}}\,,
\end{equation}
which match with the coefficients \eqref{eq:SinftyB_higher} and \eqref{eq:SinftyC_higher} of the reductions of the higher-dimensional type I and II solutions respectively. Therefore we can identify the lower-dimensional type I and II solutions with the higher-dimensional ones in the IR region $\varphi\rightarrow \infty$.

We now expand the rest of the functions as
\begin{equation}
	W(\varphi) = W_{\infty}\,e^{b\varphi}+ \ldots\,, \quad T(\varphi) = T_{\infty}\,e^{2b\varphi}+ \ldots\, .
\end{equation}
For the type II solutions, we  obtain  from \eqref{Wtext} that
\begin{equation}
	W_{\infty}^{\text{II}} = b^{-1}S_{\infty}^{\text{II}}\,, \quad T_{\infty}^{\text{II}} = 0\,,
	\label{typeBleading}
\end{equation}
while for the type I solution, we obtain
\begin{equation}
	W_{\infty}^{\text{I}}  = 2b\,(d-1)\,S_{\infty}^{\text{I}} \,, \quad T_{\infty}^{\text{I}}  = -\frac{d}{2}\,(1-2\,(d-1)\,b^{2})\,(S_{\infty}^{\text{I}})^{2}\,.
	\label{typeCleading}
\end{equation}
For type 0 solutions, we instead have the expansions
\begin{equation}
	W(\varphi) = W_{\infty}^{0}\,e^{b_{\text{G}}\varphi}+W_{\infty,1}^{0}\,e^{(b_{\text{G}}-b_{\text{G}}^{-1})\,\varphi}+\ldots \,, \quad T(\varphi) = T_{\infty}^{0}\,e^{(2b_{\text{G}}-b_{\text{G}}^{-1})\,\varphi}+ \ldots\,,
 \label{eq:type0_WT_expansions}
\end{equation}
where the coefficients are given by
\begin{equation}
    W_{\infty}^{0} = b_{\text{G}}^{-1}\,S_{\infty}^{0}\,,\quad  W_{\infty,1}^{0} =\frac{b_{\text{G}}^2-1}{b_{\text{G}}^3}\,,\quad T_{\infty}^{0}=-\frac{1}{b_{\text{G}}^2}\,S_{\infty}^{0}\,S_{\infty,1}^{0}\,.
\end{equation}
We consider the function $ U $ which satisfies the first-order differential equation \eqref{Uequationtext}. At leading order using $S,W\sim e^{b\varphi}$, we obtain the equation (in the type 0 case, we simply set $b=b_{\text{G}}$ on the right-hand side)
\begin{equation}
	S_{\infty}\,U'(\varphi) -\frac{(d-2)\,W_{\infty}}{2\,(d-1)}\,U(\varphi) = -\frac{2}{d}\,e^{-b\varphi}\,.
 \label{eq:leadingUequation_typeBC}
\end{equation}
The homogeneous equation has the solution
\begin{equation}
	U(\varphi) =  \mathcal{U}\,e^{\frac{(d-2)\,W_{\infty}}{2(d-1)\,S_{\infty}}\,\varphi} + \ldots\,,
\end{equation}
where $ \mathcal{U} $ is an integration constant. The full solution of \eqref{eq:leadingUequation_typeBC} also includes a particular solution of the non-homogeneous equation which leads to the expansion
\begin{equation}
	U(\varphi) =
	\begin{dcases}
		\mathcal{U}\,e^{\frac{d-2}{2b\,(d-1)}\,\varphi} + U_{\infty}^{\text{II}}\,e^{-b\varphi} + \ldots\,, \quad &\text{type II}\\
		\mathcal{U}\,e^{(d-2)\,b\varphi}+ U_{\infty}^{\text{I}}\,e^{-b\varphi} + \ldots\,, \quad &\text{type I}\\
	\end{dcases}
	\,, \quad \varphi \rightarrow \infty\,,
\label{eq:typeBC_U_expansion}
\end{equation}
where the coefficients are
\begin{equation}
	U_{\infty}^{\text{II}} =\frac{4\,(d-1)\,b}{d\,(d-2+2\,(d-1)\,b^{2})}\,\frac{1}{S_{\infty}^{\text{II}}}\,, \quad U_{\infty}^{\text{I}} =\frac{2}{d\,(d-1)\,b}\frac{1}{S_{\infty}^{\text{I}}}\,,
	\label{UinftytypeBC}
\end{equation}
and the further subleading terms are determined by subleading terms in $S$ and $W$.

\paragraph{Subleading terms.}

We now look for subleading corrections to the leading solution \eqref{leadingS} of the form
\begin{equation}
	S(\varphi) = S_{\infty}\,e^{b\varphi} + \delta S(\varphi)\,,
\end{equation}
where $\delta S(\varphi)\slash e^{b\varphi}\rightarrow 0 $ when $\varphi\rightarrow \infty$. We  write the equation \eqref{Sequation} as $\mathcal{E}(S) = 0$. Defining $ S_{(0)}(\varphi) \equiv S_{\infty}\,e^{b\varphi} $ and expanding to leading order in $ \delta S $ gives the equation
\begin{equation}
	0=\mathcal{E}(S)=\mathcal{E}(S_{(0)}) + \mathcal{E}_1(\delta S) + \mathcal{O}(\delta S^{2})\,,
\end{equation}
where the first order term is linear in $ \delta S$ and takes the form.
\begin{equation}
	\mathcal{E}_1(\delta S) = B(\varphi)\left[\delta S''(\varphi) +C(\varphi)\,\delta S'(\varphi)+ D(\varphi)\,\delta S(\varphi)\right]\,.
\end{equation}
The overall factor is explicitly
\begin{equation}
	B(\varphi) = d\,S_{\infty}^{3}\,e^{3b\varphi}\,.
\end{equation}
Hence, at linear order we obtain the non-homogeneous differential equation
\begin{equation}
	\delta S''(\varphi) +C(\varphi)\,\delta S'(\varphi)+ D(\varphi)\,\delta S(\varphi) = K(\varphi)\,,
	\label{linearSequation1}
\end{equation}
where the non-homogenous piece
\begin{equation}
	K(\varphi) = -\frac{\mathcal{E}(S_{(0)})}{B(\varphi)}\,.
	\label{Kphi}
\end{equation}
The coefficient functions are given by\footnote{The subleading $ \mathcal{O}(e^{(2\gamma-2b)\,\varphi}) $ terms vanish when $ V_{\infty,1} = 0 $.}
\begin{align}
	C(\varphi) &= C_0 + \mathcal{O}(e^{(2\gamma-2b)\,\varphi}) \\
	D(\varphi) &= D_0+ \mathcal{O}(e^{(2\gamma-2b)\,\varphi})\\
	K(\varphi) &= K_0\,e^{(2\gamma - b)\,\varphi}+ \mathcal{O}(e^{(4\gamma-3b)\,\varphi})
\end{align}
so that at leading order in $ \varphi\rightarrow \infty $ we obtain the equation
\begin{equation}
	\delta S''(\varphi) +C_0\,\delta S'(\varphi)+ D_0\,\delta S(\varphi) = K_0\,e^{(2\gamma - b)\,\varphi}\,,
	\label{leadingSequation}
\end{equation}
where the coefficients are explicitly
\begin{align}
	C_0 &= -\frac{2\,b}{d}\left(1-\frac{(d+2)V_{\infty}}{S_{\infty}^{2}}\right)\,,\nonumber\\
	D_0 &= \frac{b^{2}(3d^{2}-5d+2)}{d(d-1)}\biggl(1-\frac{2V_{\infty}}{S_{\infty}^{2}}\biggr)-2\,\biggl(1+\frac{V_{\infty}}{(d-1)S_{\infty}^{2}}\biggr)\,,\nonumber\\
	K_0 &= -\biggl[\frac{4b\gamma}{d}\left(1-\frac{2\,V_{\infty}}{S_{\infty}^{2}}\right)-\frac{1}{d-1}-2\gamma\,(2\gamma-b)\biggr]\frac{V_{\infty,1}}{S_{\infty}}\,.
	\label{linearScoefficients}
\end{align}
Assuming\footnote{If $2\gamma -b $ equals $ \beta_+ $ or $ \beta_- $, the particular solution is modified, however, this will not be necessary for our type I and II solutions below.} $ \beta_{\pm} \neq 2\gamma-b $, the general solution of \eqref{leadingSequation} is given by
\begin{equation}
	\delta S(\varphi) =S_{*}\,e^{(2\gamma-b)\,\varphi}+ S_+\,e^{\beta_+ \varphi} + S_-\,e^{\beta_- \varphi}\,,
	\label{gendeltaS}
\end{equation}
where the first term is a particular solution of \eqref{leadingSequation} with the fixed coefficient
\begin{equation}
	S_{*} = \frac{K_0}{(2\gamma - b)^{2}+(2\gamma-b)\,C_0 + D_0}\,,
	\label{Sstar}
\end{equation}
the coefficients $ S_{\pm} $ are two free integration constants and the exponents $ \beta_{\pm} $ are determined by the homogeneous ODE (with $ K_0 = 0 $) to be
\begin{equation}
	\beta_{\pm} = \frac{1}{2}\Bigl(-C_0 \pm \sqrt{C_0^{2}-4D_0}\,\Bigr)\,,
	\label{betapm}
\end{equation}
such that $ \text{Re}\,\beta_+ > \text{Re}\,\beta_- $. Requiring the solution \eqref{gendeltaS} to be a subleading correction to \eqref{leadingS} gives the bound
\begin{equation}
	\text{Re}\,\beta_{\pm} < b\,.
	\label{betapmcondition}
\end{equation}
If this condition is violated by an exponent, we have to set the corresponding integration constant to zero. The particular solution $ S_{*}\,e^{(2\gamma-b)\varphi} $ always appears since it is subleading with respect to $S_{\infty}\,e^{b\varphi}$ due to the assumption $ b > \gamma $.

We  also calculate subleading corrections to the rest of the functions as
\begin{equation}
	W(\varphi) = W_{\infty}\,e^{b\varphi} + \delta W(\varphi), \quad T(\varphi) = T_{\infty}\,e^{2b\varphi} + \delta T(\varphi)\,,
	\label{expansionotherfunctions}
\end{equation}
where $ W_{\infty} $ and $ T_{\infty} $ are given by the type II \eqref{typeBleading} or type I \eqref{typeCleading} expressions. At leading order, we find
\begin{align}
	\delta W(\varphi) &= \frac{2(d-1)}{d}\,\delta S'(\varphi) + \frac{4b(d-1)}{d}\biggl(1+\frac{\gamma}{b}\frac{V_{\infty,1}}{V_{\infty}}\,e^{-2(b-\gamma)\,\varphi}\biggr)\frac{V_{\infty}}{S_{\infty}^{2}}\,\delta S(\varphi) +  \mathcal{O}(\delta S^{2})\,,\nonumber \\
	\delta T(\varphi) &= -de^{b\varphi}\biggl[\biggl(1-\frac{bW_{\infty}}{2S_{\infty}}\biggr)\,\delta S(\varphi) - \frac{1}{2}\,\delta W(\varphi)\biggr]\,S_{\infty}+ \mathcal{O}(\delta S^{2})\,.
\end{align}
Using the expansion \eqref{gendeltaS} of $ \delta S $ gives at leading order in $ \varphi\rightarrow \infty $,
\begin{align}
	\delta W(\varphi) &= W_{*}\,e^{(2\gamma -b)\,\varphi}+W_{+}\,e^{\beta_+ \varphi} + W_{-}\,e^{\beta_- \varphi} + \ldots\,,\label{eq:deltaWapp}\\
	\delta T(\varphi) &= T_{*}\,e^{2\gamma \varphi}+T_{+}\,e^{(\beta_++b)\, \varphi} + T_{-}\,e^{(\beta_-+b)\, \varphi} + \ldots\,,
	\label{functionslinearized}
\end{align}
where the coefficients relevant for calculations below are
\begin{align}
    W_- &= \frac{2b\,(d-1)}{d}\biggl(\frac{2V_{\infty}}{S_{\infty}^2}+\frac{\beta_-}{b}\biggr)\,S_-\,,\\
    W_* &= \frac{2b\,(d-1)}{d}\biggl[\biggl(\frac{2V_{\infty}}{S_{\infty}^2}-1+\frac{2\gamma}{b}\biggr)\,S_*-\frac{2\gamma\,V_{\infty,1}}{b\,S_{\infty}}\biggr]\,,\\
    T_-&= \frac{d}{2}(b\,W_\infty S_--2S_\infty S_-+\beta_- S_\infty W_-)\,,\\
    T_* &= \frac{d}{2}(b\,W_\infty S_*-2S_\infty S_*+(2\gamma-b)\,S_\infty W_*)\,.
\label{eq:Ws_Ts}
\end{align}
For the $ U $ function, we consider a similar ansatz
\begin{equation}
	U(\varphi) = \mathcal{U}\,e^{\frac{(d-2)\,W_{\infty}}{2(d-1)\,S_{\infty}}\,\varphi} +U_{\infty}\,e^{-b\varphi} + \delta U(\varphi)\,,
\end{equation}
where $U_{\infty}$ is given by \eqref{UinftytypeBC}. The linearized equation \eqref{Uequationtext} for $\delta U$ can be shown to have a solution of the form
\begin{equation}
	\delta U(\varphi) =  U_{*}\,e^{(2\gamma -3b)\,\varphi}+U_{+}\,e^{(\beta_+-2b)\, \varphi} + U_{-}\,e^{(\beta_--2b)\, \varphi} + \ldots\,,\quad \varphi\rightarrow \infty\,.
	\label{eq:deltaU_app}
\end{equation}
The expressions for $ W_{\pm} $, $ T_{\pm} $ and $ U_{\pm} $ in terms of $ S_{\pm} $, $ S_{\infty} $, $ V_{\infty} $ and $ V_{\infty,1} $ can be found explicitly. Assuming the condition \eqref{betapmcondition} is satisfied, \eqref{eq:deltaWapp}, \eqref{functionslinearized} and \eqref{eq:deltaU_app} are subleading corrections to the leading solutions.

\paragraph{Type II solutions.}

For type II solutions, the coefficients \eqref{linearScoefficients} are given by
\begin{align}
	C_0 &= \frac{2b^{2}-1}{2b}-\frac{3}{2\,(d-1)\,b}\,,\\
	D_0 &= \frac{d}{2\,(d-1)\,b}\biggl(\frac{1}{(d-1)\,b}-b\biggr)\,,\\
	K_0 &= -(2\gamma - b)\biggl(\frac{1}{(d-1)\,b}-2\gamma\biggr)\frac{V_{\infty,1}}{S_{\infty}^{\text{II}}}\,,
	\label{typeBcoefficients}
\end{align}
and the coefficient \eqref{Sstar} of the particular solution is given by
\begin{equation}
	S_{*}^{\text{II}}= -\frac{(2b^{2}(d-1)-d)(2\gamma-b)}{2b(d-2b(d-1)(2\gamma-b))}\frac{V_{\infty,1}}{V_{\infty}}\,S_{\infty}^{\text{II}}\,.
	\label{SstypeB}
\end{equation}
The exponents \eqref{betapm} are given by
\begin{equation}
	\beta_- = \frac{1}{(d-1)\,b}-b = 2\gamma_{\text{c}} - b, \quad \beta_+ = \frac{d}{2\,(d-1)\,b} = \frac{b_{\text{G}}^{2}}{b}\,.
 \label{eq:typeB_betapm}
\end{equation}
where $ \gamma_{\text{c}}= \frac{1}{2\,(d-1)\,b} $ is the same as the subleading exponent \eqref{eq:reduction_exponents} of the potential coming from dimensional reduction, but it arises completely independently. Above the confinement bound $ b> b_{\text{c}} $, we have $ \gamma_{\text{c}}< b $ implying $ \beta_-< b $ gives a subleading correction to $ S\sim e^{b\varphi} $. Due to the Gubser bound $ b < b_{\text{G}} $, we observe that $ \beta_+ >b $ is not  subleading, and we have to set $ S_+ = 0 $. Hence, above the confinement bound $ b_{\text{c}}< b < b_{\text{G}} $ the expansion of the type II solution to leading order in $ \varphi \rightarrow \infty $ is given by
\begin{equation}
	S_{\text{II}}(\varphi) = S_{\infty}^{\text{II}}\,e^{b\varphi} + S_{\infty,1}^{\text{II}}\,e^{(2\gamma_{\text{c}} - b)\,\varphi}+ S_{*}^{\text{II}}\,e^{(2\gamma - b)\,\varphi} + \ldots\,,\quad \varphi\rightarrow \infty\,,
\end{equation}
where $ S_{\infty,1}^{\text{II}} \equiv S_- $ is the single free integration constant of type II solutions. At the confinement bound $ b = b_{\text{c}} $, the exponent $ \beta_- = b_{\text{c}} $ gives the same solution as the leading solution. Hence, when $ b = b_{\text{c}} $ the solution reduces to
\begin{equation}
	S_{\text{II}}(\varphi) =S_{\infty}^{\text{I}}\,e^{b_{\text{c}}\varphi}+ S_{*}^{\text{II}}\big\lvert_{b = b_{\text{c}}} \,e^{(2\gamma - b_{\text{c}})\,\varphi} + \ldots\,,\quad \varphi\rightarrow \infty\,,
\end{equation}
without free integration constants and with the type I coefficient $ S_{\infty}^{\text{I}} = S_{\infty}^{\text{II}}\big\lvert_{b = b_{\text{c}}} $ in the leading term. The coefficient of the subleading term is given by
\begin{equation}
	S_{*}^{\text{II}}\big\lvert_{b = b_{\text{c}}} = -\frac{2\gamma-b_{\text{c}}}{2b_{\text{c}}(4b_{\text{c}}\,(\gamma-b_{\text{c}})-1)}\frac{V_{\infty,1}}{V_{\infty}}\,S_{\infty}^{\text{II}}\,.
	\label{eq:SstarB_bc}
\end{equation}
We shall see momentarily that $S_{*}^{\text{II}}\lvert_{b = b_{\text{c}}} = S_{*}^{\text{I}}\lvert_{b = b_{\text{c}}} $ so that for $b = b_{\text{c}}$ the type II solution coincides with the type I one at subleading order in $\varphi\rightarrow \infty$ as well.

Setting $S_+ = 0$ and substituting the exponent $\beta_-$ \eqref{eq:typeB_betapm} to \eqref{eq:deltaWapp}, \eqref{functionslinearized} and \eqref{eq:deltaU_app}, we obtain that rest of the functions have the $\varphi\rightarrow \infty$ expansions
\begin{align}
	W_{\text{II}}(\varphi) &= W_{\infty}^{\text{II}}\,e^{b\varphi}+W_{\infty,1}^{\text{II}}\,e^{(2\gamma_{\text{c}} - b)\,\varphi}+ W_{*}^{\text{II}}\,e^{(2\gamma-b)\,\varphi} + \ldots\,,\\
	T_{\text{II}}(\varphi) &= T_{\infty,1}^{\text{II}}\,e^{2\gamma_{\text{c}}\varphi} + T_*^{\text{II}}\,e^{2\gamma \varphi}+ \ldots\,,\\
	U_{\text{II}}(\varphi) &= \mathcal{U}\,e^{\frac{d-2}{2b\,(d-1)}\,\varphi}+U_{\infty}^{\text{II}}\,e^{-b\varphi} +U_{\infty,1}^{\text{II}}\,e^{(2\gamma_{\text{c}} - 3b)\, \varphi} + U_*^{\text{II}}\,e^{(2\gamma - 3b)\,\varphi} + \ldots\,,
	\label{functionstypeB}
\end{align}
where all the coefficients can be computed explicitly. The coefficients in $W$ are given by
\begin{equation}
	W_{\infty,1}^{\text{II}} = -\frac{d-2}{d\,b}\,S_{\infty,1}^{\text{II}}\,, \quad W_{*}^{\text{II}} = -\frac{2b^2(d-1)-d}{2b(2b^2(d-1)-4b(d-1)\gamma+d)}\frac{V_{\infty,1}}{V_{\infty}}\,S_\infty^{\text{II}}\,,
\end{equation}
while the coefficients in $T$ are
\begin{equation}
	T_{\infty,1}^{\text{II}} = -\biggl( 1+\frac{d-2}{2\,(d-1)\,b^{2}} \biggr)\,S_{\infty}^{\text{II}}\,S_{\infty,1}^{\text{II}}\,,\quad T_*^{\text{II}} = 0\,.
 \label{eq:typeB_T_coefficients}
\end{equation}
Consider now the special case of a potential $V(\varphi)$ coming from dimensional reduction \eqref{reducedpotentialasymptoticsapp} with $b = \sqrt{\frac{d+N-1}{2N(d-1)}}$ and $\gamma = \gamma_{\text{c}} = \frac{1}{2\,(d-1)\,b} $. In this case, we obtain the expansion
\begin{equation}
	S_{\text{II}}(\varphi) = S_{\infty}^{\text{II}}\,e^{b\varphi} + (S_{\infty,1}^{\text{II}} + S_{*}^{\text{II}})\,e^{(2\gamma_{\text{c}} - b)\varphi} + \ldots\,,\quad \varphi\rightarrow \infty\,,
	\label{eq:typeB_expansion_lower_S}
\end{equation}
where \eqref{SstypeB} is now given by
\begin{equation}
	S_{*}^{\text{II}} = \frac{(b^{2}-2b_{\text{c}}^{2})(2b^2-2b_{\text{c}}^2-1)}{2b^2\,(2b^{2}-2b_{\text{c}}^{2}+1)+1}\frac{V_{\infty,1}}{V_{\infty}}\,S_{\infty}^{\text{II}} = -\frac{1}{2}\frac{(N-1)(d-N-1)}{(N+1)(d+N-1)}\frac{V_{\infty,1}}{V_{\infty}}\,S_{\infty}^{\text{II}}\,,
\end{equation}
which matches with \eqref{eq:Sstar_reduction} obtained from dimensional reduction.

\paragraph{The type I solution.} For the type I solution, the coefficients \eqref{linearScoefficients} are given by
\begin{align}
	C_0 &= -(d+1)\,b\,,\\
	D_0 &= (3d-2)\,b^{2}-1\,,\\
	K_0 &= -\biggl(2\gamma\,(3b-2\gamma)-\frac{1}{d-1}\biggr)\frac{V_{\infty,1}}{S_{\infty}^{\text{I}}}\,,
\end{align}
so that the exponents \eqref{betapm} become
\begin{equation}
	\beta_{\pm} = \frac{b}{2}\left(d+1\pm\sqrt{d^{2}-10d+9+4b^{-2}}\right)\,.
\end{equation}
It is useful to note that these exponents can also be written as
\begin{equation}
	\beta_{\pm} = \frac{(d+1)\,b}{2}\pm \sqrt{1-\frac{b^{2}}{b_{\text{E}}^{2}}}\,,
	\label{typeCbetapm}
\end{equation}
where $ b_{\text{E}} $ is the Efimov bound \eqref{bsandbE}. The coefficient \eqref{Sstar} of the particular solution is given by
\begin{equation}
	S_{*}^{\text{I}} = -\frac{(d-1)(2\gamma-3b)\gamma+1\slash 2}{4\gamma^{2}-2(d+3)b\gamma+4db^{2}-1}\frac{V_{\infty,1}}{V_{\infty}}\,S_{\infty}^{\text{I}}\,.
	\label{SstartypeC}
\end{equation}
For $b = b_{\text{c}}$, this reduces to \eqref{eq:SstarB_bc}.

Below the Efimov bound $ b_{\text{c}}< b \leq b_{\text{E}} $, the exponents are real, but satisfy $ \beta_{\pm} > b $ so they are not subleading. Above the Efimov bound $ b_{\text{E}}< b < b_{\text{G}} $, the exponents have a non-zero imaginary part, but the real parts still satisfy $ \text{Re}\,\beta_{\pm} > b $ so that they are not subleading. Hence, we must set $ S_{\pm} = 0 $ and the type I solution is given by
\begin{equation}
	S(\varphi) = S_{\infty}^{\text{I}}\,e^{b\varphi} + S_{*}^{\text{I}}\,e^{(2\gamma -b)\varphi} + \ldots, \quad \varphi \rightarrow \infty\,,
\end{equation}
which does not contain any free integration constants. The rest of the functions have the $\varphi\rightarrow \infty$ expansions
\begin{align}
	W_{\text{I}}(\varphi) &= W_{\infty}^{\text{I}}\,e^{b\varphi}+ W_{*}^{\text{I}}\,e^{(2\gamma-b)\,\varphi} + \ldots\,,\\
	T_{\text{I}}(\varphi) &= T_{\infty}^{\text{I}}\,e^{2b\varphi}+ T_*^{\text{I}}\,e^{2\gamma \varphi}+ \ldots\,,\\
	U_{\text{I}}(\varphi) &= \mathcal{U}\,e^{\frac{d-2}{2b\,(d-1)}\,\varphi}+U_{\infty}^{\text{I}}\,e^{-b\varphi}  + U_*^{\text{I}}\,e^{(2\gamma - 3b)\,\varphi} + \ldots\,.
	\label{functionstypeC}
\end{align}
For the special case of a potential $V(\varphi)$ coming from dimensional reduction \eqref{reducedpotentialasymptoticsapp} with $b = \sqrt{\frac{d+N-1}{2N(d-1)}}$ and $\gamma = \gamma_{\text{c}} = \frac{1}{2\,(d-1)\,b} $, \eqref{SstartypeC} becomes
\begin{equation}
	S_{*}^{\text{I}} = -\frac{1}{2}\frac{d-1}{1-2d\,(d-1)\,b^{2}}\frac{V_{\infty,1}}{V_{\infty}}\,S_{\infty}^{\text{I}} = \frac{1}{2}\frac{N}{d+N}\frac{V_{\infty,1}}{V_{\infty}}\,S_{\infty}^{\text{I}}\,,
\end{equation}
which matches with the coefficient $S_{*}^{\text{I}}$ \eqref{eq:SinftyC_higher} obtained from dimensional reduction.

\section{Perturbative solutions and their dimensional reduction}\label{app:Efimov}

In this Appendix, we  study perturbations around the type I solution in the higher-dimensional Einstein theory and dimensionally reduce them to solutions of Einstein-scalar theory.

\subsection{Perturbative solutions}\label{subapp:higher_Efimov}

We consider perturbations
\begin{equation}
	A_k(\tilde{u}) = A_k^{\text{I}}(\tilde{u}) + \delta A_k(\tilde{u})\,,\quad k=1,2\,,
\end{equation}
around the type I solution \eqref{eq:typeChigherdimensional} under the assumption $ \delta A_k(\tilde{u})\slash A_k^{\text{I}}(\tilde{u}) \ll 1 $. Expanding Einstein's equations \eqref{eq:appEinstein1} -- \eqref{eq:appEinstein3} gives at leading order ($D = d+ N$)
\begin{align}
	\delta A_1'' -\coth{\biggl(\frac{\tilde{u}_0-\tilde{u}}{\tilde{\ell}}\biggr)}\,\biggl( \frac{d+D}{\tilde{\ell}}\,\delta A_1'+\frac{N}{\tilde{\ell}}\,\delta A_2'\biggr)+\frac{2\,(D-1)}{\tilde{\ell}^{2}}\csch^{2}{\biggl(\frac{\tilde{u}_0-\tilde{u}}{\tilde{\ell}}\biggr)}\,\delta A_1&= 0\,,\nonumber\\
	\delta A_2'' -\coth{\biggl(\frac{\tilde{u}_0-\tilde{u}}{\tilde{\ell}}\biggr)}\,\biggl( \frac{N+D}{\tilde{\ell}}\,\delta A_2'+\frac{d}{\tilde{\ell}}\,\delta A_1'\biggr)+\frac{2\,(D-1)}{\tilde{\ell}^{2}}\csch^{2}{\biggl(\frac{\tilde{u}_0-\tilde{u}}{\tilde{\ell}}\biggr)}\,\delta A_2&= 0\,,\nonumber\\
	d\,\delta A_1''+N\,\delta A_2''-\frac{2}{\tilde{\ell}}\coth{\biggl(\frac{\tilde{u}_0-\tilde{u}}{\tilde{\ell}}\biggr)}\,(d\,\delta A_1' + N\,\delta A_2')&= 0\,.\label{eq:einsteineqperturbation}
\end{align}
This set of equations is symmetric under the simultaneous exchange $d\leftrightarrow N$ and $\delta A_1 \leftrightarrow \delta A_2$ (the third equation is invariant while the first two are exchanged). Hence, we define
\begin{equation}
	A_{\pm} \equiv d\,A_1 \pm N\,A_2\,.
 \label{eq:Apm_app}
\end{equation}
and parametrize the perturbations in terms of $\delta A_{\pm}$. It follows that the third equation in \eqref{eq:einsteineqperturbation} becomes
\begin{equation}
	\delta A_+''-\frac{2}{\tilde{\ell}}\coth{\biggl(\frac{\tilde{u}_0-\tilde{u}}{\tilde{\ell}}\biggr)}\,\delta A_+' = 0\,,
\end{equation}
which has the general solution
\begin{equation}
	\delta A_{+} = a_1\coth{\biggl(\frac{\tilde{u}_0-\tilde{u}}{\tilde{\ell}}\biggr)} + a_2\,,
\end{equation}
where $a_{1,2}$ are two integration constants. Substituting into the first two equations in \eqref{eq:einsteineqperturbation} and taking their difference gives $a_2 = 0$. The constant $a_1$ can be cancelled by a redefinition $\tilde{u}\rightarrow \tilde{u}-\delta \tilde{u}$ of the radial coordinate in the type I solution \eqref{eq:typeChigherdimensional},
\begin{equation}
	A_{+}^{\text{I}}(\tilde{u}-\delta \tilde{u}) = 	A_{+}^{\text{I}}(\tilde{u})+\frac{D}{\tilde{\ell}}\,\coth{\biggl(\frac{\tilde{u}_0-\tilde{u}}{\tilde{\ell}}\biggr)}\,\delta \tilde{u} + \mathcal{O}(\delta \tilde{u}^{2})\,,
\end{equation}
by setting $\delta \tilde{u} = \frac{\tilde{\ell}}{D}\,a_1$. Therefore without loss of generality we may set $\delta A_+ = 0$. The remaining equation for $\delta A_-$ is obtained by multiplying the first two equations in \eqref{eq:einsteineqperturbation} by $d$, $N$ respectively and then taking their difference. The result is
\begin{equation}
	\delta A_-'' - \frac{D}{\tilde{\ell}}\coth{\biggl(\frac{\tilde{u}_0-\tilde{u}}{\tilde{\ell}}\biggr)}\,\delta A_-'+\frac{2\,(D-1)}{\tilde{\ell}^{2}}\csch^{2}{\biggl(\frac{\tilde{u}_0-\tilde{u}}{\tilde{\ell}}\biggr)}\,\delta A_- = 0\,,
\end{equation}
which has the general solution (this solution is also given in \cite{Aharony:2019vgs})
\begin{equation}
	\delta A_- = \sum_{\pm} C_{\pm}\; _2 F_1\biggl[\frac{\tilde{\beta}_{\pm}}{2},\frac{\tilde{\beta}_{\pm}+1}{2},\tilde{\beta}_{\pm}+\frac{D+1}{2};\,\tanh^{2}{\biggl(\frac{\tilde{u}_0-\tilde{u}}{\tilde{\ell}}\biggr)}\biggr]\tanh^{\tilde{\beta}_{\pm}}{\biggl(\frac{\tilde{u}_0-\tilde{u}}{\tilde{\ell}}\biggr)}\,,
	\label{eq:deltaAmsolutiontext}
\end{equation}
where $C_{\pm}$ are two integration constants and
\begin{equation}
	\tilde{\beta}_{\pm} = -\frac{1}{2}\left(D-1\pm\sqrt{(D-1)(D-9)}\right)
\end{equation}
with $D = d+N$. Expanding \eqref{eq:deltaAmsolutiontext} near the infrared $\tilde{u} = \tilde{u}_0$ gives
\begin{equation}
	\delta A_- =  \sum_{\pm} C_{\pm}\,\biggl(\frac{\tilde{u}_0-\tilde{u}}{\tilde{\ell}}\biggr)^{\tilde{\beta}_{\pm}} + \ldots\,,\quad \tilde{u}\rightarrow \tilde{u}_0^{-}\,,
	\label{eq:deltaAmIRtext}
\end{equation}
which is divergent since $\text{Re}\,\tilde{\beta}_{\pm} < 0$. Therefore the perturbation is not small near the IR and the perturbative approximation breaks down.

A perturbation of a type I solution must match with either a type II or III solution in the regime of validity of the perturbative approximation. In other words, the perturbed solution $A_{\pm}^{C} + \delta A_{\pm} $ coincides with a type II or III solution that have an infinitesimally small $\bar{\alpha}_{\IR}$ or $\alpha_{\IR}$ respectively. The sign of $\delta A_-$ determines which solution $A_{\pm}^{C} + \delta A_{\pm} $ coincides with: it is type III when $\delta A_-$ is positive and type II when $\delta A_-$ is negative. This follows from the IR $\tilde{u}\rightarrow \tilde{u}_0^-$ asymptotics
\begin{equation}
    A_{-}^{\text{III}} - A_{-}^{\text{I}} = N\log{\biggl(\frac{\tilde{u}_0-\tilde{u}}{\bar{\alpha}}\biggr)} + \mathcal{O}(1)\,,\quad A_{-}^{\text{II}} - A_{-}^{\text{I}} = -d\log{\biggl(\frac{\tilde{u}_0-\tilde{u}}{\bar{\alpha}}\biggr)} + \mathcal{O}(1)\,,
\label{eq:deltaAm_A_B_IR}
\end{equation}
which are positive and negative when respectively. By the $\mathbb{Z}_2$-transformation of the Einstein's equations, the type II and III perturbations are related due to \eqref{eq:typeAB_exchange_relation}. Since $A_{-}^{\text{II},\text{III}} \equiv dA_{1}^{\text{II},\text{III}}-NA_{2}^{\text{II},\text{III}}$, we obtain $A_{-}^{\text{II}} =  -A_{-}^{\text{III}}\big\vert_{E}$. Denoting $\delta A_{-}^{\text{III}} \equiv A_{-}^{\text{III}} - A_{-}^{\text{I}}$, $A_{-}^{\text{II}} \equiv A_{-}^{\text{II}} - A_{-}^{\text{I}}$ and using $A_{-}^{\text{I}}\big\vert_{E} = -A_{-}^{\text{I}}$, we observe  that the perturbation towards a type II solution has the opposite sign than a perturbation towards a type III solution with the same parameters,
\begin{equation}
	\delta A_{-}^{\text{II}} =  -\delta  A_{-}^{\text{III}}\big\vert_{E}\,.
	\label{eq:deltaA_AB_relation}
\end{equation}
Now for type II and III solutions, $\delta A_-^{\text{II,III}}$ are given by \eqref{eq:deltaAmIRtext} with two sets of different integration constants $ C_{\pm}^{\text{II,III}}$. By the identity \eqref{eq:deltaA_AB_relation} they are related as $  C_{\pm}^{\text{II}} = - C_{\pm}^{\text{III}}\big\vert_{E}$ since $\tilde{\beta}_{\pm}$ depends only on $D = d+N$ which is symmetric under the exchange $E$. Since type II and III solutions are parametrized by the parameters $\bar{\alpha}_{\scriptscriptstyle\text{IR}}$ and $\alpha_{\scriptscriptstyle\text{IR}}$ respectively, it follows the integration constants must be functions of them and take the form
\begin{equation}
    C_{\pm}^{\text{III}} = C_{\pm}^{\text{III}}(\bar{\alpha}_{\IR}\slash \alpha,\bar{\alpha}_{\IR}\slash \tilde{\ell},\alpha\slash \bar{\alpha})\,,\quad C_{\pm}^{\text{II}} = C_{\pm}^{\text{II}}(\bar{\alpha}_{\IR}\slash \alpha,\bar{\alpha}_{\IR}\slash \tilde{\ell},\alpha\slash \bar{\alpha})\,.
\end{equation}
In the flat space limit $\tilde{\ell}\rightarrow \infty$, the solution \eqref{eq:deltaAmsolutiontext} behaves as
\begin{equation}
    \delta A_-^{\text{II,III}} =  \sum_{\pm} C_{\pm}^{\text{II,III}}\,\biggl(\frac{\tilde{u}_0-\tilde{u}}{\tilde{\ell}}\biggr)^{\tilde{\beta}_{\pm}} + \ldots\,,\quad \tilde{\ell}\rightarrow \infty\,.
\end{equation}
For this limit to be finite, the integration constants must scale with the AdS radius $\tilde{\ell}$ in the same limit as
\begin{equation}
	C_{\pm}^{\text{III}} = \tilde{a}_{\pm}^{\text{III}}\,\biggl(\frac{\tilde{\ell}}{\bar{\alpha}_{\IR}}\biggr)^{\tilde{\beta}_{\pm}}\,,\quad  C_{\pm}^{\text{II}} = \tilde{a}_{\pm}^{\text{II}}\,\biggl(\frac{\tilde{\ell}}{\alpha_{\IR}}\biggr)^{\tilde{\beta}_{\pm}}\,,\quad \tilde{\ell}\rightarrow \infty\,,
	\label{eq:Cpm_alphaIR_1}
\end{equation}
where the dimensionless coefficients are independent of $\tilde{\ell}$ and take the form
\begin{equation}
    \tilde{a}_{\pm}^{\text{III}} = \tilde{a}_{\pm}^{\text{III}}(\bar{\alpha}_{\IR}\slash \bar{\alpha},\alpha\slash \bar{\alpha})\,,\quad \tilde{a}_{\pm}^{\text{II}} = \tilde{a}_{\pm}^{\text{II}}(\alpha_{\IR}\slash \alpha,\alpha\slash \bar{\alpha})\,.
\label{eq:tildea_coefficients}
\end{equation}
It follows that the flat space solution is given by
\begin{equation}
	\lim_{\tilde{\ell}\rightarrow \infty}\delta A_-^{\text{III}} =  \sum_{\pm} \tilde{a}_{\pm}^{\text{III}}\,\biggl(\frac{\tilde{u}_0-\tilde{u}}{\bar{\alpha}_{\IR}}\biggr)^{\tilde{\beta}_{\pm}}\,,\quad \lim_{\tilde{\ell}\rightarrow \infty}\delta A_-^{\text{II}} =  \sum_{\pm} \tilde{a}_{\pm}^{\text{II}}\,\biggl(\frac{\tilde{u}_0-\tilde{u}}{\alpha_{\IR}}\biggr)^{\tilde{\beta}_{\pm}}\,,
 \label{eq:deltaA_flat_space}
\end{equation}
which coincides with \cite{Kol:2002xz}.\footnote{It also coincides with the IR expansion \eqref{eq:deltaAmIRtext}, which shows that the geometry near the endpoint is approximately flat as expected.} The flat space solution controls the near type I limit $\alpha_{\IR},\bar{\alpha}_{\IR}\rightarrow 0$ of the perturbations, because in both $\alpha_{\IR},\bar{\alpha}_{\IR}\rightarrow 0$ and $\tilde{\ell}\rightarrow \infty$ limits, the ratios $\alpha_{\IR}\slash \tilde{\ell}$, $\bar{\alpha}_{\IR}\slash \tilde{\ell}$ go to zero. Hence, from \eqref{eq:Cpm_alphaIR_1} we obtain that
\begin{equation}
	C_{\pm}^{\text{III}} = a_{\pm}^{\text{III}}\,\biggl(\frac{\tilde{\ell}}{\bar{\alpha}_{\IR}}\biggr)^{\tilde{\beta}_{\pm}}+\ldots\,,\quad C_{\pm}^{\text{II}} = a_{\pm}^{\text{II}}\,\biggl(\frac{\tilde{\ell}}{\alpha_{\IR}}\biggr)^{\tilde{\beta}_{\pm}}+\ldots\,,\quad \alpha_{\IR},\bar{\alpha}_{\IR}\rightarrow 0\,,
	\label{eq:Cpm_alphaIR}
\end{equation}
where we have assumed\footnote{This is in agreement with numerics.} that the coefficients \eqref{eq:tildea_coefficients} have the finite limit
\begin{equation}
    a_\pm^{\text{III}} =  a_\pm^{\text{III}}(\alpha\slash \bar{\alpha}) \equiv \lim_{\bar{\alpha}_{\IR}\rightarrow 0}\tilde{a}_{\pm}^{\text{III}}(\bar{\alpha}_{\IR}\slash \bar{\alpha},\alpha\slash \bar{\alpha})\,,\quad a_\pm^{\text{II}} =  a_\pm^{\text{II}}(\alpha\slash \bar{\alpha}) \equiv \lim_{\alpha_{\IR}\rightarrow 0}\tilde{a}_{\pm}^{\text{III}}(\alpha_{\IR}\slash \alpha,\alpha\slash \bar{\alpha})\,.
\end{equation}

\subsection{Dimensional reduction}\label{subapp:Efimov_reduction}

We now dimensionally reduce the perturbative solution \eqref{eq:deltaAmsolutiontext} to a perturbative solution $\delta S(\varphi)$ of \eqref{Sequation} of the Einstein-scalar theory. Since $\delta A_+ = 0$, perturbations around the type I solution are determined by $\delta A_-$ so that from \eqref{eq:Apm_app} we obtain
\begin{equation}
	\delta A_1(\tilde{u}) = \frac{1}{2d}\,\delta A_-(\tilde{u})\,,\quad \delta A_2(\tilde{u}) = -\frac{1}{2N}\,\delta A_-(\tilde{u})\,,
\end{equation}
which combined with \eqref{SWA1A2app} gives
\begin{equation}
    \delta S(\varphi) =b\,\frac{d}{du}\delta A_-(\tilde{u}) = b\,e^{\gamma_{\text{c}}\varphi}\,\delta A_-'(\tilde{u})\,,
    \label{eq:deltaSWhigh}
\end{equation}
where we have used the relation $\frac{d\tilde{u}}{du} = e^{\gamma_{\text{c}}\varphi}$ \eqref{dutugammac}. We now expand this in the IR region $\tilde{u}\rightarrow \tilde{u}_0^-$ which according to the relation \eqref{eq:varphiutilderelation} corresponds to $\varphi\rightarrow \infty$. By substituting the IR expansion \eqref{eq:deltaAmIRtext}, we obtain
\begin{equation}
	\delta A_-'(\tilde{u}) =  \frac{\tilde{\beta}_{\pm}}{\tilde{\ell}}\sum_{\pm} C_{\pm}\,\biggl(\frac{\tilde{u}_0-\tilde{u}}{\tilde{\ell}}\biggr)^{\tilde{\beta}_{\pm}-1} + \ldots\,,\quad \tilde{u}\rightarrow \tilde{u}_0^{-}\,
	\label{eq:deltaAmIRtext_der}
\end{equation}
By expanding \eqref{eq:varphiutilderelation}, we obtain
\begin{equation}
	\biggl(\frac{\tilde{u}-\tilde{u}_0}{\tilde{\ell}}\biggr)^{-1} = \frac{\tilde{\ell}}{\bar{\alpha}}\sqrt{\frac{N-1}{d+N-1}}\,e^{(b-\gamma_{\text{c}})\varphi}+\ldots\,,\quad \tilde{u}\rightarrow \tilde{u}_0^-\,,
	\label{eq:varphiutilderelation_expanded}
\end{equation}
and by substituting to \eqref{eq:deltaAmIRtext_der}, we obtain
\begin{equation}
	\delta A_{-}'(\tilde{u}) = \sum_{\pm}B_{\pm}C_{\pm}\,e^{-(\tilde{\beta}_{\pm}-1)(b-\gamma_{\text{c}})\,\varphi} + \ldots\,,\quad \varphi\rightarrow \infty\,,
	\label{eq:deltaApprimeexp}
\end{equation}
where we have defined
\begin{equation}
	B_{\pm} \equiv \frac{b\tilde{\beta}_{\pm}}{\tilde{\ell}}\left(\frac{d+N-1}{N-1}\,\frac{\bar{\alpha}^{2}}{\tilde{\ell}^2}\right)^{(\tilde{\beta}_{\pm}-1)\slash 2}\,.
\end{equation}
Now one can prove the identity
\begin{equation}
	(\tilde{\beta}_{\pm} -1)(b-\gamma_{\text{c}}) = \gamma_{\text{c}}-\beta_{\pm}\,,
	\label{eq:betat_beta_relation}
\end{equation}
where $\beta_{\pm}$ are the exponents \eqref{eq:betapmtext} found by solving the equations of motion of the lower-dimensional theory. Therefore substituting \eqref{eq:deltaApprimeexp} to \eqref{eq:deltaSWhigh}, we obtain
\begin{equation}
	\delta S(\varphi) = S_{+}\,e^{\beta_{+}\varphi} +S_{-}\,e^{\beta_{-}\varphi}+ \ldots\,,\quad \varphi\rightarrow \infty\,,
	\label{eq:deltaSWhighexp}
\end{equation}
where we have defined
\begin{equation}
	S_{\pm} \equiv B_{\pm}C_{\pm} = \frac{b\tilde{\beta}_{\pm}}{\bar{\alpha}}\biggl(\frac{d+N-1}{N-1}\biggr)^{(\tilde{\beta}_{\pm}-1)\slash 2}\,\biggl(\frac{\bar{\alpha}}{\tilde{\ell}}\biggr)^{\tilde{\beta}_{\pm}}C_{\pm}\,.
	\label{eq:SpmCpmrelation}
\end{equation}
This equation relates the integration constants $S_{\pm}$ in the equation for $\delta S$ to the integration constants $C_{\pm}$ of the higher-dimensional theory. By substituting \eqref{eq:Cpm_alphaIR} to \eqref{eq:SpmCpmrelation} gives
\begin{equation}
	S_{\pm}^{\text{III}} \propto\frac{1}{\bar{\alpha}}\biggl(\frac{\bar{\alpha}}{\bar{\alpha}_{\text{IR}}}\biggr)^{\tilde{\beta}_{\pm}} \,,\quad  S_{\pm}^{\text{II}}  \propto \frac{1}{\bar{\alpha}}\biggl(\frac{\bar{\alpha}}{\alpha}\biggr)^{\tilde{\beta}_{\pm}}\left(\frac{\alpha}{\alpha_{\IR}}\right)^{\tilde{\beta}_{\pm}}\,,\quad \alpha_{\IR},\bar{\alpha}_{\IR}\rightarrow 0\,.
	\label{eq:SpmIRparameters}
\end{equation}
For type III solutions, we have by \eqref{phi0upstairs} that $ \log{\frac{\bar{\alpha}}{\bar{\alpha}_{\IR}}} = \frac{b^{2}-b_{\text{c}}^{2}}{b}\,\varphi_0$. Substituting to \eqref{eq:SpmIRparameters} and using the identity $\frac{b^{2}-b_{\text{c}}^{2}}{b}\,\tilde{\beta}_{\pm} = b-\beta_{\pm}$ (following from \eqref{eq:betat_beta_relation}) gives
\begin{equation}
	S_{\pm}^{\text{III}} \propto  e^{(b-\beta_{\pm})\,\varphi_0}\,,\quad \varphi_0\rightarrow \infty\,.
\end{equation}
Similarly for type II solutions,  we obtain  from \eqref{eq:typeB_reduction_coefficients} that $\frac{\alpha}{\alpha_{\IR}} \propto (- S_{\infty,1}^{\text{II}})^{1\slash 2}$ which when substituted to \eqref{eq:SpmIRparameters} gives
\begin{equation}
	S_{\pm}^{\text{II}}\propto (-S_{\infty,1}^{\text{II}})^{P_{\pm}}\,,\quad P_{\pm} =  \frac{1}{2b}\frac{b-\beta_{\pm}}{1-(b_{\text{c}}\slash b)^{2}}\,,\quad S_{\infty,1}^{\text{II}}\rightarrow -\infty\,.
\end{equation}

\section{On-shell UV asymptotics in the Einstein-scalar theory}\label{app:UVasymptotics}

In this appendix, we derive the on-shell UV asymptotic of the functions $S,W,T$ and $U$ by directly solving the second-order equation \eqref{Sequation} for $S$. This provides an independent verification of the expansions derived in \cite{Ghosh:2017big,Ghosh:2020qsx}.

We  assume that the potential has a maximum at $ \varphi = 0 $ where it has the expansion
\begin{equation}
V(\varphi) = -\frac{d(d-1)}{\ell^{2}}-\frac{\Delta_-(d-\Delta_-)}{2\ell^{2}}\,\varphi^{2} + \frac{1}{3!}\,\frac{V_3}{\ell^2}\,\varphi^3 + \frac{1}{4!}\,\frac{V_4}{\ell^2}\,\varphi^4 + \mathcal{O}(\varphi^{5})\,.
\label{eq:potential_UV_app}
\end{equation}
Wesolve the equation \eqref{Sequation} for $S(\varphi)$ in this regime. We search for a solution that satisfies $S(0) = 0$ and first consider a Taylor series in $\varphi$ with integer coefficients. In this way, we find the particular solution\footnote{There is also a second solution which corresponds to the alternative quantization not considered in this paper (see the discussion above equation \eqref{eq:final_uv_S_app}).}
\begin{equation}
	S_{(0)}(\varphi) = \frac{\Delta_-}{\ell}\,\varphi\,\biggl(1 +\frac{V_3}{2\Delta_-(3\Delta_--d)} \,\varphi+\mathcal{O}(\varphi^2)\biggr)\,,
	\label{eq:S0UVexp}
\end{equation}
where the $\varphi^2$ term is also determined by $V_3,V_4$ while further subleading terms depend on $\mathcal{O}(\varphi^{5})$ terms of the potential. We now look for other subleading corrections to this solution using the ansatz
\begin{equation}
	S(\varphi) = S_{(0)}(\varphi) + \delta S(\varphi)\,,
\end{equation}
where $\delta S(\varphi)$ is treated as a small perturbation relative to $S_{(0)}(\varphi) $ and linearize the equation \eqref{Sequation} in $\delta S$. The linearization is done in Appendix \ref{app:solutionsunboundedscalar} and the result is the equation \eqref{linearSequation1}. In the current case $S_{(0)}(\varphi)$ \eqref{eq:S0UVexp} is a solution so that the non-homogeneous term vanishes $K(\varphi) = 0$. At leading order, the equation \eqref{linearSequation1} therefore  becomes explicitly
\begin{equation}
	\delta S''(\varphi) - \biggl(\frac{d-\Delta_-+2}{\Delta_-}\,\varphi^{-1} + \mathcal{O}(\varphi)\biggr)\,\delta S'(\varphi) + \biggl(\frac{(d-\Delta_-)(\Delta_-+2)}{\Delta_-^{2}}\,\varphi^{-2}+\mathcal{O}(1)\biggr)\,\delta S(\varphi) = 0\,,
	\label{eq:linS_UV}
\end{equation}
where the subleading terms are determined by \eqref{eq:S0UVexp}. By substituting the ansatz
\begin{equation}
    \delta S = X_0\,\varphi^{\beta}\,[1 + X_1\,\varphi + X_2\,\varphi^2 + \mathcal{O}(\varphi^3)]
\end{equation}
with $\beta>0$, we obtain the equation
\begin{equation}
	X_0\,\biggl(\beta+1-\frac{d}{\Delta_-}\biggr)\biggl(\beta-1-\frac{2}{\Delta_-}\biggr)\,\varphi^{\beta-2}+\mathcal{O}(\varphi^{\beta}) = 0\,,
	\label{eq:linSequation_UV_midstep}
\end{equation}
which up to order $\mathcal{O}(\varphi^{\beta})$ has the two solutions $\beta = \beta_{1,2}$ where
\begin{equation}
	\beta_{1} = -1 + \frac{d}{\Delta_-}\,,\quad \beta_{2} = 1 + \frac{2}{\Delta_-}\,.
\end{equation}
The overall normalization $X_0$ is an integration constant while the subleading terms in \eqref{eq:linSequation_UV_midstep} determine $X_1,X_2$ in terms of coefficients of the potential appearing in \eqref{eq:potential_UV_app}. Because equation \eqref{eq:linS_UV} is linear, the general solution is therefore given by
\begin{align}
	\delta S(\varphi) &= s_1\,\varphi^{-1+d\slash \Delta_-}\,\biggl(1 - \frac{(\Delta_-+d)\,V_3}{2\Delta_-^2(3\Delta_--d)}\,\varphi+ \mathcal{O}(\varphi^2)\biggr)\nonumber\\
 &+ s_2\,\varphi^{1+2\slash \Delta_-}\,\biggl(1 - \frac{(d\,(\Delta_--2)+(\Delta_-+2)^2)\,V_3}{2\Delta_-^2(3\Delta_--d)(3\Delta_--d+2)}\,\varphi+ \mathcal{O}(\varphi^2)\biggr)\,,
	\label{eq:deltaS_UV_solution}
\end{align}
where $s_{1,2}$ are two integration constants (corresponding to $X_0$). We have included the leading $\varphi$-corrections and we may also compute the $\varphi^2$-corrections in terms of $V_{3,4}$. Notice that the power-laws in \eqref{eq:deltaS_UV_solution} are subleading to $S_{(0)}(\varphi)= \frac{\Delta_-}{\ell}\,\varphi + \mathcal{O}(\varphi^2)$ when $0 < \Delta_- < \frac{d}{2}$, which for the dimension $\Delta = d-\Delta_-$ of the dual operator, amounts to $\frac{d}{2} < \Delta < d$. The lower bound is always valid in the standard quantization $ \Delta>\frac{d}{2}$ assumed in this paper, which implies that the first integration constant $s_1$ is free. However, when $\Delta > d$ corresponding to $\Delta_-< 0 $, the second integration constant must be set to zero $s_2 = 0$.

Higher-order $\mathcal{O}(\delta S^2)$ will modify the solution at subleading orders and they appear as corrections that are second-order in products of $s_1$ and $s_2$. Assuming $0 < \Delta_- < \frac{d}{2}$, up to first order in $s_{1}$ and $s_2$, we therefore  obtain the expansion
\begin{align}
	S(\varphi) &= \frac{\Delta_-}{\ell}\,\varphi\,\biggl(1 +\frac{V_3}{2\Delta_-(3\Delta_--d)} \,\varphi+\mathcal{O}(\varphi^2)\biggr)\nonumber\\
 &+s_1\,\varphi^{-1+d\slash \Delta_-}\,\biggl(1 - \frac{(\Delta_-+d)\,V_3}{2\Delta_-^2(3\Delta_--d)}\,\varphi+ \mathcal{O}(\varphi^2)\biggr)\nonumber\\
 &+ s_2\,\varphi^{1+2\slash \Delta_-}\,\biggl(1 - \frac{(d\,(\Delta_--2)+(\Delta_-+2)^2)\,V_3}{2\Delta_-^2(3\Delta_--d)(3\Delta_--d+2)}\,\varphi+ \mathcal{O}(\varphi^2)\biggr)\nonumber\\
 &+\mathcal{O}(s_1^2)+\mathcal{O}(s_2^2)+\mathcal{O}(s_1s_2)\,,\quad \varphi\rightarrow 0\,.
 \label{eq:final_uv_S_app}
\end{align}
Using this expansion, we may calculate expansions for rest of the functions $W$ and $T$. We obtain
\begin{align}
	W(\varphi) &= \frac{2(d-1)}{\ell}+\frac{\Delta_-}{2\ell}\,\varphi^{2} +\mathcal{O}(\varphi^{3})+\frac{\Delta_-}{d}\,s_1\,\varphi^{d\slash \Delta_-}\,[1 + \mathcal{O}(\varphi)]\nonumber\\
 &+\frac{2(d-1)(2\Delta_--d+2)}{d\Delta_-}\,s_2\,\varphi^{2\slash \Delta_-}[1 + \mathcal{O}(\varphi)]\nonumber\\
 &+\mathcal{O}(s_1^2)+\mathcal{O}(s_2^2)+\mathcal{O}(s_1s_2)\,,\quad \varphi\rightarrow 0\,.
  \label{eq:final_uv_W_app}
\end{align}
The terms appearing explicitly in this expression are determined by terms of order $\mathcal{O}(\varphi^2)$ inside the brackets in the expansion of $S$ \eqref{eq:final_uv_S_app} and involve non-trivial cancellations. Similarly, we obtain
\begin{align}
    T(\varphi) &= \frac{2(d-1)(2\Delta_--d+2)}{\ell\Delta_-}\,s_2\,\varphi^{2\slash \Delta_-}[1 + \mathcal{O}(\varphi)]+\mathcal{O}(\varphi^{2+d\slash \Delta_-})+\mathcal{O}(\varphi^3)\nonumber\\
 &+\mathcal{O}(s_1^2)+\mathcal{O}(s_2^2)+\mathcal{O}(s_1s_2)\,,\quad \varphi\rightarrow 0\,.
    \label{eq:final_uv_T_app}
\end{align}
We now relate the integration constants $s_{1,2}$ to the parameters appearing in the UV expansions of $(\varphi,A)$ given in \eqref{Aasymp} and \eqref{phyasymp}. At leading order, we obtain from \eqref{phyasymp} the expansion
\begin{equation}
	e^{u\slash \ell} = \frac{\varphi_-^{-1\slash \Delta_-}}{\ell}\,\varphi^{1\slash \Delta_-}\biggl(1-\frac{1}{\Delta_-}\,\varphi_{+}\varphi_-^{-\Delta_+\slash \Delta_-}\varphi^{-2+d\slash \Delta_-} + \ldots\biggr)\,,\quad \varphi \rightarrow 0\,,
	\label{eq:phi_UV_inverted}
\end{equation}
which implies that
\begin{equation}
	S(\varphi) = \dot{\varphi}(u) = \frac{\Delta_-}{\ell}\,\varphi + \frac{d-2\Delta_-}{\ell}\,\varphi_{+}\varphi_{-}^{-\Delta_+\slash \Delta_-}\,\varphi^{-1+d\slash \Delta_-}  +\ldots\,,\quad \varphi \rightarrow 0\,,
\end{equation}
where we used $\Delta_+ = d-\Delta_-$ to write $\Delta_+\slash \Delta_- = -1+d\slash \Delta_-$. Therefore,  comparing with \eqref{eq:S0UVexp}, we observe that the coefficient of $\varphi$ matches. Comparing with \eqref{eq:deltaS_UV_solution} gives
\begin{equation}
	s_1 = \frac{2\Delta_+-d}{\ell}\,\varphi_{+}\varphi_{-}^{-\Delta_+\slash \Delta_-} = \frac{d}{\Delta_-}\frac{\mathcal{C}}{\ell}\,,
\end{equation}
where $\mathcal{C}$ is defined in \eqref{eq:curlyC}.

To fix $s_2$, we   compute the expansion of $W(\varphi)$. After fixing the metric on the conformal boundary, the expansion \eqref{Aasymp} takes the form
\begin{equation}
	A(u) = -\frac{u}{\ell} -\frac{\kappa \ell^{2}}{4(d-1)} \,e^{2u\slash \ell} + \ldots\,,\quad u\rightarrow -\infty\,,
\end{equation}
where we used $e^{-2c\slash \ell} = \frac{\ell^{2}}{4\alpha^{2}} = \frac{\kappa \ell^{2}}{4(d-1)} $. By using \eqref{eq:phi_UV_inverted}, we obtain
\begin{equation}
	W(\varphi) = -2(d-1)\,\dot{A}(u) = \frac{2(d-1)}{\ell} + \frac{\mathcal{R}}{d\ell}\, \varphi^{2\slash \Delta_-} +\ldots\,,\quad \varphi\rightarrow 0\,,
\end{equation}
with $\mathcal{R} = d\kappa\,\varphi_{-}^{-2\slash \Delta_-}$. Comparing with the coefficient of $\varphi^{2\slash \Delta_-}$ in \eqref{eq:final_uv_W_app}, we can identify
\begin{equation}
	s_2 = \frac{\Delta_-}{2(d-1)(2\Delta_--d+2)}\frac{\mathcal{R}}{\ell}\,.
\end{equation}
Therefore with the above identifications of integration constants, we obtain the expansion
\begin{align}
	S(\varphi) &= \frac{\Delta_-}{\ell}\,\varphi\,[1 + \mathcal{O}(\varphi)]+\frac{d}{\Delta_-}\frac{\mathcal{C}}{\ell}\,\varphi^{-1+d\slash \Delta_-}\,[1 + \mathcal{O}(\varphi)]\\
	&\;\;\;\; +\frac{\Delta_-}{2(d-1)(2\Delta_--d+2)}\frac{\mathcal{R}}{\ell}\,\varphi^{1+2\slash \Delta_-}[1+\mathcal{O}(\varphi)]\nonumber\\
 &\;\;\;\;+\mathcal{O}(\mathcal{C}^2)+\mathcal{O}(\mathcal{R}^2)+\mathcal{O}(\mathcal{C}\,\mathcal{R})\,,\quad \varphi\rightarrow 0\,.
\end{align}
Similarly, we obtain
\begin{align}
    W(\varphi) &= \frac{2(d-1)}{\ell}+\frac{\Delta_-}{2\ell}\,\varphi^{2}+\frac{\mathcal{R}}{d\ell}\,\varphi^{2\slash \Delta_-}[1 + \mathcal{O}(\varphi)] +\frac{\mathcal{C}}{\ell}\,\varphi^{d\slash \Delta_-}\,[1 + \mathcal{O}(\varphi)]+\mathcal{O}(\varphi^{3})\nonumber\\
 &\;\;\;\;+\mathcal{O}(\mathcal{C}^2)+\mathcal{O}(\mathcal{R}^2)+\mathcal{O}(\mathcal{C}\,\mathcal{R})\,,\quad \varphi\rightarrow 0\,,\\
    T(\varphi) &=  \frac{\mathcal{R}}{\ell^{2}}\,\varphi^{2\slash \Delta_-}[1 + \mathcal{O}(\varphi)]+\mathcal{O}(\varphi^{2+d\slash \Delta_-})+\mathcal{O}(\varphi^3)\nonumber\\
 &\;\;\;\;+\mathcal{O}(\mathcal{C}^2)+\mathcal{O}(\mathcal{R}^2)+\mathcal{O}(\mathcal{C}\,\mathcal{R})\,,\quad \varphi\rightarrow 0\,.
\end{align}
These expansion match with \cite{Ghosh:2017big}, but we have also determined the coefficient of the $\varphi^{1+2\slash \Delta_-}$ term in $S(\varphi)$ explicitly in terms of $\mathcal{R}$.

The terms involving higher powers of $\mathcal{R}$ and $\mathcal{C}$ are multiplied by appropriate subleading powers of $\varphi^{2\slash \Delta_-}$ and $\varphi^{d\slash \Delta_-}$ respectively, because they arise from $\mathcal{O}(\delta S^2)$ terms where $\delta S$ is given by \eqref{eq:deltaS_UV_solution}. In particular, $W(\varphi)$ contains a quadratic term in $\mathcal{R}$ of the form $\mathcal{R}^2\,\varphi^{4\slash \Delta_-}$, which in $d=4$ comes at the same order as the $\mathcal{C}\,\varphi^{d\slash \Delta_-}$ term.

\paragraph{Expansion of $U$.} The expansion of $U(\varphi)$ is obtained by solving the first-order equation \eqref{Uequationtext} which may be written as
\begin{equation}
	U'(\varphi)-\frac{d-2}{2\,(d-1)}\,\frac{W(\varphi)}{S(\varphi)}\,U(\varphi)+\frac{2}{d}\frac{1}{S(\varphi)}=0\,.
 \label{eq:Ueq_app}
\end{equation}
The general solution is given by
\begin{equation}
    U(\varphi) = \ell\,\widetilde{\mathcal{B}}\,e^{f(\varphi)}+h(\varphi)\,,
    \label{eq:gen_U}
\end{equation}
where $\widetilde{\mathcal{B}}$ is a free integration constant and
\begin{equation}
    f(\varphi) = \frac{d-2}{2\,(d-1)}\int^\varphi_1 ds\,\frac{W(s)}{S(s)}\,,\quad h(\varphi) = -\frac{2}{d}\,e^{f(\varphi)}\int^\varphi_1 ds\,\frac{e^{-f(s)}}{S(s)}\,.
    \label{eq:f_h_functions}
\end{equation}
Using the on-shell UV asymptotics above, we can expand
\begin{align}
    &\frac{d-2}{2\,(d-1)}\,\frac{W(\varphi)}{S(\varphi)}\nonumber\\
    &= \frac{d-2}{\Delta_-}\,\varphi^{-1}\,[1 + \mathcal{O}(\varphi)] + \frac{(d-2)(\Delta_--d+1)}{d\Delta_-\,(d-1)(2\Delta_--d+2)}\,\mathcal{R}\,\varphi^{-1+2\slash \Delta_-}\,[1 + \mathcal{O}(\varphi)]\nonumber\\
    &+\frac{d-2}{2\,(d-1)\,\Delta_-}\,\mathcal{C}\,\varphi^{-1+d\slash \Delta_-}\,[1 + \mathcal{O}(\varphi)]\,.
\end{align}
Substituting to \eqref{eq:f_h_functions} gives
\begin{equation}
    e^{f(\varphi)} = f_0\,\varphi^{(d-2)\slash \Delta_-}\,\biggl(1 +\frac{(d-2)(\Delta_--d+1)}{2d\,(d-1)(2\Delta_--d+2)}\,\mathcal{R}\,\varphi^{2\slash \Delta_-} + \mathcal{O}(\mathcal{C}\,\varphi^{d\slash \Delta_-})+ \mathcal{O}(\varphi)\biggr)\,,
    \label{eq:f_function_exp}
\end{equation}
where $f_0$ is a constant (the integral function evaluated at the lower endpoint of the integral) that will be absorbed to $\widetilde{\mathcal{B}}$. Similarly, we have the expansion
\begin{align}
        \frac{2}{d}\frac{1}{S(\varphi)} &= \frac{2\ell}{d\Delta_-}\,\varphi^{-1}\,[1 + \mathcal{O}(\varphi)]  -\frac{\ell\,\mathcal{R}}{d(d-1)(2\Delta_--d+2)\Delta_-}\,\varphi^{-1+2\slash \Delta_-} \,[1 + \mathcal{O}(\varphi)]\\
        &+\mathcal{O}(\mathcal{C}\,\varphi^{-1+d\slash \Delta_-})\,,
        \label{eq:nonhom_expansion}
\end{align}
which together with \eqref{eq:f_function_exp} gives for $d\neq 4$ the expansion
\begin{equation}
    h(\varphi) = \ell\,\biggl(\frac{2}{d(d-2)}-\frac{\mathcal{R}}{d^2(d^2-5d+4)}\,\varphi^{2\slash \Delta_-} + \mathcal{O}(\varphi^2)+\mathcal{O}(\varphi^{4\slash \Delta_-})+\mathcal{O}(\varphi^{d\slash \Delta_-})\biggr)\,,\quad d \neq 4\,.
\end{equation}
When $d = 4$, the expansion includes a logarithmic term and takes the form
\begin{equation}
    h(\varphi) = \ell\,\biggl(\frac{1}{4}+\frac{\mathcal{R}}{48\Delta_-}\,\varphi^{2\slash \Delta_-}\log{\varphi} + \frac{(\Delta_--3)\,\mathcal{R}}{96\,(\Delta_--1)}\,\varphi^{2\slash \Delta_-} + \mathcal{O}(\varphi^2)+\mathcal{O}(\varphi^{4\slash \Delta_-})\biggr)\,,\quad d = 4\,.
    \label{eq:h_exp_d_is_4}
\end{equation}
In $d = 4$ the expansion of $e^{f(\varphi)}$ \eqref{eq:f_function_exp} includes a $\varphi^{2\slash \Delta_-}$-term as well. Therefore in the expansion of $U(\varphi)$ \eqref{eq:gen_U}, the $\varphi^{2\slash \Delta_-}$-term coming from $h(\varphi)$ \eqref{eq:h_exp_d_is_4} can be absorbed to the integration constant by defining $\mathcal{B} \equiv \widetilde{\mathcal{B}}f_0 + \frac{(\Delta_--3)\,\mathcal{R}}{96\,(\Delta_--1)} $. In $d = 4$, we therefore obtain the total expansion
\begin{align}
    U(\varphi) &=  \frac{\ell}{4}\,[1 +  \mathcal{O}(\varphi^2)]+\ell\,\mathcal{B}\,\varphi^{2\slash \Delta_-}[1+\mathcal{O}(\varphi) + \mathcal{O}(\varphi^{2\slash \Delta_-})+ \mathcal{O}(\varphi^{d\slash \Delta_-})]\nonumber\\
    &+ \frac{\ell\,\mathcal{R}}{48\Delta_-}\,\varphi^{2\slash \Delta_-}[\log{\varphi} + \mathcal{O}(\varphi^{2\slash \Delta_-})]\,.
\end{align}

\section{On-shell action of the Einstein-scalar theory}\label{app:onshelllowerdimensional}

In this appendix, we provide details for the computation of the on-shell action of the Einstein-scalar theory.

\paragraph{Regularized free energy.}

Consider the Euclidean Einstein-scalar action \eqref{einsteinscalar} supplemented with a Gibbons--Hawking--York term at the cut-off surface $ u = \log{\epsilon} $ which makes the variational principle for the metric well defined. For a metric of the form \eqref{bulkansatz}, the regularized on-shell action takes the form \cite{Ghosh:2017big,Ghosh:2020qsx}
\begin{equation}
	I_{\text{on-shell}}^{\text{reg}} = M_{\text{p}}^{d-1}\int d^{d}x\sqrt{\zeta}\,\biggl( \frac{2\kappa}{d}\int_{\log{\epsilon}}^{u_0}du\,e^{(d-2)A(u)}+2\,(d-1)\bigl[e^{dA(u)}\,\dot{A}(u)\bigr]\bigg\lvert^{u = u_0}_{u = \log{\epsilon}}\,\biggr)\,,
	\label{onshellactionstart}
\end{equation}
where $\epsilon\rightarrow 0^+$ is the regulator and we use the notation $f(u)\lvert_{u=u_1}^{u=u_2}\, \equiv f(u_2)-f(u_1)$. Now notice that the integrand and the boundary term may be written in terms of the functions \eqref{WST} as
\begin{equation}
	du\,e^{(d-2)A(u)} = d\varphi\,S(\varphi)^{-1}\,T(\varphi)^{-\frac{d}{2}+1}\,,\quad 2\,(d-1)\,e^{dA(u)}\dot{A}(u) = -W(\varphi)\,\biggl(\frac{T(\varphi)}{d\kappa}\biggr)^{-\frac{d}{2}}\,.
\end{equation}
The regularized free energy $F_{\text{reg}} = -I_{\text{on-shell}}^{\text{reg}}$ becomes
\begin{equation}
	F_{\text{reg}} = M_{\text{p}}^{d-1}\,\widetilde{\Omega}_d\,\biggl[- \frac{2}{d}\int_{\varphi_{\UV}}^{\varphi_{\scriptscriptstyle\text{IR}}}d\varphi\,S(\varphi)^{-1}\,T(\varphi)^{-\frac{d}{2}+1}+\Bigl(W(\varphi)\,T(\varphi)^{-\frac{d}{2}}\Bigr)\bigg\lvert^{\varphi = \varphi_{\IR}}_{\varphi = \varphi_{\UV}}\,\biggr]\,,
 \label{eq:Freg_midstep_app}
\end{equation}
where the shorthand $\varphi_{\UV} \equiv \varphi(\log{\epsilon}) $, $\varphi_{\IR} \equiv \varphi(u_0) $ and the dimensionless coefficient
\begin{equation}
	\widetilde{\Omega}_d \equiv \frac{2\,d^{\frac{d}{2}}(d-1)^{\frac{d}{2}}\pi^{\frac{d+1}{2}}}{\Gamma\bigl(\frac{d+1}{2}\bigr)}\,.
\label{eq:Omega_tilde_d}
\end{equation}
The remaining $\varphi$-integral can be written as a boundary term \cite{Ghosh:2017big,Ghosh:2018qtg}
\begin{equation}
	-\frac{2}{d}\int_{\varphi_{\UV}}^{\varphi_{\scriptscriptstyle\text{IR}}}d\varphi\,S(\varphi)^{-1}\,T(\varphi)^{-\frac{d}{2}+1} = U(\varphi)\,T(\varphi)^{-\frac{d}{2}+1}\bigg\lvert^{\varphi = \varphi_{\scriptscriptstyle\text{IR}}}_{\varphi = \varphi_{\UV}},
	\label{varphiintegralboundary}
\end{equation}
where the function $ U(\varphi) $ is defined as the solution of the first order differential equation
\begin{equation}
	SU'-\frac{d-2}{2(d-1)}\,WU +\frac{2}{d} = 0\,.
	\label{Uequation}
\end{equation}
Substituting to \eqref{eq:Freg_midstep_app}, the regularized free energy takes the form
\begin{equation}
	F_{\text{reg}} = M_{\text{p}}^{d-1}\,\widetilde{\Omega}_d\,\biggl(W(\varphi)\,T(\varphi)^{-\frac{d}{2}}+ U(\varphi)\,T(\varphi)^{-\frac{d}{2}+1}\biggr)\bigg\lvert^{\varphi = \varphi_{\scriptscriptstyle\text{IR}}}_{\varphi = \varphi_{\UV}}\ .
	\label{Freg1}
\end{equation}

\paragraph{Vanishing of IR boundary terms.}

We now show that the IR boundary term at $ \varphi = \varphi_{\IR} $ in \eqref{Freg1} vanishes for type I, II and III solutions given a potential with exponential asymptotics
\begin{equation}
	V(\varphi) = V_{\infty}\,e^{2b\varphi} +V_{\infty,1}\,e^{2\gamma\varphi} + \ldots\,, \quad \varphi \rightarrow \infty\,.
\end{equation}
In Appendix \ref{reduced:typeA}, we show that the on-shell IR asymptotics of type III solutions are given by
\begin{equation}
	W(\varphi) = \frac{W_0}{\sqrt{\varphi_{0}-\varphi}} + \mathcal{O}(\sqrt{\varphi_{0}-\varphi})\,, \quad T(\varphi) = \frac{T_0}{\varphi_{0}-\varphi} + \mathcal{O}(\sqrt{\varphi_{0}-\varphi})\,,\quad \varphi \rightarrow \varphi_0^{-}\,,
\end{equation}
where the expressions for $ W_0 $ and $ T_0 $ are given in \eqref{W0expressionapp} and \eqref{T0U0expressionapp}. For the type I solution, we have derived in Appendix \ref{app:solutionsunboundedscalar} the asymptotics
\begin{equation}
	W(\varphi) = W_{\infty}^{\text{I}}\,e^{b\varphi} + \ldots\,, \quad T(\varphi) = T_{\infty}^{\text{I}}\,e^{2b\varphi} + \ldots\,, \quad \varphi \rightarrow \infty\,,
	\label{WTasymptoticsC}
\end{equation}
where $W_{\infty}^{\text{I}}$ and $T_{\infty}^{\text{I}}$ are given in \eqref{typeCleading}. Hence, the first IR term in \eqref{Freg1} has the expansion
\begin{equation}
  W(\varphi)\,T(\varphi)^{-\frac{d}{2}} = \left\{\begin{alignedat}{3}
    &W_0\,T_0^{-\frac{d}{2}}\,(\varphi_0 - \varphi)^{\frac{d-1}{2}} + \ldots\,, \quad &&\varphi \rightarrow \varphi^{-}_{0}\,,\quad  &&\text{type III}\\
		&W_{\infty}^{\text{I}}\,T_{\infty}^{\text{I}}\,e^{-(d-1)\,b\varphi}+ \ldots\,, \quad &&\varphi \rightarrow \infty\,,\quad  &&\text{type I}
  \end{alignedat}\right.
  \label{eq:U_IR_expansion}
\end{equation}
and since $ d> 1 $, this always vanishes for type I and III solutions at the IR endpoint.

For type II solutions, the coefficient $T_{\infty}$ of $e^{2b\varphi}$ vanishes $T_{\infty}^{\text{II}} = 0$ \eqref{typeBleading} so that we must consider it separately.\footnote{The coefficient vanishes also for the type I solution at the confinement bound $b = b_{\text{c}}$ as can be seen from \eqref{typeCleading}, but we always assume that $b>b_{\text{c}}$ so we do not have to consider this case.} The explicit expansions for type II solutions are
\begin{equation}
	W(\varphi) = W_{\infty}^{\text{II}}\,e^{b\varphi} + \ldots\,, \quad T(\varphi) =
	\begin{dcases}		T_{\infty,1}^{\text{II}}\,e^{2\gamma_{\text{c}}\varphi} + \ldots\,, \quad &\gamma < \gamma_{\text{c}}\\
		T_{*}^{\text{II}}\,e^{2\gamma\varphi} + \ldots\,, \quad &\gamma > \gamma_{\text{c}}
	\end{dcases}\,, \quad \varphi \rightarrow \infty\,,
	\label{WTasymptoticsB}
\end{equation}
where $W_{\infty}^{\text{II}}$ is given in \eqref{typeBleading} while $T_{\infty,1}^{\text{II}}$ and $T_*^{\text{II}}$ are given in \eqref{eq:typeB_T_coefficients}. Therefore, for type II solutions, we obtain
\begin{equation}
	W(\varphi)\,T(\varphi)^{-\frac{d}{2}} =
	\begin{dcases}		W_{\infty}^{\text{II}}\,T_{\infty,1}^{\text{II}}\,e^{(b-d\gamma_{\text{c}})\,\varphi} + \ldots\,, \quad &0 < \gamma < \gamma_{\text{c}}\\
		W_{\infty}^{\text{II}}\,T_{*}^{\text{II}}\,e^{(b-d\gamma)\,\varphi} + \ldots\,, \quad & \gamma_{\text{c}} <\gamma < b
	\end{dcases}\,,\quad \varphi \rightarrow \infty\,, \quad \text{type II}\,.
 \label{eq:WtimesT_typeB}
\end{equation}
For \eqref{eq:WtimesT_typeB} to vanish at the IR endpoint $ \varphi \rightarrow \infty $ when $0 < \gamma < \gamma_{\text{c}}$, we must have $ b < d\gamma_{\text{c}}= b_{\text{G}}^2\slash b $ where $ b_{\text{G}} = \sqrt{\frac{d}{2\,(d-1)}} $. Therefore for $0 < \gamma < \gamma_{\text{c}}$ the IR boundary term vanishes if $b$ is below the Gubser bound $b< b_{\text{G}}$. In the second case $ \gamma_{\text{c}} <\gamma < b $, the vanishing of the IR term imposes the upper bound $ b < d\gamma $. This bound is automatically satisfied below the Gubser bound $b< b_{\text{G}}$ by the fact that $b < b_{\text{G}}^2\slash b = d\gamma_{\text{c}} < d\gamma$. Hence, the IR boundary term of type II solutions always vanishes below the Gubser bound $ b < b_{\text{G}} $. For $\gamma = \gamma_{\text{c}}$ the leading coefficient in the expansion of $T$ is $T_{\infty,1}^{\text{II}}+T_{*}^{\text{II}}$ and the conclusion of the analysis is the same.

We now look at the second IR term in \eqref{Freg1}. In Appendix \ref{reduced:typeA}, we show for type III solutions, the function $ U(\varphi) $ has the IR asymptotics
\begin{equation}
	U(\varphi) = \frac{\mathcal{U}}{(\varphi_{0}-\varphi)^{\frac{d}{2}-1}} + U_0\,\sqrt{\varphi_0-\varphi} + \ldots\,, \quad \varphi \rightarrow \varphi^{-}_{0}\,
\end{equation}
where $ \mathcal{U} $ is the integration constant of the equation \eqref{Uequation} and $ U_0 $ is given in \eqref{eq:typeA_U0}. Therefore, for type III solutions, the second IR term goes as
\begin{equation}
	U(\varphi)\,T(\varphi)^{-\frac{d}{2}+1} = \mathcal{U}\,T_0^{-\frac{d}{2}+1} + \ldots\,, \quad \varphi\rightarrow \varphi^{-}_{0}\,
\end{equation}
where ellipsis denote terms that go to zero as $ \varphi\rightarrow \varphi^{-}_{0} $. Hence, we observe  that the IR boundary term vanishes if and only if we set the integration constant to zero $ \mathcal{U} = 0 $ as found in \cite{Ghosh:2017big}.

As shown in Appendix \ref{app:solutionsunboundedscalar}, the function $ U(\varphi) $ behaves for type I and II solutions as \eqref{eq:typeBC_U_expansion} which we recall here for convenience
\begin{equation}
	U(\varphi) =
	\begin{dcases}
		\mathcal{U}\,e^{(d-2)\,\gamma_{\text{c}}\varphi} + U_{\infty}^{\text{II}}\,e^{-b\varphi} + \ldots\,, \quad &\text{type II}\\
		\mathcal{U}\,e^{(d-2)\,b\varphi}+ U_{\infty}^{\text{I}}\,e^{-b\varphi} + \ldots\,, \quad &\text{type I}\\
	\end{dcases}
	\,, \quad \varphi \rightarrow \infty\,.
\label{eq:typeBC_U_expansion_2}
\end{equation}
Combining with the asymptotics of $ T $ in \eqref{WTasymptoticsC},  we obtain  for type I solutions
\begin{equation}
	U(\varphi)\,T(\varphi)^{-\frac{d}{2}+1} = \mathcal{U}\,(T_{\infty}^{\text{I}})^{-\frac{d}{2}+1} + \ldots\,, \quad \varphi\rightarrow \infty\,, \quad \text{type I}\,,
\end{equation}
where ellipsis denote terms that vanish in the limit $ \varphi \rightarrow \infty $. For type II solutions,  we obtain  by using \eqref{WTasymptoticsB} that
\begin{equation}
	U(\varphi)\,T(\varphi)^{-\frac{d}{2}+1}=
	\begin{dcases}
		\mathcal{U}\,(T_{\infty,1}^{\text{II}})^{-\frac{d}{2}+1} + \ldots\,, \quad &0 < \gamma <\gamma_{\text{c}}\\
		\mathcal{U}\,(T_{*}^{\text{II}})^{-\frac{d}{2}+1}\,e^{-(d-2)(\gamma-\gamma_{\text{c}})\,\varphi} + \ldots\,, \quad & \gamma_{\text{c}} <\gamma < b
	\end{dcases},\quad \varphi \rightarrow \infty\,,
\end{equation}
where ellipsis denote terms that vanish in the limit $ \varphi \rightarrow \infty $. For $\gamma_{\text{c}} <\gamma < b$, this vanishes automatically, while for $0 < \gamma <\gamma_{\text{c}}$ it vanishes if we set $\mathcal{U} = 0$. Therefore, we observe  that for both type I and II solutions, the second IR boundary term vanishes under the choice $ \mathcal{U} = 0 $ required by type III solutions. To conclude, we have shown that the IR boundary term in the renormalized free energy vanishes when $b_{\text{c}} <b < b_{\text{G}}$ if we set $ \mathcal{U} = 0 $.

For completeness, we also evaluate the boundary terms for type 0 solutions. Using \eqref{eq:type0_WT_expansions}, we obtain
\begin{equation}
    W(\varphi)\,T(\varphi)^{-\frac{d}{2}} =W_{\infty}^{0}\,(T_{\infty}^{0})^{-\frac{d}{2}} \,e^{b_{\text{G}}\varphi}+\ldots\,,\quad \varphi\rightarrow \infty\,,
\end{equation}
which is always divergent. Therefore the on-shell action of type 0 solutions is infinite due to a divergent IR boundary term.

\paragraph{Holographic renormalization.}

We  now fix the integration constant appearing in $ U(\varphi) $ to zero $ \mathcal{U} = 0 $. As shown, then all IR boundary terms in the regularized free energy \eqref{Freg1} vanish and we are left with a pure UV boundary term
\begin{equation}
	F_{\text{reg}} = -M_{\text{p}}^{d-1}\,\widetilde{\Omega}_d\,\biggl( U(\varphi_{\UV})\,T(\varphi_{\UV})^{-\frac{d}{2}+1}+W(\varphi_{\UV})\,T(\varphi_{\UV})^{-\frac{d}{2}}\biggr)\ ,
	\label{Freg2}
\end{equation}
where $ \varphi_{\UV} = \varphi(\log{\epsilon}) \rightarrow 0 $ as $ \epsilon \rightarrow 0 $. The divergences must be renormalized by including the counterterm $ F_{\text{ct}} $ defined in \eqref{eq:Fcttext}. It involves three functions $ W_{\text{ct}} $, $ U_{\text{ct}} $, $ Y_{\text{ct}} $ that satisfy the first-order differential equations (in general $ d $)
\begin{align}
	(W_{\text{ct}}')^{2}-\frac{d}{2\,(d-1)}\,W_{\text{ct}}^{2} -2V&= 0\,,\label{eq:Wct}\\
	W_{\text{ct}}'\,U_{\text{ct}}'-\frac{d-2}{2\,(d-1)}\,W_{\text{ct}}\,U_{\text{ct}}+\frac{2}{d} &= 0\,,\label{eq:Uct}\\
	W_{\text{ct}}'\,Y_{\text{ct}}'-\frac{d-4}{2\,(d-1)}\,W_{\text{ct}}\,Y_{\text{ct}} - \frac{(d-2)^{2}}{4\,(d-1)}\,U_{\text{ct}}^{2}+\frac{d}{2}\,(U_{\text{ct}}')^{2}&=0\,.\label{eq:Yct}
\end{align}
For any $d$, the asymptotics of the first two counterterms are given by
\begin{align}
W_{\text{ct}}(\varphi) &= \frac{2(d-1)}{\ell}+\frac{\Delta_-}{2\ell}\,\varphi^{2} +\frac{\mathcal{C}_{\text{ct}}}{\ell}\,\varphi^{d\slash \Delta_-}+\ldots\,,\\
U_{\text{ct}}(\varphi) &= \ell\,\biggl(\frac{2}{d(d-2)}+\mathcal{B}_{\text{ct}}\,\varphi^{(d-2)\slash \Delta_-}+\ldots\biggr)\,,
\end{align}
where $\mathcal{C}_{\text{ct}} $, $\mathcal{B}_{\text{ct}} $ are integration constants (see also \cite{Ghosh:2020qsx}).\footnote{Note that the $U_{\text{ct}}$ in \cite{Ghosh:2020qsx} differs from ours by a factor of $d\slash 2$.} These are the same expansions as the expansions of $W$ and $U$ derived in Appendix \ref{app:UVasymptotics}, but with $\mathcal{R} = 0$. In $d = 4$, the third counterterm includes a logarithmic term in its expansion
\begin{equation}
    Y_{\text{ct}}(\varphi) = \ell^{3}\,\biggl(\frac{1}{48\Delta_{-}}\log{\varphi}+\mathcal{A}_{\text{ct}} + \ldots\biggr)\,,
\end{equation}
where $ \mathcal{A}_{\text{ct}}$ is an integration constant.

\addcontentsline{toc}{section}{References}
\bibliography{References.bib}
\bibliographystyle{JHEP}

\end{document}